\documentclass{aa}  

\usepackage{graphicx}
\usepackage{txfonts}
\usepackage{hyperref}
\hypersetup{colorlinks=true,urlcolor=black, citecolor=blue, linkcolor=black}

\usepackage{multirow}
\usepackage{todonotes}
\usepackage{soul}
\usepackage{comment}
\usepackage{orcidlink}
\usepackage{subcaption,booktabs}

\usepackage{natbib}

\bibpunct{(}{)}{;}{a}{}{,}

\definecolor{steelblue}{rgb}{0.274 0.510 0.706}

\defcitealias{Sana2012}{S12}
\defcitealias{Moe_DiStefano2017}{MDS17}

\begin{document} 

\title{Population synthesis of Be X-ray binaries in the Small Magellanic Cloud: Angular momentum recycling and stable mass transfer} 

    \titlerunning{BeXRBs in the SMC}

   \subtitle{}
\author{Víctor López Oller\inst{1}\orcidlink{0009-0009-8472-6872} \thanks{e-mail: \href{mailto:victor.lopez.physics@gmail.com}{victor.lopez.physics@gmail.com}}, Boyuan Liu\inst{1, 2}\orcidlink{0000-0002-4966-7450} \thanks{e-mail: \href{mailto:boyuan.liu.astro@gmail.com}{boyuan.liu.astro@gmail.com}}, Michela Mapelli
\inst{1,3,4,5,6}\orcidlink{0000-0001-8799-2548}, Stefano Rinaldi \inst{1}\orcidlink{0000-0001-5799-4155}, Cecilia Sgalletta\inst{1}\orcidlink{0009-0003-7951-4820}, Julia\,Bodensteiner\inst{7}\orcidlink{0000-0002-9552-7010}, Giuliano Iorio\inst{8}\orcidlink{0000-0003-0293-503X}, Rebekka Schupp \inst{1}\orcidlink{0009-0007-1546-4157}}
    \authorrunning{López et al.}
    
    \institute{
    $^{1}$Universit\"at Heidelberg, Zentrum f\"ur Astronomie (ZAH), Institut f\"ur Theoretische Astrophysik, Albert-Ueberle-Str. 2, 69120, Heidelberg, Germany\\
    $^{2}$Institute of Astronomy, University of Cambridge, Madingley Road, Cambridge, CB3 0HA, UK\\
    $^{3}$Physics and Astronomy Department Galileo Galilei, University of Padova, Vicolo dell'Osservatorio 3, I--35122, Padova, Italy\\
    $^{4}$Gran Sasso Science Institute (GSSI), Viale Francesco Crispi 7, 67100 L’Aquila, Italy\\
    $^{5}$Universit\"at Heidelberg, Interdisziplin\"ares Zentrum f\"ur Wissenschaftliches Rechnen, Heidelberg, Germany\\
    $^{6}$INFN, Sezione di Padova, Via Marzolo 8, I--35131 Padova, Italy\\
    $^{7}$Anton Pannekoek Institute for Astronomy, University of Amsterdam, Science Park 904, 1098 XH Amsterdam, The Netherlands\\
    $^{8}$Institut de Ciències del Cosmos (ICCUB), Universitat de Barcelona (UB), c. Martí i Franquès 1, 08028 Barcelona, Spain\\
    }

   \date{Received 04/05/2026; accepted 31/07/2026}

% \abstract{}{}{}{}{} 
% 5 {} token are mandatory
 
\abstract
{Be X-ray binaries (BeXRBs) are key laboratories for constraining binary interaction processes such as mass transfer, angular-momentum transport, and  natal kicks. The Small Magellanic Cloud (SMC), hosting a nearly complete and well-characterized BeXRB population, offers a unique opportunity to test these physical processes at low metallicity. We aim to identify the combination of binary-evolution parameters that simultaneously reproduces the observed number and the joint distribution of orbital period and optical magnitude of SMC BeXRBs. We performed an extensive grid analysis of binary population-synthesis models exploring different mass transfer efficiencies, angular-momentum transport prescriptions and Roche-lobe overflow stability criteria. We also considered the impact of natal kicks, and that of the propeller effect of rotating magnetic fields of neutron stars. Synthetic populations obtained with the binary population synthesis code \textsc{sevn} were statistically compared to observations using likelihood-based methods applied to the orbital period and $V$-band magnitude distributions, together with requirements on the total number of systems. We find that models in which mass transfer via Roche-lobe overflow is assumed to be always stable and angular momentum is recycled back into the orbit through tides when the accretor approaches critical rotation provide the best match to observations.
Our best-fitting models favor low  natal kicks ($\lesssim 100\ \rm km\ s^{-1}$), a moderate mass transfer efficiency ($f_{\rm MT} \simeq 0.6$), a minimum Be threshold spin close to critical rotation, and a strong suppression of accretion onto neutron stars due to the propeller effect. Specifically, the observable population is highly sensitive to the treatment of the propeller effect, which regulates the X-ray luminosity of wide, low-accretion-rate systems. 
Our results have direct implications for the evolutionary pathways leading to  binary compact objects.}

   \maketitle

\section{Introduction}\label{sec:introduction}

The publication of the fourth Gravitational-Wave Transient Catalog \citep[GWTC-4.0,][]{Abac2025} has once again underscored the need for a deeper and more quantitative understanding of the physics governing the evolution of massive binary systems (see e.g., \citealp{Marchant2024} for a recent review). The formation pathways of compact-object binaries, particularly those involving neutron stars (NSs) and black holes, are regulated by a sequence of poorly constrained processes (see e.g., \citealp{Mapelli2021} for a  review), including non-conservative mass transfer (MT), the transport and loss of angular momentum (AM), common-envelope (CE) evolution, and the physics of core-collapse supernovae  \citep[SNe,][]{Postnov2014, Kruckow2016, Schneider2021, Bavera2021}. Each of these processes leaves measurable imprints on the orbital architecture, mass distribution, and spin evolution of the binary system, and collectively they determine the likelihood that a massive binary emerges as a gravitational wave source \citep{Santoliquido2021, Mandel2022, Boesky2024}. As a result, constraining the internal physics of binary evolution at earlier evolutionary processes is essential for interpreting present and future gravitational wave detections.

In this context, Be X-ray binaries (BeXRBs) constitute a uniquely informative population for probing the physical mechanisms shaping massive binaries \citep[see e.g.,][]{Vinciguerra2020}. In the classical picture, a BeXRB consists of a rapidly rotating B-type star surrounded by a viscous decretion disk (VDD) and an accreting compact object (\citealp{Negueruela1998, Okazaki2001}; see \citealp{Reig2011} for a review). These systems are expected to arise from binary stars that have undergone stable Roche-lobe overflow (RLOF), during which AM transport spins up the accretor to near-critical rotation \citep{Pols1991,Shao2014,Hastings2021,Chaty2022}. 
This MT-induced spin-up  allows for the formation of a VDD that emits Balmer emission lines, thereby turning the accretor into a Be star. 
If the former donor star subsequently collapses into a compact object and the SN explosion does not unbind the binary, the resulting system can become a BeXRB, in which the VDD extends beyond the Roche lobe while the star underfills it  \citep{Reig2011}. The  properties of BeXRBs are thus extremely sensitive to the physics of MT  (e.g., accretion efficiency, MT stability, AM transport) and  SN explosions (e.g., mass ejection and natal kicks). 

The Small Magellanic Cloud (SMC) is particularly valuable 
for the study of BeXRBs. Owing to its low metallicity, relatively well constrained star formation history (SFH), and relative proximity, the SMC hosts the richest known population of BeXRBs, with 102 candidates \citep{Coe2015,Haberl16}. Extensive surveys have produced catalogs that are both homogeneous and nearly complete (but new candidates have recently been proposed, e.g., \citealt{Gaudin25}). A significant fraction ($\sim 46\%$) of these systems have measured orbital periods, $P_{\rm orb}$, mostly based on their long-term light curves in X-rays and the optical, and for the majority ($\sim 97\%$), optical magnitude or color-magnitude information is available \citep{Haberl16}. Moreover, the limited spatial extent and relatively uniform chemical composition of the SMC allow the stellar population to be approximately modeled with a single representative metallicity \citep{Rubele2015, Choudhury2018}, unlike the Milky Way (MW) 
with its complex star formation and merger history \citep{Douna2015, Lian2023}. 
This homogeneity significantly reduces degeneracies when exploring the impact of metallicity-dependent processes such as stellar winds, AM evolution, and the stability of MT. As a result, constraints derived from the SMC BeXRB population can be more directly interpreted in terms of underlying binary physics. 

Previous population-synthesis studies \citep[e.g.,][]{Zhang2004,Belczynski2009,Linden2009,Shao2014,Shao2020,Zuo2014,Vinciguerra2020,Mondal2020,Xing2021,Misra2023b,Romero2023,Liu2023,Rocha2024,Schurmann2025} have succeeded in generating synthetic BeXRB populations that are qualitatively consistent with observed catalogs. These works have typically used $P_{\rm orb}$ and optical magnitude, $m_V$, as the main comparison quantities in the SMC \citep[see e.g.,][]{Vinciguerra2020,Schurmann2025}, whereas in the MW the availability of dynamical and spectroscopic constrains has allowed analysis based on the mass of the Be star, $M_{\star}$, and the post-SN eccentricity, $e$ \citep[see e.g.,][]{Rocha2024}. Although these comparisons have provided important insights such as the preference for stable MT or the need for moderate SN natal kicks \citep{Vinciguerra2020,Xing2021,Rocha2024,Schurmann2025}, most studies have relied on visual or qualitative assessments of similarity between models and observations. As a consequence, the parameter space of MT efficiency, $f_{\rm MT}$, AM loss prescriptions, and initial binary conditions remains only loosely constrained.

In this work, we aim to advance beyond qualitative comparisons 
by developing a quantitative statistical methodology to evaluate the ability of different binary-evolution models to reproduce the observed SMC BeXRB population. We construct a multidimensional likelihood framework that simultaneously compares the synthetic and observed distributions of $P_{\rm orb}$ and $m_V$, the latter computed for each synthetic system using metallicity and effective temperature-appropriate bolometric corrections and extinction maps. By ranking models according to their likelihood, we are able to identify which combinations of physical parameters 
are statistically favored by the observations. 

 The model grid explored in this work spans the key physical processes that shape the progenitor evolution of BeXRBs as shown by previous studies \citep[see e.g.,][]{Vinciguerra2020}. We vary: the accretion efficiency, $f_{\rm MT}$, which controls the fraction of material accreted by the companion during stable RLOF; the mode of AM loss associated with non-conservative MT, considering prescriptions ranging from Jeans-like mass ejection from the donor to isotropic re-emission from the accretor and circumbinary outflows \citep{Tauris2023}; and the treatment of AM feedback, which governs whether the transferred AM is recycled back into the orbit or simply removed from the system if the accretor is about to be overspun. We also examine different criteria for the stability of RLOF, determining whether MT always proceeds stably or can lead to a CE phase under certain criteria. In addition, we consider two  sets of initial conditions, 
 which differ in the underlying distributions of primary masses, mass ratios, and orbital separations. 
 We also investigate the effects of natal kicks and NS propeller effect on BeXRB properties. 
 
Here, we only consider BeXRBs with accreting NSs, because no BeXRBs with an accreting black hole have been confirmed to date and a previous candidate has recently been ruled out \citep{Muller2026}. Moreover, although black hole high-mass X-ray binaries are observed in nearby galaxies, no confirmed systems of this type have been identified in the SMC. 
As an additional test of model robustness, we check whether 
the models that best match SMC BeXRBs are also capable of reproducing the observed BeXRB population in the MW, where 74 BeXRB candidates are observed from a total of 152 high-mass X-ray binaries \citep{Fortin23}. 

This paper is structured as follows. Section~\ref{sec:methods} describes the methodology; namely, our population-synthesis simulations (Section~\ref{sec:sevn}) and the available observational constraints (Section~\ref{sec:obs_constrains}). 
Section~\ref{sec:results} presents the main results for the SMC, and Section~\ref{sec:discussion} discusses their physical implications. Section~\ref{sec:summary} summarizes the main conclusions and outlines remaining caveats. 

\section{Methods}\label{sec:methods}

\subsection{Population synthesis} \label{sec:sevn}

To obtain our synthetic population of BeXRBs, we used \textsc{sevn} \citep{Spera2017,Spera2019,Mapelli2020,Iorio2023}, a binary population synthesis code that simulates the coevolution of large populations of binary stars by interpolation of precomputed single stellar evolutionary tracks together with semi-analytic prescriptions to model binary processes such as tides, MT, CE evolution, and SN explosions. For the present study, we used stellar tracks representative of SMC metallicity ($Z=0.0035$, \citealp{Davies2015}; \citealp{Costa2025}). We refer to \cite{Iorio2023} for a complete description of the code.

\subsubsection{Treatment of BeXRBs in \textsc{sevn}} \label{sec:sevn_BeXRBs}

We extend \textsc{sevn} to study BeXRBs based on \citet{Liu2023}. The BeXRB model consists of two aspects: a criterion to decide when a binary can be classified as a BeXRB and a reliable mapping from NS accretion rates and Be star properties to X-ray and optical observables.

To classify a simulated binary as a BeXRB, we require that it satisfies the following conditions \citep{Liu2023}:

\begin{enumerate}
    \item The binary is made of an O/B main-sequence star with $M_{\star}>3\, \mathrm{M_\odot}$ and a NS; 
    systems containing white dwarf accretors are excluded from the BeXRB sample because they are intrinsically faint, and therefore observationally rare \citep{Coe2020,Kennea2021,Gaudin2024}. While Be stars in observed BeXRBs have minimum masses of $\sim 6\, \mathrm{M_\odot}$ \citep{Hohle2010,Fortin23}, isolated classical Be stars have a lower minimum mass of $\sim 3\, \mathrm{M_\odot}$ \citep{Vieira2017}. We therefore adopt a minimum Be star mass of $3\, \mathrm{M_\odot}$, thereby allowing the Be phenomenon to occur over the full empirically established mass range. Within this approach, we expect the higher minimum mass observed in BeXRB systems to arise naturally from binary evolution, without the need to invoke additional or ad hoc interaction mechanisms to reconcile the two regimes. In contrast to previous studies \citep[e.g.,][]{Misra2023b,Liu2023} that impose a higher mass threshold by construction, our choice treats this constraint as an emergent property of the population, providing 
    additional constraints on the underlying binary-evolution physics. Although high-mass X-ray binaries with accreting black holes have been observed \citep{Tetarenko2016,Orosz2007}, there are no confirmed black hole BeXRBs in observations, possibly explained by stronger VDD truncation or CE mergers \citep{Belczynski2009}. Moreover, the formation of such systems is highly uncertain, since poorly understood processes are involved in their formation, such as the black hole natal kick \citep[see e.g.,][]{Vigna2025,Nagarajan2025}. For this reason, during this work we only consider BeXRBs with accreting NSs.
    \item The donor must support a VDD. We adopted the following proxy for the presence of a VDD. First, the donor must rotate sufficiently fast, with a dimensionless spin, $\omega = \Omega_{\rm rot}/\Omega_{\rm crit}>\omega_{\rm min}$, where $\Omega_{\rm rot}$ is the angular rotation rate, $\Omega_{\rm crit}$ is the critical angular rotation rate, and $\omega_{\rm min}$ is the threshold spin for 
    the formation of a VDD, which is treated as a free parameter that is adjusted to match the observed number of BeXRBs (see Section~\ref{sec:obs_constrains}). Second, the orbital period must be larger than one day; shorter orbits hinder disk formation through tidal truncation \citep{Panoglou2016, Panoglou2018,Rivinius2019}. Almost all of the $\sim 200$ detected BeXRBs so far have $P_{\rm orb}>10$ days \citep{Raguzova2005, Coe2015,Haberl16,Antoniou2016,Fortin23,Kaltenbrunner2026}.
    \item The donor itself must not fill its Roche lobe  at periastron, since systems undergoing RLOF are considered RLOF X-ray binaries \citep{Reig2011}.
\end{enumerate}

Regarding the observable X-ray luminosity, we first followed the prescription in \citet{Liu2023} to calculate the peak accretion rate of a NS from the VDD of the Be star given the NS mass, binary orbital parameters, and the empirical relation between VDD density and Be stellar mass from observations of classical Be stars in the MW. We refer to \citet{Liu2023} for further details.
The bolometric luminosity was then computed as $L_{\rm bol}=\epsilon\dot{M}_{\rm acc}c^2$, assuming a canonical radiative efficiency $\epsilon=0.2$. To obtain the observable X-ray luminosity, we applied a single bolometric correction factor, $f_{\rm X}$, appropriate for the typical energy range of X-ray observations, such that $L_{\rm X}=f_{\rm X}L_{\rm bol}$ with $f_{\rm X}=0.5$ for our default model. 

This adopted Be star accretion prescription is further justified empirically, since it can reproduce the observed $L_{\rm X}$–$P_{\rm orb}$ relation for BeXRBs in the MW \citep{Dai2006, Raguzova2005, Liu2023}. 
However, as highlighted by \citet{Nova2025}, VDDs around Be stars are variable structures, so it is very difficult to model the duty cycles of BeXRBs considering other sources of variability such as accretion physics, disk wrapping and truncation, and the intrinsic variability of the Be star due to non-radial pulsations \citep{Osaki1986, Baade1988, Semaan2018}. The variability can introduce errors in the calculation of the peak X-ray luminosity. Such errors can be absorbed into the bolometric correction factor, $f_{\rm X}$. In order to account for the variability, we considered different values of this correction factor in the range $f_{\rm X}\in [0.05,1]$ for select models and found that our results are unaffected by the choice of $f_{\rm X}$. This indicates that our results are insensitive to VDD variability and that most simulated BeXRBs are luminous enough to be detected with $L_{\rm bol} \gtrsim 2\times{}10^{35}$ erg/s (see Section~\ref{sec:obs_constrains}). We do not consider the duty cycles of BeXRBs in our analysis, since the SMC sample of BeXRBs is obtained from long-term observations rather than one snapshot.

BeXRBs are promising candidates 
to evolve into ultra-luminous X-ray sources \citep[with $L_{\rm X}\gtrsim 10^{39}\ \rm erg\ s^{-1}$, e.g.,][]{Kaaret2017,Fabrika2021,King2023}, 
as shown by previous studies \citep[e.g.,][]{Marchant2017,Wiktorowicz2017,Wiktorowicz2019,Wiktorowicz2021,Shao2019,Shao2020,Abdusalam2020,Misra2020,Mondal2020,Kuranov2021,Liu2023}. We capped the accretion rate at 100 times the Eddington limit, where the adopted super-Eddington factor was calibrated to reproduce the observed high-end luminosities of high-mass X-ray binaries, reaching up to $\sim 10^{41}\ \mathrm{erg\ s^{-1}}$ \citep{Gilfanov2022}. 
As all X-ray binaries, BeXRBs 
might be subject to beaming effects during super-Eddington accretion,  
reducing their detectability 
and potentially biasing our analysis
\citep{King2001,King2009, King2017,King2019,King2023,King2024,Mao2026}. In order to quantify the impact of this effect, we considered the beaming models by \citet{Mao2026} for selected cases and found that the beaming effect only affects $\lesssim 5\%$ of all simulated BeXRBs with negligible differences in our results, which shows that most BeXRBs in our simulations have $L_{\rm X} < 10^{38}$ erg s$^{-1}$. Since there is also a lack of features around the Eddington limits of NSs in the observed luminosity function \citep{Gilfanov2022}, we ignored the beaming effect in general.

For comparison with optical observations, we converted the effective temperature, $T_{\rm eff}$, and luminosity of the simulated Be star to $V$-band magnitudes, $m_V$, using bolometric-correction tables:  
\begin{equation}
m_V=M_{\rm bol, \odot}-2.5\log(L/\text{L}_\odot)-BC_{V}\left( T_{\rm eff},\rm [Fe/H] \right)+MD+A_V,
\end{equation}
where $M_{\rm bol, \odot}=4.74$ is the solar absolute bolometric magnitude \citep{Schootemeijer2021}, $L$ is the luminosity of the star, $MD=18.9$ is the average distance modulus of the SMC \citep{Graczyk2020}, $A_V=0.53$ is the extinction in the $V$-band regarding the SMC \citep{Rubele2015} and $BC_V$ is the bolometric correction in the $V$-band as a function of $T_{\rm eff}$ and the metallicity throughout the stellar iron abundance relative to solar $\rm [Fe/H]$. The $BC_V$ was taken from the MIST \citep{Dotter2016,Choi2016} website\footnote{\url{https://waps.cfa.harvard.edu/MIST/model_grids.html}, where we took the $BC_V$ table with the label `UBV(RI)c + 2MASS + Kepler + Hipparcos + Tycho + Gaia'.} and we considered $\rm [Fe/H]=-0.75$ for the SMC \citep{Rubele2015}. We ignore the effects of the VDD on the magnitude, which are small ($\Delta m_V\lesssim 0.7$) according to observations of classical Be stars \citep[][]{deWit2006}.

We verified that our conclusions are insensitive to reasonable variations in the adopted bolometric corrections and extinction prescriptions, as the resulting changes in the predicted $m_V$ distributions are negligible for the statistical comparisons presented below. For instance, we reanalyzed select models adopting an extinction value of $A_V = 0.35$ \citep{Schootemeijer2021} and a stellar iron abundance of $\rm [Fe/H]=-0.5$. The lower extinction simply produces a uniform shift of the $m_V$ distribution, while the change in Fe abundance has a negligible impact on the overall properties of the simulated population, 
supporting our choice of adopting a single metallicity for the SMC, despite the age–metallicity relation reported by \citet{Rubele2015}.

\subsubsection{Initial binary conditions}\label{sec:IC}

The initial binary population is defined by a set of initial conditions assigning probability distributions to the primary mass, mass ratio $q$, orbital period, and eccentricity. Such probability distribution functions are  inferred from observations. In our simulations, primary stellar masses, $M_1$, were randomly drawn from the \citet{Kroupa2001} initial mass function (IMF) over the range $5-100$ M$_\odot$, and we only considered secondary stars with $M_2\ge 2\ \rm M_\odot$ . We restricted our analysis to systems with $M_1\geq5$ M$_\odot$, since only sufficiently massive primaries can evolve into NSs, and thus act as progenitors of BeXRBs; this choice is supported by numerical tests showing that lower-mass primaries do not contribute to the BeXRB population. 

To assess the sensitivity of our results to empirical assumptions about binary properties, we adopted two alternative prescriptions for the mass ratio and orbital parameter distributions. The first follows the observationally motivated, independent distributions reported by \citet{Sana2012} for Galactic massive stars (hereafter \citetalias{Sana2012}), while the second is based on the correlated, multi-dataset parametrization of \citet{Moe_DiStefano2017} (hereafter \citetalias{Moe_DiStefano2017}).
For each combination of physical assumptions, we evolved $10^{6}$ binaries in order to achieve adequate statistics in the rare BeXRB regime.

For \citetalias{Sana2012}, we assumed that the whole stellar population (in terms of mass) is composed of $90\%$ binary stars and $10\%$ single stars. This binary fraction was derived by extrapolating the power-law fit to the orbital period distribution function of the binaries considered by \citet[]{Sana2012} with a binary fraction of 70\% in the range $P_{\rm orb}\in [10^{0.15},10^{3.5}]\ \rm days$ to $P_{\rm orb}\in [10^{0.15},10^{5.5}]\ \rm days$. We checked that considering the original range changes the total number of BeXRBs by $\lesssim10\%$ with little effect on the predicted distributions so we considered this extrapolation for higher $P_{\rm orb}$ in order to also study the influence of wide orbits on the BeXRB population. For the underlying stellar population, we assumed that both single stars and the primary components of binaries are drawn from the \citet{Kroupa2001} IMF but in the range $M_1\in [0.01,100]\ \rm M_{\odot}$. Although this prescription does not yield an exactly \citet{Kroupa2001} IMF when secondaries are included, the resulting deviations are negligible for the mass range relevant to BeXRBs \citep[$ \gtrsim 1\, \rm M_\odot, $][]{Liu2023}. Under these assumptions, we estimate that the total stellar mass represented by our binary sample corresponds to a fraction, $f_{\rm imf}\simeq0.24$,
of the total mass of the full stellar population, following the formalism of Appendix A in \citet{Bavera2020}. For \citetalias{Moe_DiStefano2017}, we derived $f_{\rm imf}$ by integrating the mass and orbital period-dependent binary (mass) fraction, which gives an almost identical value of $f_{\rm imf}$. This factor, $f_{\rm imf}$, is used to normalize the simulated number of BeXRBs \citep{Liu2023}.

Finally, the initial stellar rotation rate constitutes an additional initial condition parameter of particular relevance for BeXRB formation \citep{Liu2023,Schurmann2025}. 
Rapid rotation is a prerequisite for the formation of OBe stars, and therefore plays a central role in our identification of BeXRBs (see Section~\ref{sec:sevn_BeXRBs}). As a consequence, stars with higher initial rotation rates are expected to have an enhanced probability of entering the Be phase. We adopted the empirical prescription for the initial rotation velocity, $v_{\rm rot,0}$, from \citet[see their Sec.~7.2]{Hurley2000}, which is based on main-sequence observations compiled by \citet{Lang1992},
\begin{align}\label{eq:v_rot_0}
v_{\rm rot,0}=330\ {\rm km\ s^{-1}}\,(M/{\rm M_\odot})^{3.3}\,\left[15+(M/{\rm M_\odot})^{3.45}\right]^{-1}\ ,
\end{align}
where $M$ is the initial stellar mass. 

To assess the impact of the initial rotational configuration, we further explored two alternative prescriptions for select models. In the first case, we assumed that the initial stellar spins are synchronized with the orbital motion \citep[as in e.g.,][]{Langer2020, Schurmann2025}, mimicking efficient tidal coupling, predicted to be achieved during the evolution toward RLOF \citep{Zahn1977}. In the second, more extreme case, we set $v_{\rm rot,0}=0$, corresponding to initially non-rotating stars. This latter prescription provides a conservative lower limit on the formation efficiency of BeXRBs, as it suppresses the direct contribution of primordial rotation to the Be phenomenon.

\subsubsection{MT efficiency} \label{sec:MT_efficiency}

\begin{table*}
\caption{Summary of the parameters explored in the model grid.}
\centering
\label{tab:parameters_grid}
\begin{tabular}{l l l l}
\hline
Parameter & Values & Description & Detailed Section \\
\hline
$f_{\rm MT}$ 
& 0.2, 0.4, 0.6, 0.8, 1.0 \tablefootmark{a}
& MT efficiency during RLOF\tablefootmark
& Section~\ref{sec:MT_efficiency} \\

AM loss 
& Jeans-mode, isotropic re-emission (ISO),
& AM loss treatment for non-accreted mass
& Section~\ref{sec:AM_loss} \\
& circumbinary ring (CIRC) & during RLOF MT & \\

AM feedback 
& Recycle of AM excess (RAM), 
& Treatment of surplus AM of critically
& Section~\ref{sec:AM_feedback}\\
& elimination of AM excess (EAM) & rotating accretors & \\

RLOF stability 
& Always-stable MT (ASMT), 
& RLOF MT stability criterion
& Section~\ref{sec:RLOF_stability}\\
& modified Hurley $q_c$-criteria (Hurley) & & \\

\hline
\end{tabular}

\tablefoot{
\tablefoottext{a}{$f_{\rm MT}=1$ corresponds to conservative MT. For models assuming efficient AM recycling and always-stable RLOF, additional values, $f_{\rm MT}=0.1, 0.3, 0.5,$ and $0.7$, are also explored. See Section~\ref{sec:results}}.

}
\end{table*}

We systematically varied the key binary-physics parameters that regulate the formation and observable properties of BeXRBs, as summarized in Table~\ref{tab:parameters_grid}. One of the most important of them is the accretion efficiency, $f_{\rm MT}$, defined as the fraction of the mass lost by the donor star that is ultimately accreted by the companion. The value of $f_{\rm MT}$ directly controls both the degree of orbital evolution during RLOF and the amount of mass and AM deposited onto the mass gainer, thereby influencing its rotational state and its likelihood of entering a Be phase. As such, $f_{\rm MT}$ represents a central parameter linking binary interaction physics to the emergence of BeXRB populations.

Two physically motivated prescriptions are commonly adopted to model MT efficiency. In rotationally limited models, accretion is curtailed once the mass gainer approaches critical rotation, leading to intrinsically low values of $f_{\rm MT}$ \citep{Packet1981, Langer2003, Petrovic2005a, Ghodla2023}. By contrast, thermally limited models restrict accretion when the thermal timescales of the donor and accretor differ substantially, typically capping the accretion rate at a fraction of the accretor’s thermal rate \citep{Paczynki1972, Tout1997, Hurley2002}. However, both approaches appear to be in tension with systems in which rapid rotation is empirically required, such as Be–sdOB binaries, consisting of a rapidly rotating Be star and a stripped hot subdwarf companion \citep{Lechien2025,Bao2025}. Motivated by this (see also \citealt{Vinciguerra2020}) and by the intrinsic similarity of the systems considered here, we therefore adopted a phenomenological approach in which $f_{\rm MT}$ is treated as a constant parameter. We explored a broad range spanning the conservative limit, $f_{\rm MT}=1$, down to highly non-conservative regimes with $f_{\rm MT}=0.1$. Our primary model grid samples $f_{\rm MT} \in \left( 1.0,\,0.8,\,0.6,\,0.4,\,0.2 \right)$, while additional values, $f_{\rm MT}=0.1, 0.3, 0.5,$ and $0.7$, are considered for select models to better capture the effects of $f_{\rm MT}$. 

\subsubsection{AM loss treatment}\label{sec:AM_loss}

The second key parameter governs the specific AM removed from the system by mass that is not accreted, and is therefore only relevant in non-conservative MT regimes $(f_{\rm MT}<1)$. We considered three physically distinct prescriptions for AM loss. In the Jeans-mode (hereafter Jeans AML), the expelled material carries away the specific AM of the donor star. In the isotropic re-emission scenario (hereafter ISO AML), %\citep[hereafter ISO AML;][]{Hurley2002}, 
mass is assumed to be first transferred to the accretor and subsequently re-emitted isotropically from its vicinity, thereby removing the specific AM of the accretor. Finally, in the circumbinary ring prescription (hereafter CIRC AML), the non-accreted mass is deposited into a circumbinary disk or ring; we assumed the ring to reside at a fixed radius equal to twice the binary semi-major axis, following \citet{Artymowicz1994}. Both the MT efficiency and the adopted AM loss prescription have been shown to strongly affect the orbital evolution and the resulting properties of BeXRB populations \citep[e.g.,][]{Vinciguerra2020}.

\subsubsection{AM feedback from critically rotating accretors}\label{sec:AM_feedback}

During MT phases, the AM conveyed from the donor to the accretor can, in principle, spin the accretor up to critical rotation \citep{Deschamps2013}, although recent studies conclude that this is not always the case \citep[e.g.,][]{Wang2026}. In \textsc{sevn}, we assume that the in-falling material carries the specific AM of a Keplerian accretion disk, such that the spin-up process is typically very efficient. How mass accretion and AM transport proceed when the accretor approaches critical rotation is still in debate. 

One widely used approach is to limit the accretion by self-consistently reducing the accretion rate to avoid over spinning the accretor \citep{Shao2014,Paxton2015} and/or enhancing stellar winds from the accretor that carry away AM \citep[][]{Langer1998}. However, this approach effectively reduces MT efficiency and tends to overproduce the abundance of low-mass ($\lesssim 6\ \rm M_\odot$) Be stars in BeXRBs \citep{Shao2014,Rocha2024,Schurmann2025,Xu2025}. Besides, as mentioned above, it has been found recently that reproducing the observed Be+sdOB star binaries requires a high MT efficiency ($\sim 50\%$) that disfavors rotation-limited accretion \citep{Bao2025,Lechien2025,Xing2026}. In light of this, we do not limit the mass accretion by any means (beyond that controlled by $f_{\rm MT}$) when the accretor approaches critical rotation. The physical interpretation is that the surplus AM of the infalling material can be transported back to the circumstellar disk via viscous stresses \citep{Popham1991,Paczynski1991}. 

The remaining question is how the disk, while receiving the surplus AM, interacts with the binary. Here, we consider two extreme regimes to bracket the possible outcomes:
\begin{itemize}
    \item Elimination of AM (EAM): The disk does not interact with the binary, and the surplus AM is eventually eliminated or ejected from the system without any orbital response. This is the default choice in \textsc{sevn}, and it is effectively used in binary-evolution models that do not track stellar spins. 
    \item Recycle of AM (RAM): The disk (or the streams of MT in general) efficiently interacts with the binary through tides, such that the surplus AM is entirely recycled to the orbit. This scheme has been adopted by \citet[see their appendix~A]{Liu2023}, leading to a good qualitative agreement with the observed orbital period distribution of BeXRBs in the SMC.
\end{itemize}
Distinguishing between these two regimes is particularly relevant for BeXRBs, as their formation requires both efficient spin-up of the secondary to near-critical rotation and orbital configurations that allow the system to survive the SN and enter a long-lived X-ray emitting phase \citep{Reig2011}. 

In addition to the circumstellar-disk-binary tidal interactions during RLOF considered in the RAM model, we also modeled tidal interactions between the two stars using the equilibrium tide formalism \citet{Hut1981}, as implemented in \textsc{sevn} by \citet[see their sec.~2.3.4]{Iorio2023} following \citet{Hurley2002}. This model accounts for the direct exchange of AM between stellar rotation and the orbit, driving synchronization and circularization on timescales that depend on the stellar structure and orbital separation. Tides are relevant for BeXRBs because Be stars are defined as fast rotators. In close systems, efficient tidal coupling can subsequently remove AM from the star and transfer it to the orbit, reducing the stellar spin and circularizing the orbit. The tidal spin-down therefore acts in competition with MT-induced spin-up and might play an important role in shaping the properties of Be stars in our models. However, 
the overall effect of tides on the simulated BeXRB population is small. For instance, in our best-fitting model of SMC BeXRBs (see Sec.~\ref{sec:results}), completely turning off stellar tides only decreases the total number of BeXRBs by $\sim 5\%$ and hardly modifies their distribution in the $P_{\rm orb}-m_V$ space.

\subsubsection{RLOF stability}\label{sec:RLOF_stability}

We explore two alternative prescriptions for the stability of MT phases. The first follows a formalism commonly adopted in binary population synthesis studies (e.g., \citealt{Hurley2002}), %; \citealt{Hobbs2005}), 
in which the donor-to-accretor mass ratio, $q\equiv M_{\rm d}/M_{\rm a}$, is compared to a critical threshold, $q_{\rm c}$, which depends on the evolutionary stage of the donor. If $q > q_{\rm c}$, MT is assumed to become dynamically unstable, possibly triggering a CE phase. In our implementation, we assumed that MT from main-sequence and Hertzsprung-gap donors is always stable, while $q < q_{\rm c}$ is required for stable MT from donors in other phases (given phase-dependent values of $q_{\rm c}$), following the fiducial option of \citet{Iorio2023} (see the QCRS model in their Table~3). 
We refer to this prescription as the (modified) Hurley model.

In addition, we considered an extreme, maximally permissive scenario in which MT is assumed to remain always stable whenever it begins in the evolutionary phases relevant for BeXRB formation, irrespective of the mass ratio. This prescription, denoted ASMT, provides an upper limit to the efficiency with which binaries can undergo prolonged mass exchange and spin-up the secondary to near-critical rotation. Note that in \textsc{sevn}, MT is unstable whenever both stars fill the Roche lobe or the Roche lobe of the donor becomes smaller than its core \citep[see their sec. 2.3.1]{Iorio2023}, which applies to both the %Hurley 
\citet{Hurley2002} and ASMT models considered here.

Systems that undergo unstable MT and enter a CE phase are assumed to merge and do not produce post-CE binaries in our synthetic catalog. In practice, we use an extremely small value, $\alpha_{\rm CE}=10^{-20}$, for the CE ejection efficiency, such that the CE phase always extracts enough orbital energy to trigger mergers. This choice is motivated by the fact that previous population-synthesis studies have found that CE evolution is hardly involved in reproducing the observed BeXRBs \citep{Shao2014,Vinciguerra2020,Rocha2024,Schurmann2025}. The same assumption is also used in the recent work by \citet{Xu2025,Schurmann2025} for binary population synthesis of the SMC. 

\subsubsection{NS evolution regulated by accretion}\label{sec:NS_accretion}

In addition to the binary interaction processes systematically explored by our model grid (Table~\ref{tab:parameters_grid}), the observable properties of BeXRBs are also affected by the physics of NS accretion and SNe. Below, we introduce our treatment for NS evolution driven by accretion in \textsc{sevn}, and the SN model is described in the following subsection.

Accretion in interacting binaries plays a central role in shaping the spin and magnetic field evolution of NSs, particularly during RLOF or BeXRB phases \citep{Marchant2024}. The interactions between the accretion flow and the NS magnetic field also leads to self-regulation of the accretion process, which can explain the observed bimodal spin distribution of NSs in BeXRBs \citep[e.g.,][]{Xu2019}. We adopted the semi-analytical model for accretion-regulated NS evolution implemented in \textsc{sevn} by \citet{Cecilia2023}. Here, we have further included the spin-up by accretion from winds and VDDs, while \citet{Cecilia2023} only considered NS accretion during RLOF. Below, we briefly discuss the key physical elements of this model, and the reader is referred to sec.~2.3.2 of \citet{Cecilia2023} for details. 

As material is transferred through an accretion flow, part of its AM is channeled to the NS, leading to spin-up when the accreting matter is able to couple efficiently to the stellar magnetosphere. In \textsc{sevn}, this accretion-induced spin-up is turned on when the Keplerian angular velocity at the magnetospheric boundary exceeds the NS spin frequency, and the spin-up rate is calculated following physically motivated prescriptions for disk-mediated accretion onto magnetized NSs \citep[e.g.,][]{kiel2008, chattopadhyay2020}. 

On the other hand, 
if the corotation radius ($r_{\rm cor}=(GM_{\rm NS}/\Omega_{\rm rot}^2)^{1/3}$) is inside the magnetospheric radius, the system enters the propeller regime where a centrifugal barrier prevents further accretion \citep{illarionov1975}. In this state, the magnetic field rotates faster than the surrounding flow and incoming matter is expelled rather than accreted, suppressing both the spin-up torque and the accretion-powered luminosity. The propeller mechanism therefore acts as a self-regulating process that limits the growth of the NS spin and strongly modulates the observable X-ray emission. 

In practice, we followed a physically motivated prescription to regulate accretion in the presence of the propeller effect: the magnetospheric radius, $R_{\rm mag}$, was parameterized as a fixed fraction of the Alfvén radius, and when this radius exceeds the corotation radius, the accretion rate was reduced by a factor of $f_{\rm prop}R/R_{\rm mag}$, given the NS radius, $R$, and the dimensionless propeller-efficiency factor, $f_{\rm prop}$. We explored the sensitivity of our results to this treatment by varying the parameters governing the propeller mechanism, adopting as fiducial values a magnetospheric-to-Alfvén radius ratio of $0.5$ \citep{chattopadhyay2020} and a fully suppressive propeller efficiency, $f_{\rm prop}=0$ \citep{illarionov1975}. We find that the propeller mechanism is essential for accurately modeling the 
observational properties of BeXRBs in the SMC (see Sec.~\ref{sec:propeller}). 

Accretion also affects the long-term evolution of the NS magnetic field. Motivated by the low magnetic field strength observed in accreting and recycled pulsars \citep{lorimer2008binary}, we assume that sustained mass accretion leads to an accretion-induced decay of the magnetic field \citep{konar1997}. As the field weakens, the magnetospheric radius decreases, altering the balance between accretion and centrifugal inhibition and, together with the spin-down of the NS, enabling transitions between spin-up and propeller-dominated phases over the binary lifetime. The coupled evolution of spin and magnetic field is followed, allowing the torque regime of the NS to respond dynamically to changes in the magnetic field strength and accretion rate. In addition to the evolution driven by accretion, we also considered the radiation-driven spin down and decay of magnetic fields over time \citep[see Sec. 2.3.1 of][]{Cecilia2023}.

\subsubsection{SN and natal kicks models}\label{sec:SN_and_natal_kicks}

Natal kicks during SN explosions have been shown to have a strong impact on BeXRBs \citep[e.g.,][]{Vinciguerra2020,Igoshev2021,Schurmann2025,Xu2025}. In this work we considered the model by \cite{Giacobbo_Mapelli2020}, in which natal kicks are drawn from a Maxwellian distribution but rescaled by a factor proportional to $M_{\rm ejecta}/M_{\rm remnant}$, where $M_{\rm remnant}$ is the mass of the NS and $M_{\rm ejecta}$ is the ejected mass during the SN. We further explored the impact of the assumed Maxwellian velocity distribution, characterized by its one-dimensional root-mean-square velocity, $\sigma$. In our fiducial model, we did not adopt the commonly used distribution proposed by \citet{Hobbs2005}, with $\sigma = 265$ km s$^{-1}$, but instead followed the revised analysis of \citet{Disberg2025}, where an error in the original fitting procedure was identified and a lower value of $\sigma=217$ km s$^{-1}$ is inferred. This choice leads to a more moderate kick distribution, which  produces a higher number of synthetic BeXRBs. The properties of the compact-object remnants themselves were determined using the delayed SN model of \citet{Fryer2012} and we set the maximum mass of a NS to $2.2$ M$_\odot$.

\subsection{Observational constrains}\label{sec:obs_constrains}

The statistical comparison between our synthetic populations and the observed BeXRB sample in the SMC was carried out in two complementary steps. First, we assessed whether the predicted distributions of key observables are statistically consistent with those inferred from the data. Second, we performed a number-count analysis to test whether the predicted formation efficiencies reproduce the observed population number, taking into account the uncertainties in the SFH of the SMC.

\subsubsection{Observed SMC sample}\label{sec:SMC_sample}

For the observational sample, we adopted the catalog of \citet{Haberl16}\footnote{An online version of the catalog is publicly available at \url{https://projects.mpe.mpg.de/heg/smc_xmmlp/smc_hmxb.txt}}, which includes a total of 102 BeXRB candidates in the SMC. Among these, 49 systems have measured orbital periods, $P_{\rm orb}$. 
In addition, the catalog provides 100 $V$-band magnitude measurements. For this work we do not consider BeXRBs in the Magellanic Bridge and the BeXRB candidate with an accreting white dwarf.

Selection filters 
accounting for observational monitoring bias were applied to both synthetic and observed catalogs. We adopted a monitoring timescale of $t_{\rm obs}=1.6\times10^3$ days that sets an effective maximum orbital period accessible to the X-ray surveys considered \citep{Haberl16,Fortin23}, and we imposed an X-ray luminosity detection threshold of $L_{\rm X} > 10^{34}$ erg/s to mimic the statistics of most well-observed BeXRBs \citep{Raguzova2005,Chen2014,Brown2018,Liu2023}. 

\subsubsection{Statistical comparison with observations}\label{sec:comparison_observations}

For each synthetic model, we constructed the joint probability density function (PDF) of the relevant observables using a Gaussian kernel density estimator (KDE) applied to the simulated BeXRB population that satisfy our identification criteria. In order to compute the PDF, each BeXRB phase, $i$, was weighted by $\Delta t_i \times {\rm SFR}(t_i)/M_{\rm tot}$, where $\Delta t_i$ is the duration of the phase, $t_i$ is the corresponding stellar age, ${\rm SFR}(t)$ is the star formation rate of the SMC as a function of look-back time $t$, and $M_{\rm tot}=M_{\rm BPS}/f_{\rm imf}$ is the total stellar mass underlying the simulated binaries that have total mass of $M_{\rm BPS}$. We took the star formation rate (SFR) of the SMC from \citet[][see their fig.~16]{Rubele2015}.

We performed the statistical comparison in two ways: 
accounting only for the one-dimensional distribution of orbital periods, $P_{\rm orb}$, and 
then considering the joint distribution in $P_{\rm orb}$ and $V$-band magnitude, $m_V$. The $P_{\rm orb}$ distribution primarily probes the AM transport regime, as different AM transport prescriptions directly affect the degree of orbital shrinkage or widening during binary evolution. The addition of $m_V$ in the two-dimensional analysis provides additional sensitivity to $f_{\rm MT}$ as the optical magnitude is 
linked to the stellar mass through its impact on luminosity and effective temperature. 

To quantify the agreement between synthetic and observed distributions, we employed a resampling likelihood approach. From each synthetic catalog, we drew at random a subsample of $N$ objects, where $N$ equals the number of observed systems in the selected comparison sample, and we evaluated the likelihood of that subsample under Gaussian KDE representing the full synthetic population. This draw-and-evaluate procedure was repeated 2000 times 
to construct an empirical distribution of likelihoods for samples of size $N$ drawn from the model. The likelihood of the real observed sample was then computed with the same Gaussian KDE and compared to the empirical distribution, yielding the corresponding p-value that measures how typical or probable the observed sample is with respect to random realizations for that model. Increasing the number of resamples beyond 2000 produced negligible changes in our results. Further details regarding how we calculate the likelihood can be found in Appendix~\ref{app:likelihood}.

\subsubsection{Total number of systems and $\omega_{\rm min}$}\label{sec:total_number_wmin}

The number of BeXRBs predicted by each model  
provides a further, independent constraint. Because model counts scale with the assumed SFR of the SMC, we compared predicted and observed counts taking into account the uncertainty in the SFR \citep{Rubele2015, Liu2023}. Concretely, we represent the expected number predicted by a model as a normal distribution centered on the model prediction with a $20\%$ standard deviation, and computed the corresponding p-value and likelihood for the observed number within this distribution. Our results do not change by changing either the specific shape or the standard deviation considered. In this case, models predicting less than 75 BeXRBs are outside a $2\sigma$ confidence interval with respect to the observations. 

A residual degree of model calibration is required because the spin threshold for 
a star to be a Be star, $\omega_{\rm min}$, is not a priori known. Previous studies \citep[e.g.,][]{Vinciguerra2020,Rocha2024,Liu2023,Schurmann2025} typically consider a constant value of $\omega_{\rm min} = 0.7-0.95$ based on observations of classical Be stars \citep[see Section 3.2. of][]{Rivinius2013}. In this work, for each model we choose $\omega_{\rm min}$ within the physically allowed interval $\left( 0.1 - 0.99 \right)$ so as to bring the predicted number of BeXRBs as close as possible to the observed count of $102$ systems. Then, the fixed value of $\omega_{\rm min}$ probes  whether a model requires physically extreme spin thresholds to match the observed multiplicity. We find that models requiring $\omega_{\rm min} \leq 0.5$ typically cannot produce enough BeXRBs as observed for plausible SMC SFRs, whereas models allowing moderate to high spin thresholds can be reconciled with the data under SFR uncertainties. 

It should be mentioned that the transition from the non-Be phase to the Be phase in reality is unlikely to occur with a sharp, universal threshold $\omega_{\rm min}$, but rather a gradual transition
that may depend on stellar properties such as mass and effective temperature \citep{Huang2010}. Observations of classical Be stars \citep[and references therein]{Rivinius2013} indicate that there exists a value of $\omega$ above which all B-type stars are observed as Be stars; this threshold value is temperature dependent, increasing from $\omega \sim 0.8$ for hotter stars to $\omega \sim 0.999$ for cooler ones. For instance, \citet{Huang2010} analyzed the rotational distribution of non-Be B stars and identified empirical upper limits to their $\omega$, finding $\omega \lesssim 0.997$ for late-type B stars $(M_\star < 4,\mathrm{M_\odot})$ and $\omega \lesssim 0.83$ for more massive stars $(M_\star > 8.6,\mathrm{M_\odot})$. Note that these limits do not preclude the existence of Be stars at lower rotation rates, and spins as low as $\omega\sim 0.7$ have also been found in Be stars \citep{Cranmer2005}. 

Following these findings, there are two observationally motivated regimes: a minimum rotation rate associated with the onset of the Be phenomenon, and a higher threshold beyond which all B stars exhibit Be characteristics. 
Our definition of the threshold spin, however, differs from both observational regimes, as we assume a sharp and universal transition for simplicity. 
Therefore, the comparison with the observational thresholds for the onset or ubiquity of the Be phase should be regarded as qualitative rather than exact. In practice, we state that a model achieves agreement with observations when the corresponding $\omega_{\rm min}$ falls in the range of $\sim 0.83-0.997$ \citep{Huang2010}. 
A detailed investigation of the precise rotational boundary is beyond the scope of the present study, especially given that our analysis relies on currently available catalogs and that recently proposed candidates, bias effects, and uncertainties in spin measurements may slightly modify the empirical constraints \citep[see e.g.,][]{Rivinius2013,Gaudin25,Lailey2026}. 

The final ranking metric for each model was obtained by combining the percentiles and likelihoods from both the PDF $\left( P_{\rm orb}\, \text{or}\, P_{\rm orb}-m_V\right)$ comparison and the number-count comparison (see Appendix~\ref{app:likelihood}). We state that a model is able to reproduce observations when the likelihood of the real observed sample is within the 98\% confidence interval of the likelihood distribution from random sampling of the synthetic catalog. 
Finally, in order to rank such models, we rely on the combined likelihood, where a higher likelihood means better agreement with observations.

\section{Results} \label{sec:results}

%%%%%%%%%%%%FIGURE%%%%%%%%%%%%%%
\begin{figure*}
    \centering
    \includegraphics[width=1\textwidth]{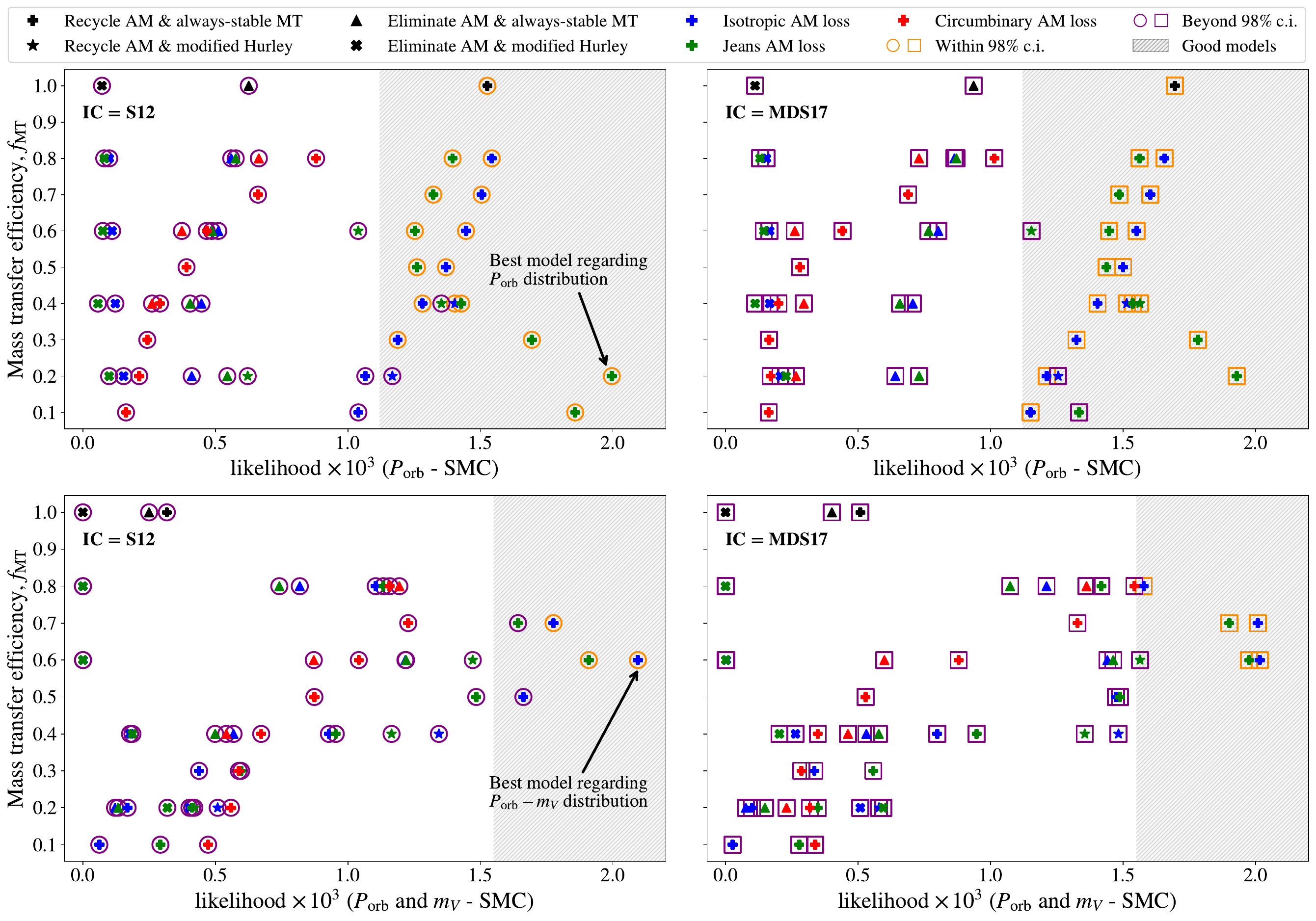}
    \caption{Combined likelihood of the $P_{\rm orb}$ distribution (top) or of the joint $P_{\rm orb}$ and $m_V$ distribution (bottom), both further combined with the total number of systems, as a function of MT efficiency, $f_{\rm MT}$, for all models producing more than 75 BeXRBs in the SMC. The marker shape encodes the combination of MT stability and AM feedback onto the orbit: plus symbols correspond to always-stable MT with AM recycle to the orbit, 
    stars to modified Hurley stability with AM recycle, 
    triangles to always-stable MT without AM recycle, 
    and crosses to modified Hurley stability without AM recycle. 
    The marker color identifies the specific AM loss channel associated with non-accreted mass, with blue indicating isotropic re-emission 
    from the vicinity of the accretor, green Jeans-mode 
    from the donor, and red AM loss through a circumbinary ring. 
    For fully conservative MT ($f_{\rm MT}=1$), the AM loss prescription is irrelevant; these models are therefore shown with black symbols.
    The outer contour color of the marker distinguishes models within (circular and square orange outlines) and outside (circular and square purple outlines) the 98\% confidence interval. Left column: Initial conditions from \citetalias{Sana2012}, denoted with circular outer symbols. Right column: Initial conditions from \citetalias{Moe_DiStefano2017}, denoted with square outer symbols. The shaded regions correspond to the regime enclosing all models able to reproduce observations following our statistical comparison (see Section~\ref{sec:obs_constrains}).}
    \label{fig:good_smc}
\end{figure*}
% %%%%%%%%%%%%%%%%%%%%%%%%%%%%%%%

\subsection{Orbital period and $V$-band magnitude}
\label{sec:porb_mv_lh}
Fig.~\ref{fig:good_smc} summarizes the statistical classification of the explored models under a $98\%$ confidence interval criterion, where we only include the models that produce more than 75 BeXRBs in the SMC. The upper panels display the likelihood 
evaluated taking into account only the orbital period distribution 
and the total number of systems, whereas the lower panels show the likelihood estimated for the joint distribution of $P_{\rm orb}$ and $m_V$, again including the total number of systems. 

In both comparison spaces, only models that combine AM recycle to the orbit at near-critical rotation (hereafter, AM recycling) with stable RLOF are able to simultaneously reproduce the observed SMC BeXRB sample in terms of both total number and distribution (highlighted in the shaded 
region of Fig.~\ref{fig:good_smc}). Models in which the excess AM is simply ejected from the system systematically fail to match the observed distribution while models in which MT can become dynamically unstable systematically fail to match either the distribution or the total number of the observed population. This indicates that efficient spin-orbit coupling during MT, together with the avoidance of CE-like phases during RLOF, constitutes a necessary condition for reproducing the SMC BeXRB population with \textsc{sevn}. The fact that the observed orbital period distribution is only reproduced assuming always-stable RLOF together with efficient recycling of excess AM justifies restricting this combination of parameters to the additional explored values, $f_{\rm MT}=0.1, 0.3, 0.5,$ and $0.7$, to further study the effect of MT efficiency.

When considering the orbital period distribution alone, the constraints primarily probe the AM loss model. No strong preference emerges between the Jeans and isotropic re-emission prescriptions, whereas the circumbinary ring mode is clearly disfavored, as it tends to overproduce 
the population of short $P_{\rm orb}$. As expected, the period distribution is only weakly sensitive to the MT efficiency $f_{\rm MT}$, since $P_{\rm orb}$ mainly reflects the integrated effect of AM loss on orbital widening or shrinkage. A robust outcome of this analysis is the requirement that AM transferred to a near-critically rotating accretor must be recycled to the orbit, rather than being entirely ejected. 

The inclusion of the $V$-band magnitude significantly strengthens the constraints. 
In the $P_{\rm orb}-m_V$ space, 
the likelihood analysis consistently favors moderately non-conservative MT, with $f_{\rm MT}\sim 0.6$, in agreement with independent constraints derived from related Be systems such as Be+sdOB binaries, consisting of a Be star and a stripped hot subdwarf companion \citep{Lechien2025}. 

The models identified as consistent with the observations in the joint two-dimensional analysis exhibit higher likelihood values than those selected using only the $P_{\rm orb}$ distribution. This reflects the stronger constraints imposed by simultaneously reproducing multiple observables. 
Within the viable region of the parameter space, there is a mild preference for isotropic re-emission AM loss, 
although this trend is less important than the assumption of stable RLOF and AM recycling. 
Our results are largely insensitive to the choice of the initial conditions, as shown in the two columns of Fig.~\ref{fig:good_smc}, where most models compatible with the observations under one initial conditions are likewise compatible under the other.

Additionally, we performed a Kolmogorov-Smirnov (KS) test \citep{Berger2014} to quantify the distance between the empirical distribution function of the sample and the cumulative distribution function (CDF) of the reference distribution, in this case the observed distribution. The KS test confirms our likelihood-based results: models with an intermediate MT efficiency, AM recycling, and stable RLOF simultaneously are preferred to reproduce the observed $P_{\rm orb}$ and $m_V$ distributions. In particular, models with a low MT efficiency, $f_{\rm MT}\sim0.2$, better match the $P_{\rm orb}$ distribution but fail for the $m_V$ distribution, while higher MT efficiency values, $f_{\rm MT}\sim0.8-1.0$, worsen both comparisons.

%%%%%%%%%%%%FIGURE%%%%%%%%%%%%%%
\begin{figure}
    \centering
    \includegraphics[width=0.5\textwidth]{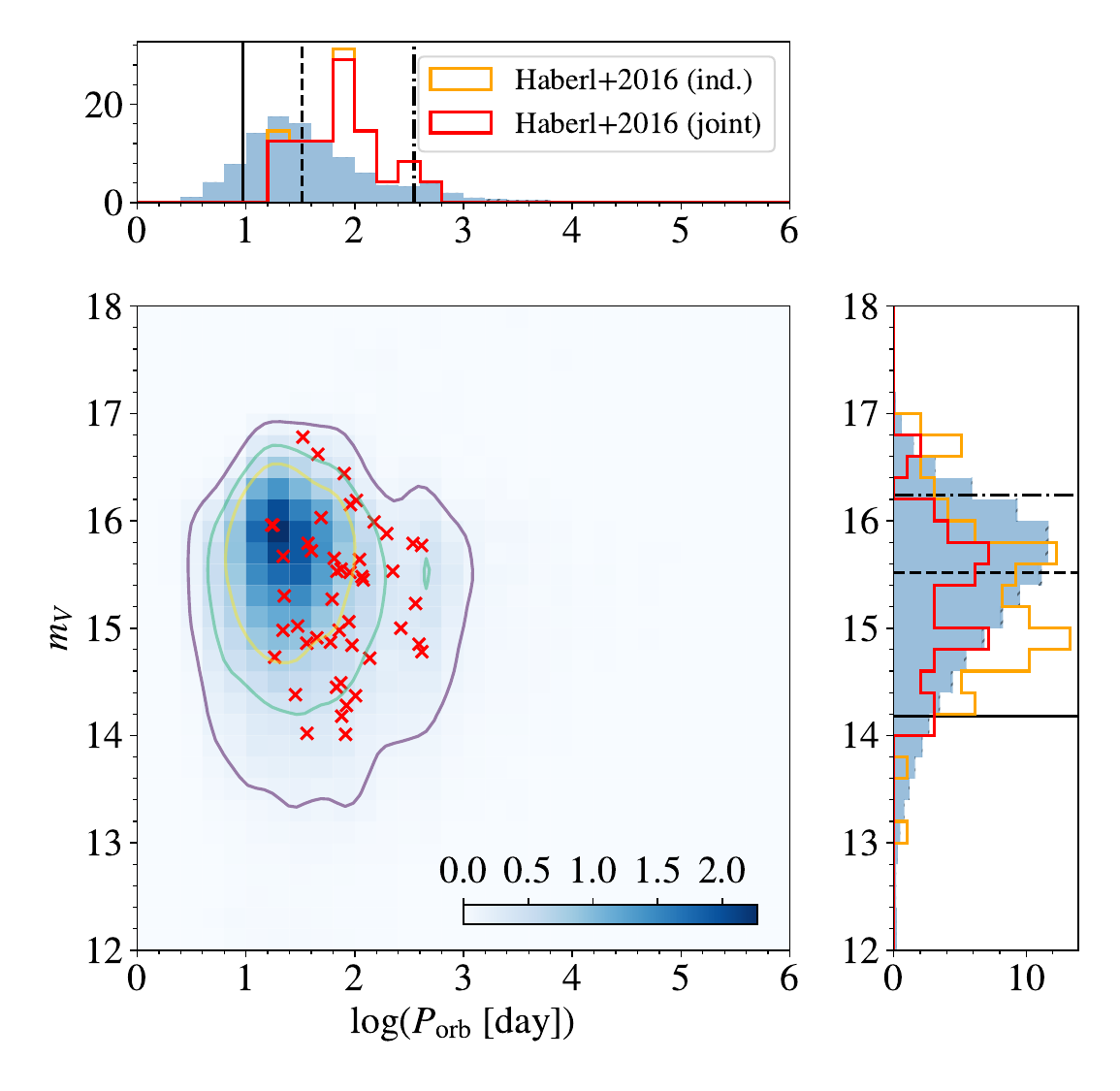}
    \caption{Orbital period ($P_{\rm orb}$) and $V$-band magnitude ($m_V$) distributions for the overall best-fitting model of the SMC BeXRB population, corresponding to a MT efficiency of $f_{\rm MT}=0.6$, isotropic re-emission AM loss, always-stable RLOF, recycling of AM to the orbit when the accretor approaches critical rotation, and initial conditions from \citetalias{Sana2012}. Each synthetic BeXRB phase, $i$, is weighted by $\Delta t_i \times {\rm SFR}(t_i)/M_{\rm tot}$, where $\Delta t_i$ is the duration of the phase, $t_i$ is the corresponding stellar age, ${\rm SFR}(t)$ is the star formation rate of the SMC as a function of look-back time, $t$, and $M_{\rm tot}=M_{\rm BPS}/f_{\rm imf}$ is the total stellar mass underlying the simulated binaries that have total mass of $M_{\rm BPS}$, such that the counts in each bin represent the expected number of BeXRBs. In the 2D maps, the blue, green, and yellow contours enclose 90\%, 50\%, and 10\% of systems from a Gaussian smoothed density field produced by the weighted data. In the marginalized distributions, the solid, dashed, and dotted lines indicate the 10th, 50th, and 90th percentiles of the model predictions, respectively, while the shaded areas correspond to BeXRB phases with $P_{\rm orb} > t_{\rm obs}=1.6\times10^3$ days, which lie beyond the adopted observational time baseline considered in our statistical approach. Observed systems with measurements of both $P_{\rm orb}$ and $m_V$ are shown as red crosses in the 2D map and as red  histograms in the marginalized distributions, while yellow  histograms further include the systems for which only one of the two quantities is available \citep{Haberl16}. The distribution of each  observational sample is renormalized to have the same number of BeXRBs as in the synthetic population with $P_{\rm orb}<t_{\rm obs}$.}
    \label{fig:P_m_V_smc}
\end{figure}
%%%%%%%%%%%%%%%%%%%%%%%%%%%%%%%%%

Fig.~\ref{fig:P_m_V_smc} presents the orbital period and $V$-band magnitude distributions for the overall best-fitting configuration, characterized by $f_{\rm MT}=0.6$, isotropic re-emission mass loss, always-stable RLOF, AM recycling to the orbit at near-critical rotation, and initial conditions from \citetalias{Sana2012}. This model reproduces both the distribution and the total number of the observed sample. Although an excess of short-period ($P_{\rm orb}\lesssim 10\ \rm days$) systems is predicted, the comparison is inherently statistical: we evaluated the probability that the observed dataset is drawn from a given synthetic distribution, rather than requiring an exact system-by-system correspondence. In this sense, the absence of short-period BeXRBs in the SMC does not imply a direct inconsistency as the real observed sample is among the $\sim60-70$\% most probable random realizations from this model considering either the $P_{\rm orb}$ distribution or the joint distribution (both combined with the number). Additional properties of the synthetic BeXRB population for this model are shown in Appendix~\ref{app:SMC}.

%%%%%%%%%%%%FIGURE%%%%%%%%%%%%%%
\begin{figure}
    \centering
    \includegraphics[width=0.5\textwidth]{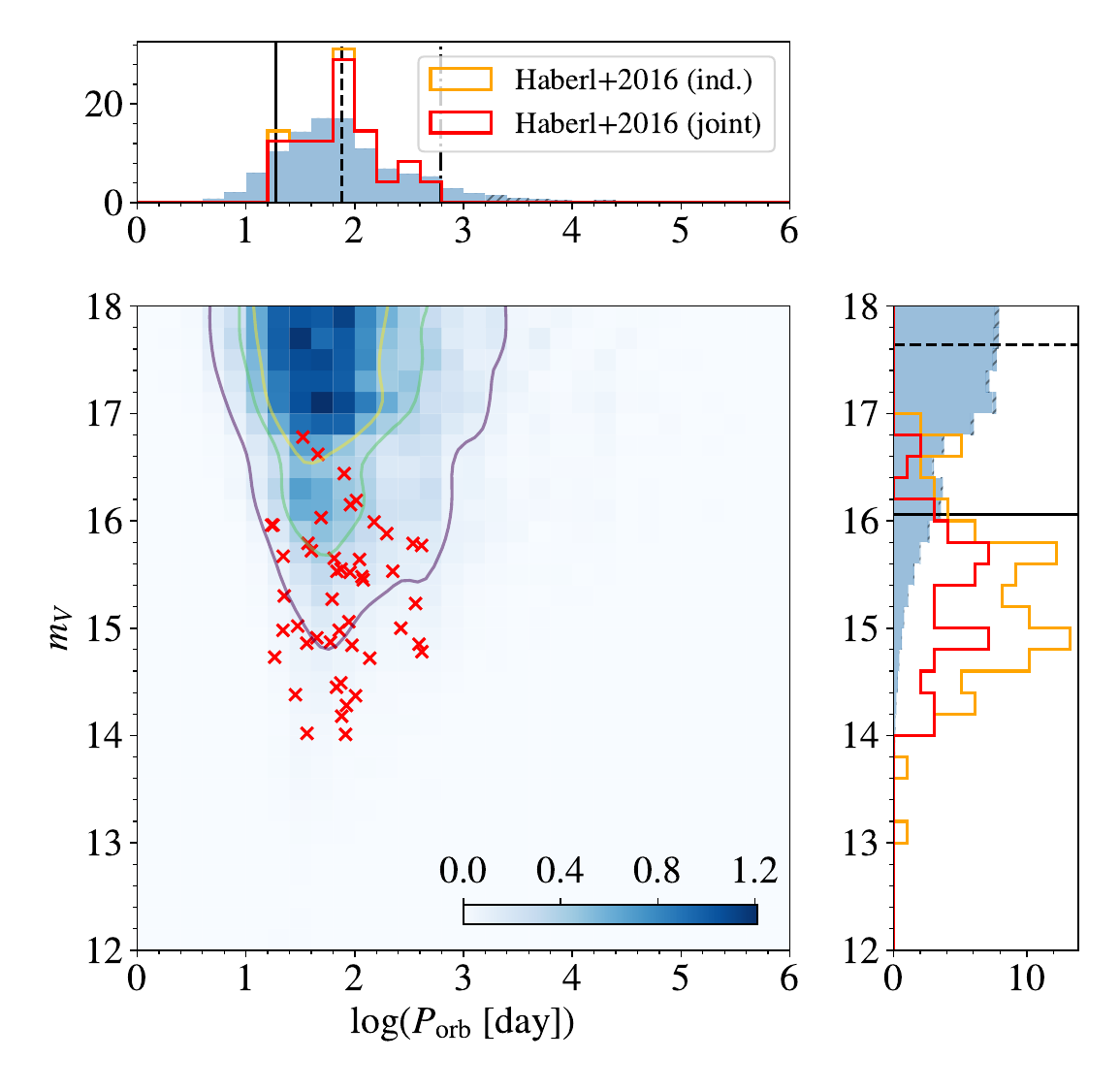}
    \caption{Same as Fig.~\ref{fig:P_m_V_smc}, but for the model that best reproduces the orbital period distribution of BeXRBs in the SMC (see upper panels of Fig.~\ref{fig:good_smc}). This model also adopts always-stable RLOF, recycling of AM to the orbit when the accretor approaches critical rotation, and initial conditions from \citetalias{Sana2012}, but it instead considers a lower MT efficiency, $f_{\rm MT}=0.2$, and Jeans-mode AM loss.}
    \label{fig:P_m_V_smc_best_Porb}
\end{figure}
%%%%%%%%%%%%%%%%%%%%%%%%%%%%%%%%%

Fig.~\ref{fig:P_m_V_smc_best_Porb} shows the $P_{\rm orb}$ and $m_V$ distributions for the best-fitting model when only the $P_{\rm orb}$ distribution is considered. This configuration differs from the overall best-fitting model by a lower MT efficiency ($f_{\rm MT}=0.2$) and the adoption of a Jeans-mode prescription for AM loss. The reduced MT efficiency leads to a population with almost no short-period systems, improving the agreement with the observed $P_{\rm orb}$ distribution. However, this same effect results in less massive Be stars, and therefore systematically fainter systems, producing a $m_V$ distribution that is inconsistent with observations.

%%%%%%%%%%%%FIGURE%%%%%%%%%%%%%%
\begin{figure}
    \centering
    \includegraphics[width=0.5\textwidth]{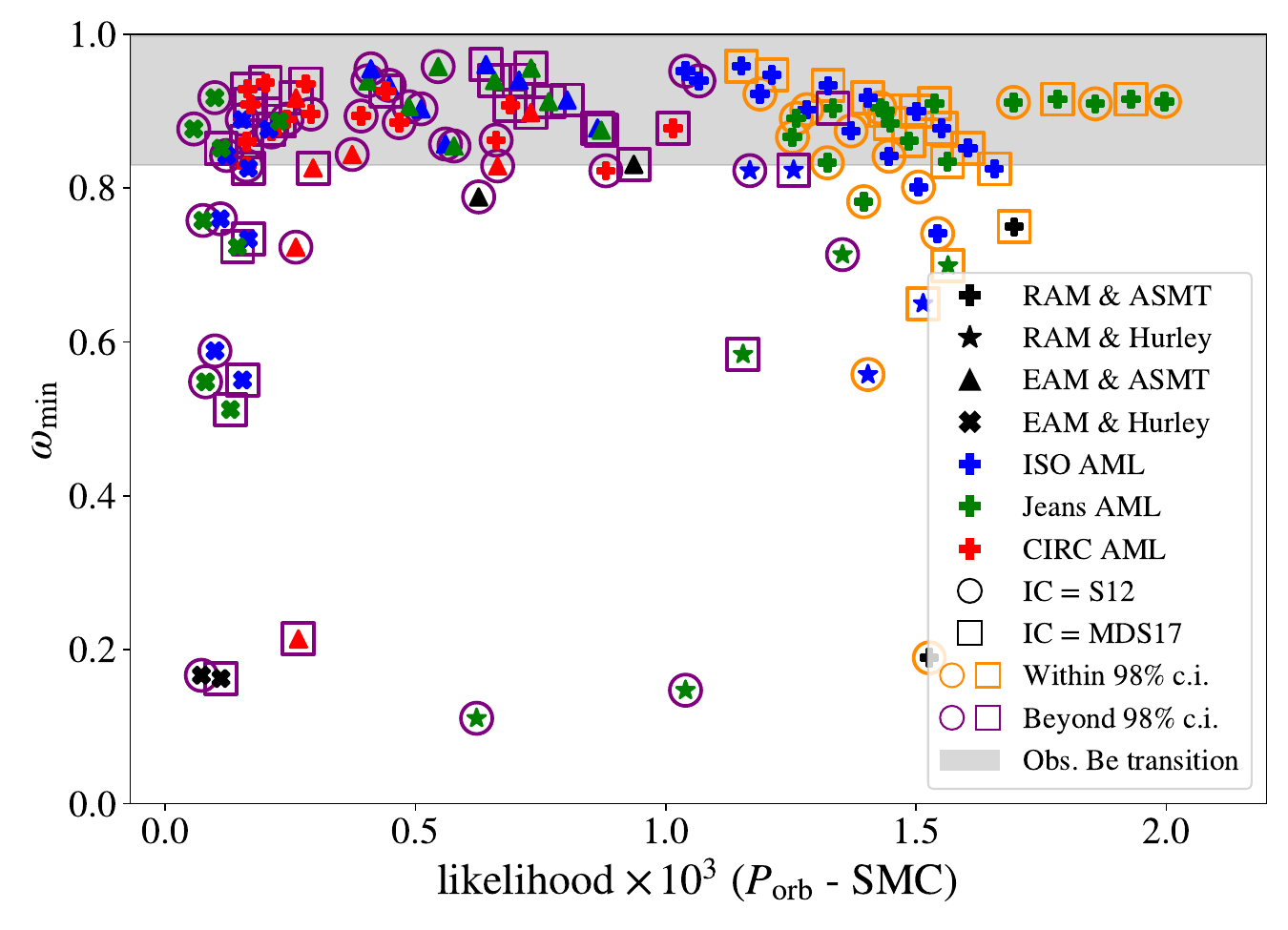}
    \caption{Combined likelihood of the $P_{\rm orb}$ distribution and the total number of systems, versus the threshold spin for Be effect $\omega_{\rm min}$, for all models producing more than 75 BeXRBs in the SMC. Here, $\omega_{\rm min}$ is chosen to match the observed number of BeXRBs, $102$ \citep{Haberl16}, as much as possible in each model. As in Fig.~\ref{fig:good_smc}, the marker shape encodes the combination of MT stability and AM feedback onto the orbit and the marker color identifies the specific AM loss channel associated with non-accreted mass. For fully conservative MT ($f_{\rm MT}=1$), the AM loss prescription is irrelevant; these models are therefore shown with black symbols. The outer symbol shape identifies the initial conditions, with circular outlines for \citetalias{Sana2012} and squared outlines for \citetalias{Moe_DiStefano2017}. 
    The outer contour color of the marker distinguishes models within (orange) and outside (purple) the 98\% confidence interval. 
    The shaded gray regions correspond to the minimum spin required to become Be stars from observations (see e.g., sec. 3.1 of \citealp{Rivinius2013} and \citealp{Huang2010}).}
    \label{fig:wmin_porb_smc}
\end{figure}
%%%%%%%%%%%%%%%%%%%%%%%%%%%%%%%%%

\subsection{Threshold spin}

The threshold spin adopted for Be star identification is also consistent with observational expectations. As shown in Fig.~\ref{fig:wmin_porb_smc}, the models that satisfy the statistical criteria cluster around $\omega_{\rm min}\sim 0.8$--$0.9$, in line with empirical determinations of the threshold spin for Be stars \citep{Huang2010}. For instance, the overall best-fitting model has $\omega_{\rm min}=0.842$ and the best $P_{\rm orb}$-fitting model has $\omega_{\rm min}=0.913$. Models outside this range, illustrated in Appendix~\ref{app:SMC_bad_models}, typically underproduce the number of BeXRBs in the SMC.

In addition to the fiducial initial rotation described by Eq.~\eqref{eq:v_rot_0}, we also considered two different models: a prescription in which the stars are initially synchronized with the orbit as tidal interaction predicts this to be achieved during evolution towards RLOF \citep{Zhang2004}, and a zero initial rotation for all stars. Our fiducial prescription produce slightly more systems $(\sim5\%)$ than the other two models, with the initially non-rotating model producing slightly less systems $(\sim1\%)$ than the initially synchronized model. This implies that the rotation needed for the Be phase to be triggered is mostly achieved via AM transport by accreted mass during RLOF \citep[see also][]{Liu2023,Staritsin2026}.

%%%%%%%%%%%%FIGURE%%%%%%%%%%%%%%
\begin{figure}
    \centering
    \includegraphics[width=0.5\textwidth]{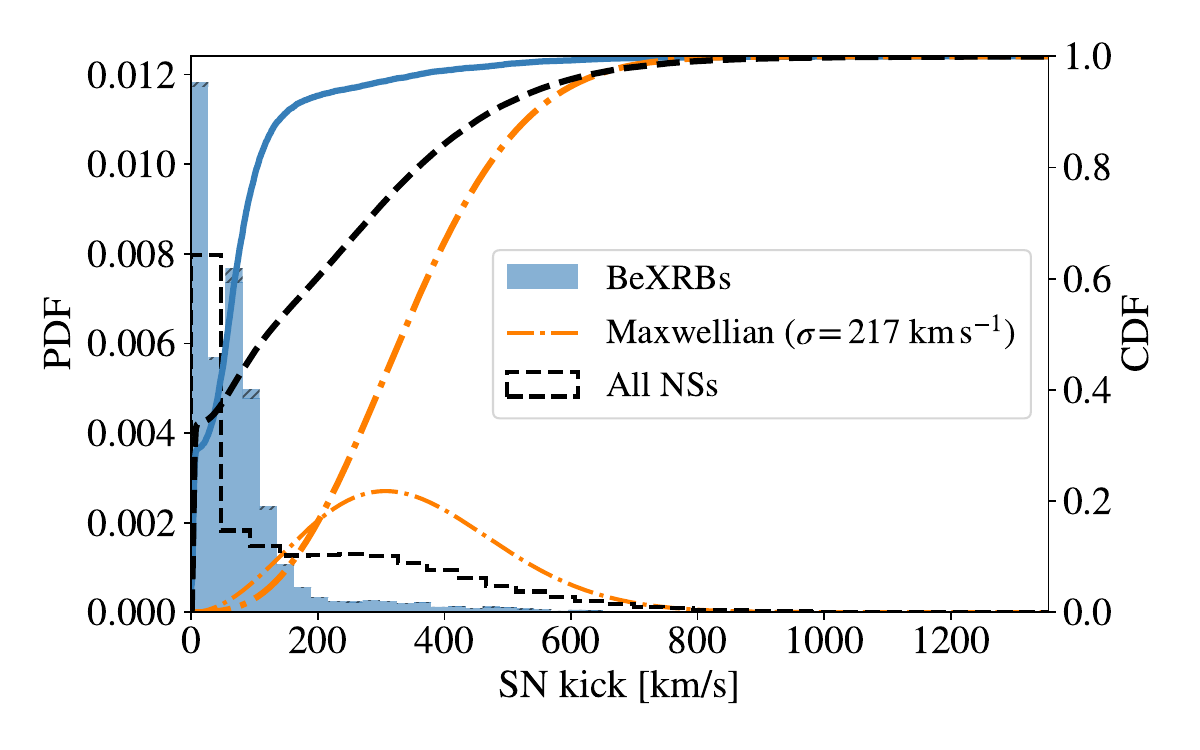}
    \caption{SN natal kick distributions in the overall best-fitting SMC model. The blue histogram shows the natal kick distribution of NSs in BeXRBs, while the hatched area corresponds to natal kicks associated to BeXRBs with $P_{\rm orb} > t_{\rm obs}=1.6\times10^3$ days. The thin dashed black contour represents the natal kick distribution of all NSs formed in the simulations, irrespective of whether the binary is disrupted or not. The thin dash-dotted orange line indicates the underlying Maxwellian distribution prior to the applied correction following \citet{Giacobbo_Mapelli2020}. The thick solid, dashed, and dash-dotted curves show the CDF (right $y$ axis) for NSs in BeXRBs, all NSs, and pre-correction kicks, respectively.}
    \label{fig:SN_kicks_smc}
\end{figure}
%%%%%%%%%%%%%%%%%%%%%%%%%%%%%%%%%

\subsection{Natal kicks}

The natal kick distribution of NSs forming BeXRBs is strongly skewed toward low velocities as shown by  Fig.~\ref{fig:SN_kicks_smc}. Approximately $40\%$ of systems experience effectively zero kick, around $80\%$ receive kicks below $100\,\mathrm{km\,s^{-1}}$, and only a small fraction reach values on the order of $200\,\mathrm{km\,s^{-1}}$. 
This indicates that only progenitor systems that form with low natal kicks  can  produce BeXRBs. This is consistent with previous results for BeXRBs \citep[e.g.,][]{Vinciguerra2020,Valli2025,Schurmann2025} and binary NSs \citep{beniamini2016b,beniamini2016,giacobbo2019,Cecilia2023}.

\section{Discussion}\label{sec:discussion}

\subsection{Orbital period distribution and the angular momentum budget}

One of the 
main observational features of the SMC BeXRB population is the 
rarity of short-period ($P_{\rm orb}\lesssim 10\ \rm days$) systems \citep{Coe2015,Haberl16}. Any viable binary-evolution scenario must therefore avoid excessive orbital shrinkage during the first MT episode and subsequent phases to reproduce this feature, unless we attribute it to other factors, such as enhanced tidal truncation of the VDD \citep[e.g.,][]{Xing2021,Xu2025}. In classical binary evolution, orbital contraction is typically mitigated by assuming a low accretion efficiency, such that most of the transferred material is lost from the system. 
Indeed, several previous population synthesis studies have adopted rather inefficient MT, with $f_{\rm MT}\sim 0.3$ \citep[e.g.,][]{Vinciguerra2020} and even $f_{\rm MT}\simeq 0.05$ \citep{Rocha2024}. Although this prescription reduces orbital shrinkage, previous studies often remain in tension with the observed period distribution \citep[e.g.,][]{Rocha2024,Schurmann2025,Xu2025} and, in many cases, with the observed properties of the Be companions.

In our analysis, when considering only the orbital period distribution (upper panels of Fig.~\ref{fig:good_smc}), we also recover an apparent preference for relatively low MT efficiency. This is expected, as the period distribution is primarily sensitive to the global AM budget rather than to the detailed structure of the accretor. However, our results indicate that 
orbit widening cannot be due only to mass loss. Instead, a key physical ingredient is the treatment of the excess AM that would otherwise spin the accretor beyond critical rotation.

In our best-fitting models, when the accretor approaches critical rotation, the surplus AM is not assumed to be simply ejected from the system. Rather, efficient tidal coupling \citep{Popham1991,Paczynski1991} allows part of this AM to be returned to the orbit. This recycling mechanism fundamentally alters the secular orbital evolution. By transferring AM back to the orbit, the system avoids excessive shrinkage without requiring highly inefficient MT. In this framework, the 
rarity of short-period BeXRBs in the SMC is marginally reproduced, while still permitting intermediate MT efficiencies. Therefore, the observational constraint on the period distribution does not necessarily imply strongly non-conservative MT; it instead constrains the combined interplay between MT efficiency, AM loss, and tidal coupling between the binary orbit and the circumstellar disk around the accretor.

We also find that circumbinary ring AM loss prescriptions are clearly disfavored. This mode extracts too much AM from the orbit and systematically drive the binaries toward shorter periods than observed. Conversely, isotropic re-emission is mildly preferred by the likelihood ranking over Jeans AM loss.

Finally, we must point out that our work only considered a limited set of models for AM transport during MT, which are not expected to fully capture the complex interactions between stars, MT streams, circumstellar/binary disks, and outflows in reality. As shown in recent hydrodynamic simulations \citep[e.g.,][]{Munoz2019,Munoz2020,Dittmann2022,Wang2022,Lai2023}, additional mechanisms such as re-accretion from circumbinary disks and torques from colliding outflows can further boost AM recycling. We plan to explore a broader range of physically motivated AM transport models in future work.

\subsection{Be star $V$-band magnitude and total BeXRB number constraints: MT efficiency and RLOF stability}

While the orbital period distribution is predominantly sensitive to AM transport, the inclusion of the $V$-band magnitude provides a direct constraint on the mass of the Be star, and therefore on the MT efficiency during the first MT phase. The Be star mass is set by the amount of material accreted from the primary and is nonlinearly connected to the observed $m_V$ through the luminosity and effective temperature. Consequently, the joint distribution in $(P_{\rm orb}, m_V)$ probes both the orbital dynamics and the accretion history.

Previous studies have frequently reported inconsistencies with the observed Be star masses, either in terms of $m_V$ \citep[e.g.,][]{Schurmann2025} or spectral type \citep[e.g.,][]{Rocha2024}. In many cases, the assumed low MT efficiencies lead to systems in which the secondary does not gain sufficient mass to match the observed Be star population. Conversely, extremely high MT efficiencies tend to overproduce high-mass Be stars.

In our framework, once the AM prescription includes orbital recycling through tides, the addition of the $m_V$ constraint clearly favors an intermediate MT efficiency, $f_{\rm MT}\simeq 0.6$. Importantly, we allow the formation of classical Be stars down to $\sim 3\,M_\odot$ \citep{Vieira2017}, unlike previous studies \citep[e.g.,][]{Misra2023b,Liu2023} that only consider early-type and/or high-mass ($M_\star\gtrsim6\ \rm M_\odot$) Be stars for BeXRBs.
Our results are consistent with independent estimates from population synthesis of Be+sdOB star binaries \citep[e.g.,][]{Lechien2025,Bao2025}, which also make no initial assumption regarding the masses of Be stars in binaries. We should note, however, that rapid rotation has recently been studied to have an impact on the observational properties of Be stars \citep{Rast2025}, which could affect our comparison regarding $m_V$. Regardless, we expect this effect to have little influence on the overall distribution.

Regarding RLOF stability, both prescriptions considered here could generate similar shapes in the joint $P_{\rm orb}-m_V$ distribution. However, the modified Hurley model produces significantly fewer BeXRBs overall, leading to an under-prediction of the observed number even when accounting for the SMC SFR uncertainty. This indicates that a predominantly stable RLOF phase is required to reproduce the observed population, reinforcing the view that BeXRB progenitors do not undergo CE evolution in most cases as previous studies concluded \citep{Shao2014, Vinciguerra2020, Rocha2024, Schurmann2025,Xu2025}. Interestingly, a preference for an always-stable RLOF has also been studied in previous work in the context of merging compact objects \citep[e.g.,][]{Inayoshi2017,Picco2024,Xu2025bbh}, among other systems \citep[e.g.,][]{Martinez2025,McNeill2025}.

For select models, we further assessed the robustness of our conclusions against alternative assumptions for the onset of unstable MT. In addition to the prescriptions adopted throughout this work, we explored models in which stability is determined by a fixed $q_{\rm c}$, either for all donor evolutionary stages or only for post-Hertzsprung-gap donors. Overall, we find that the predicted total number of BeXRBs is highly sensitive to the considered $q_{\rm c}$ as most of BeXRB progenitors are highly unequal-mass systems. 
For $q_{\rm c}=5$, these new models yield results similar to those obtained with the always-stable MT model, reducing the predicted number of systems by less than 20\%. Such high critical mass ratios have been found in detailed and rapid binary evolution models \citep{Ge2015,Ge2020,Langer2020,Echeveste2026}, which support the viability of the stable evolutionary scenario favored by our analysis.
Further details are given in Appendix~\ref{app:RLOF_stability}.

\subsection{The propeller effect and the emergence of a long-period subpopulation}
\label{sec:propeller}

The treatment of the propeller effect has a profound impact on the predicted BeXRB population. Physically, the propeller regime occurs when the magnetospheric radius exceeds the corotation radius, preventing material from accreting onto the NS. In wide systems, where the MT rate is intrinsically low, NSs are typically in the propeller regime. As a consequence, their X-ray luminosity depends sensitively on whether residual accretion is considered.

In our fiducial implementation, a strict propeller prescription characterized by $f_{\rm prop}=0$ strongly suppresses accretion, 
effectively delaying or avoiding the X-ray emission phase when centrifugal and magnetic inhibition sets in. 
In this case, wide systems with low MT rates fall below the adopted detection threshold ($L_X>10^{34}\ \rm erg\ s^{-1}$) and are not classified as BeXRBs. The resulting synthetic population is therefore dominated by systems with moderate orbital periods and sufficiently high accretion rates. The BeXRB phase is shorter, and the parameter space occupied by the simulated systems closely matches the observational sample.

Allowing for non-negligible residual accretion with $f_{\rm prop}\gtrsim10^{-6}$ modifies this picture in two ways. First, the BeXRB lifetime is modestly extended, as systems that would otherwise be fully suppressed can now accrete, which overall increases the total number of predicted systems. 

Second, and more importantly, an additional subpopulation emerges at long orbital periods ($P_{\rm orb}\sim10^{3}$ days) and fainter Be stars, the latter being related to lower Be star masses (see Appendix~\ref{app:SMC_prop}). 
Once a small fraction of the inflow is allowed to penetrate through the centrifugal barrier and be accreted, many long-period systems can produce weak X-ray emission above the detection threshold (and be classified as BeXRBs) for extended times. 
By tracing their progenitors, we find that these systems originate preferentially from binaries with lower initial primary masses, $M_1$. For a fixed $f_{\rm MT}=0.6$, reduced primary masses naturally translate into less massive Be stars after the first MT phase. The partial lifting of the propeller barrier thus allows systems that would be marginal or X-ray faint under a strict prescription to contribute to the observable population \citep[see e.g.,][]{Zainab2026}.

The emergence of this subpopulation is not governed only by the residual accretion itself; 
rather, it also depends on the adopted luminosity threshold for detection. 
The interplay between the propeller prescription and the observational cutoff therefore introduces a selection effect that directly shapes the inferred population. This sensitivity has two important implications and overall reinforces the importance of centrifugal inhibition in NS evolution and observational signatures \citep{Cecilia2023}. First, it demonstrates that the observed absence of very long-period BeXRBs could partly reflect detection biases linked to the X-ray luminosity threshold. Second, it indicates that future surveys reaching lower $L_X$ limits may uncover a tail of long-period, low-luminosity systems. The existence of such a population is a direct consequence of marginally accreting binaries near the propeller boundary, which can potentially provide a probe of the accretion physics in magnetized spinning NSs as well as binary evolution physics. For instance, the recent work by \citet{Xu2026} shows that the mass ejection at periastron passage during the luminous blue variable phase can widen and unbind wide binaries involving O/B stars. This effects can potentially shape the long-period tail of the $P_{\rm orb}$ distribution of BeXRBs, which we plan to investigate in future work.

\subsection{Threshold spin for Be phase and completeness}

Finally, our best-fitting models require a minimum spin threshold of $\omega_{\rm min}\sim 0.84-0.91$ close to the optimistic lower bound inferred from observations \citep{Huang2010}, which is consistent with the high completeness of the SMC BeXRB catalog. However, this conclusion is not entirely decoupled from the propeller treatment. When adopting a lower $L_X$ threshold and allowing residual accretion, a long tail of low-luminosity systems emerges, some of which correspond to less rapidly rotating Be stars. Therefore, the inferred constraint on $\omega_{\rm min}$ depends on the observational sensitivity to faint systems and the observational time, as very long-period BeXRBs, should they exist, do not have the chance to show X-ray bursts in the effective duration of monitoring. Future X-ray surveys capable of probing lower luminosities and with longer monitoring time could thus refine the allowed range of $\omega_{\rm min}$ and, indirectly, the accretion efficiency onto the NS.

Several additional physical ingredients may in principle influence the formation and properties of BeXRBs but are not explored in detail in this work. For instance, \citet{Langer1998} introduced a prescription for rotation-induced mass loss that could affect rapidly rotating stars and has been suggested to be relevant in the context of BeXRBs \citep{Rocha2024}. However, this model becomes ill defined as stars approach the Eddington luminosity, whereas observational evidence indicates the existence of super-Eddington systems \citep{Reynolds2021}. For a limited subset of models, we therefore tested the \citet{Langer1998} prescription by adopting a linear extrapolation for Eddington-corrected effective spin above 0.8, finding that rotationally induced mass loss has a negligible impact on the resulting BeXRB populations. A more comprehensive investigation of this effect 
is beyond the scope of the present study.

\subsection{Comparison with the MW}

We have also studied the BeXRB population of the MW to assess the robustness of our results in different host galaxies. Given the lack of homogeneous visual magnitude measurements, this comparison is restricted to the orbital period distribution. We find that the models that best reproduce the SMC population are likewise able to reproduce the observed $P_{\rm orb}$ distribution of Galactic BeXRBs, despite the larger observational uncertainties and incompleteness affecting the MW sample. Further details of this comparison are provided in Appendix~\ref{app:MW}.

\section{Summary and conclusions} \label{sec:summary}

We have performed a comprehensive statistical comparison between synthetic BeXRB populations and the observed sample in the SMC, combining statistical distribution-based and number-count constraints within a unified likelihood framework. For each evolutionary model, we constructed PDFs using a Gaussian KDE applied to the synthetic catalogs that pass our observational filters. We considered both the one-dimensional orbital period distribution and the two-dimensional $P_{\rm orb}-m_V$ space, and complemented these with a number-count likelihood that accounts for uncertainties in the SMC SFR. This approach allows us to assess not only whether a model reproduces the observed distributions, but also whether it predicts the correct formation efficiency.

We find that models that avoid excessive orbital shrinking during the first MT phase are associated with higher likelihood values. 
For instance, circumbinary AM loss prescriptions are  disfavored, as they overproduce short-period systems that are not observed in the SMC. Instead, models in which excess AM -- that would otherwise lead to critical rotation of the accretor -- is recycled back into the orbit through efficient (circumstellar-disk-orbit) tidal coupling successfully reproduce the observed period distribution. This mechanism 
demonstrates that the orbital constraints primarily probe the global AM budget rather than MT efficiency alone. Moreover, we find a strong preference for models in which RLOF is always stable, as these scenarios are the only ones capable of reproducing the observed total number of BeXRBs in the SMC.

Accounting for the distribution of visual magnitudes provides additional constraints on the accretion efficiency. Specifically, our results support a moderately non-conservative MT efficiency, $f_{\rm MT}\simeq 0.6$. Lower efficiencies tend to produce Be stars that are too faint, whereas higher efficiencies yield systems that are systematically over-massive and too luminous compared to the observations. 

We have also explored the impact of the propeller effect on the predicted BeXRB population. A strict propeller prescription ($f_{\rm prop}=0$) suppresses accretion in wide systems, keeping them below the adopted $L_X\gtrsim 10^{34}\ \rm erg\ s^{-1}$ detection threshold, and thus excluding them from the observable sample. Allowing a small residual accretion fraction ($f_{\rm prop}\gtrsim 10^{-6}$) reveals an additional subpopulation of long-period, low-mass BeXRBs that would otherwise remain undetected. We conclude that this population 
emerges from the interplay between marginal accretion and the observational luminosity cutoff. This highlights the importance of selection effects and suggests that deeper X-ray surveys may uncover a faint, long-period tail if the suppression of accretion by the propeller effect is not very strong.

Finally, our best-fitting models favor a low natal kick regime for NSs forming BeXRBs, consistent with a substantial fraction of systems receiving negligible or modest kicks ($\lesssim 100\ \rm km\ s^{-1}$). 
We also find that the preferred models require Be stars to approach critical rotation with $\omega\equiv \Omega_{\rm rot}/\Omega_{\rm crit}\sim 0.8-0.9$, in agreement with observational constraints and the high completeness of the SMC catalog.

Overall, the SMC BeXRB population is best reproduced by models that combine (i) intermediate MT efficiency, (ii) AM recycling to the orbit via tides, (iii) predominantly stable MT via RLOF, (iv) small natal kicks, and (v) a realistic treatment of the propeller regime. The joint use of distribution-based likelihoods and number counts proves essential to disentangle degeneracies and to isolate the key physical processes governing the formation and evolution of BeXRBs in low-metallicity environments. In particular, our exploratory results highlight the importance of AM recycling and call for future studies on AM transport in binary stellar evolution. 

\begin{acknowledgements}
The authors thank the referee for useful comments and suggestions.
The authors acknowledge financial support from the German Excellence Strategy via the Heidelberg Cluster of Excellence (EXC 2181 - 390900948) STRUCTURES, from the European Research Council for the ERC Consolidator grant DEMOBLACK, under contract no. 770017. 
The authors also acknowledge support from the state of Baden-W\"urttemberg through bwHPC and the German Research Foundation (DFG) through grants INST 35/1597-1 FUGG and INST 35/1503-1 FUGG.
S.R. is funded by the Deut\-sche For\-schungs\-ge\-mein\-schaft (DFG, German Research Foundation) – project number 546677095. CS acknowledges financial support from the Alexander von Humboldt Foundation for the Humboldt Research Fellowship.
GI is supported by a fellowship grant from la Caixa Foundation (ID 100010434). The fellowship code is LCF/BQ/PI24/12040020.
The authors thanks the DEMOBlack group, MFV and MMA for useful discussion and comments.
We used \textsc{sevn} (\url{https://gitlab.com/sevncodes/sevn}) to generate our BeXRBs catalogs \citep{Spera2019,Mapelli2020,Iorio2023}. We adopted the stellar evolution tracks \texttt{SEVNtracks\_parsec\_ov05\_AGB} for regular stars and \texttt{SEVNtracks\_parsec\_pureHe36} for pure helium stars \citep{Costa2025}. Our results are based on \textsc{sevn} version 2.16.0, branch \texttt{SEVN\_liutest\_2.15\_Reff}, commit \texttt{015e45978859465719ae400b6d6f2ff2354548ea}.
This research made use of \textsc{NumPy} \citep{Harris20} and \textsc{SciPy} \citep{SciPy2020}. For the plots we used \textsc{Matplotlib} \citep{Hunter2007}.

\end{acknowledgements}

\bibliographystyle{aa}
\bibliography{ref}

@ARTICLE{giacobbo2019,
       author = {{Giacobbo}, Nicola and {Mapelli}, Michela},
        title = "{The impact of electron-capture supernovae on merging double neutron stars}",
      journal = {\mnras},
         year = 2019,
        month = jan,
       volume = {482},
       number = {2},
        pages = {2234-2243},
          doi = {10.1093/mnras/sty2848},
archivePrefix = {arXiv},
       eprint = {1805.11100},
 primaryClass = {astro-ph.SR},
       adsurl = {https://ui.adsabs.harvard.edu/abs/2019MNRAS.482.2234G}
}

@ARTICLE{beniamini2016b,
       author = {{Beniamini}, Paz and {Piran}, Tsvi},
        title = "{Formation of double neutron star systems as implied by observations}",
      journal = {\mnras},
         year = 2016,
        month = mar,
       volume = {456},
       number = {4},
        pages = {4089-4099},
          doi = {10.1093/mnras/stv2903},
archivePrefix = {arXiv},
       eprint = {1510.03111},
 primaryClass = {astro-ph.HE},
       adsurl = {https://ui.adsabs.harvard.edu/abs/2016MNRAS.456.4089B}
}

@ARTICLE{beniamini2016,
       author = {{Beniamini}, Paz and {Hotokezaka}, Kenta and {Piran}, Tsvi},
        title = "{Natal Kicks and Time Delays in Merging Neutron Star Binaries: Implications for r-process Nucleosynthesis in Ultra-faint Dwarfs and in the Milky Way}",
      journal = {\apjl},
         year = 2016,
        month = sep,
       volume = {829},
       number = {1},
          eid = {L13},
        pages = {L13},
          doi = {10.3847/2041-8205/829/1/L13},
archivePrefix = {arXiv},
       eprint = {1607.02148},
 primaryClass = {astro-ph.HE},
       adsurl = {https://ui.adsabs.harvard.edu/abs/2016ApJ...829L..13B}
}

@ARTICLE{Paxton2015,
       author = {{Paxton}, Bill and {Marchant}, Pablo and {Schwab}, Josiah and {Bauer}, Evan B. and {Bildsten}, Lars and {Cantiello}, Matteo and {Dessart}, Luc and {Farmer}, R. and {Hu}, H. and {Langer}, N. and {Townsend}, R.~H.~D. and {Townsley}, Dean M. and {Timmes}, F.~X.},
        title = "{Modules for Experiments in Stellar Astrophysics (MESA): Binaries, Pulsations, and Explosions}",
      journal = {\apjs},
         year = 2015,
        month = sep,
       volume = {220},
       number = {1},
          eid = {15},
        pages = {15},
          doi = {10.1088/0067-0049/220/1/15},
archivePrefix = {arXiv},
       eprint = {1506.03146},
 primaryClass = {astro-ph.SR},
       adsurl = {https://ui.adsabs.harvard.edu/abs/2015ApJS..220...15P}
}

@ARTICLE{Spera2017,
       author = {{Spera}, Mario and {Mapelli}, Michela},
        title = "{Very massive stars, pair-instability supernovae and intermediate-mass black holes with the sevn code}",
      journal = {\mnras},
         year = 2017,
        month = oct,
       volume = {470},
       number = {4},
        pages = {4739-4749},
          doi = {10.1093/mnras/stx1576},
archivePrefix = {arXiv},
       eprint = {1706.06109},
 primaryClass = {astro-ph.SR},
       adsurl = {https://ui.adsabs.harvard.edu/abs/2017MNRAS.470.4739S}
}

@ARTICLE{Spera2019,
       author = {{Spera}, Mario and {Mapelli}, Michela and {Giacobbo}, Nicola and {Trani}, Alessandro A. and {Bressan}, Alessandro and {Costa}, Guglielmo},
        title = "{Merging black hole binaries with the SEVN code}",
      journal = {\mnras},
         year = 2019,
        month = may,
       volume = {485},
       number = {1},
        pages = {889-907},
          doi = {10.1093/mnras/stz359},
archivePrefix = {arXiv},
       eprint = {1809.04605},
 primaryClass = {astro-ph.HE},
       adsurl = {https://ui.adsabs.harvard.edu/abs/2019MNRAS.485..889S}
}

@ARTICLE{Mapelli2020,
       author = {{Mapelli}, Michela and {Spera}, Mario and {Montanari}, Enrico and {Limongi}, Marco and {Chieffi}, Alessandro and {Giacobbo}, Nicola and {Bressan}, Alessandro and {Bouffanais}, Yann},
        title = "{Impact of the Rotation and Compactness of Progenitors on the Mass of Black Holes}",
      journal = {\apj},
         year = 2020,
        month = jan,
       volume = {888},
       number = {2},
          eid = {76},
        pages = {76},
          doi = {10.3847/1538-4357/ab584d},
archivePrefix = {arXiv},
       eprint = {1909.01371},
 primaryClass = {astro-ph.HE},
       adsurl = {https://ui.adsabs.harvard.edu/abs/2020ApJ...888...76M}
}

@ARTICLE{Iorio2023,
       author = {{Iorio}, Giuliano and {Mapelli}, Michela and {Costa}, Guglielmo and {Spera}, Mario and {Escobar}, Gast{\'o}n J. and {Sgalletta}, Cecilia and {Trani}, Alessandro A. and {Korb}, Erika and {Santoliquido}, Filippo and {Dall'Amico}, Marco and {Gaspari}, Nicola and {Bressan}, Alessandro},
        title = "{Compact object mergers: exploring uncertainties from stellar and binary evolution with SEVN}",
      journal = {\mnras},
         year = 2023,
        month = sep,
       volume = {524},
       number = {1},
        pages = {426-470},
          doi = {10.1093/mnras/stad1630},
archivePrefix = {arXiv},
       eprint = {2211.11774},
 primaryClass = {astro-ph.HE},
       adsurl = {https://ui.adsabs.harvard.edu/abs/2023MNRAS.524..426I}
}

@ARTICLE{Santoliquido2021,
       author = {{Santoliquido}, Filippo and {Mapelli}, Michela and {Giacobbo}, Nicola and {Bouffanais}, Yann and {Artale}, M. Celeste},
        title = "{The cosmic merger rate density of compact objects: impact of star formation, metallicity, initial mass function, and binary evolution}",
      journal = {\mnras},
         year = 2021,
        month = apr,
       volume = {502},
       number = {4},
        pages = {4877-4889},
          doi = {10.1093/mnras/stab280},
archivePrefix = {arXiv},
       eprint = {2009.03911},
 primaryClass = {astro-ph.HE},
       adsurl = {https://ui.adsabs.harvard.edu/abs/2021MNRAS.502.4877S}
}

@ARTICLE{Kroupa2001,
       author = {{Kroupa}, Pavel},
        title = "{On the variation of the initial mass function}",
      journal = {\mnras},
         year = 2001,
        month = apr,
       volume = {322},
       number = {2},
        pages = {231-246},
          doi = {10.1046/j.1365-8711.2001.04022.x},
archivePrefix = {arXiv},
       eprint = {astro-ph/0009005},
 primaryClass = {astro-ph},
       adsurl = {https://ui.adsabs.harvard.edu/abs/2001MNRAS.322..231K}
}

@ARTICLE{Fryer2012,
       author = {{Fryer}, Chris L. and {Belczynski}, Krzysztof and {Wiktorowicz}, Grzegorz and {Dominik}, Michal and {Kalogera}, Vicky and {Holz}, Daniel E.},
        title = "{Compact Remnant Mass Function: Dependence on the Explosion Mechanism and Metallicity}",
      journal = {\apj},
         year = 2012,
        month = apr,
       volume = {749},
       number = {1},
          eid = {91},
        pages = {91},
          doi = {10.1088/0004-637X/749/1/91},
archivePrefix = {arXiv},
       eprint = {1110.1726},
 primaryClass = {astro-ph.SR},
       adsurl = {https://ui.adsabs.harvard.edu/abs/2012ApJ...749...91F}
}

@ARTICLE{Chen2014,
       author = {{Chen}, Ke-Jung and {Woosley}, Stan and {Heger}, Alexander and {Almgren}, Ann and {Whalen}, Daniel J.},
        title = "{Two-dimensional Simulations of Pulsational Pair-instability Supernovae}",
      journal = {\apj},
         year = 2014,
        month = sep,
       volume = {792},
       number = {1},
          eid = {28},
        pages = {28},
          doi = {10.1088/0004-637X/792/1/28},
archivePrefix = {arXiv},
       eprint = {1402.4134},
 primaryClass = {astro-ph.HE},
       adsurl = {https://ui.adsabs.harvard.edu/abs/2014ApJ...792...28C}
}

@ARTICLE{Dotter2016,
       author = {{Dotter}, Aaron},
        title = "{MESA Isochrones and Stellar Tracks (MIST) 0: Methods for the Construction of Stellar Isochrones}",
      journal = {\apjs},
         year = 2016,
        month = jan,
       volume = {222},
       number = {1},
          eid = {8},
        pages = {8},
          doi = {10.3847/0067-0049/222/1/8},
archivePrefix = {arXiv},
       eprint = {1601.05144},
 primaryClass = {astro-ph.SR},
       adsurl = {https://ui.adsabs.harvard.edu/abs/2016ApJS..222....8D}
}

@ARTICLE{Choi2016,
       author = {{Choi}, Jieun and {Dotter}, Aaron and {Conroy}, Charlie and {Cantiello}, Matteo and {Paxton}, Bill and {Johnson}, Benjamin D.},
        title = "{Mesa Isochrones and Stellar Tracks (MIST). I. Solar-scaled Models}",
      journal = {\apj},
         year = 2016,
        month = jun,
       volume = {823},
       number = {2},
          eid = {102},
        pages = {102},
          doi = {10.3847/0004-637X/823/2/102},
archivePrefix = {arXiv},
       eprint = {1604.08592},
 primaryClass = {astro-ph.SR},
       adsurl = {https://ui.adsabs.harvard.edu/abs/2016ApJ...823..102C}
}

@ARTICLE{Harris20,
       author = {{Harris}, Charles R. and {Millman}, K. Jarrod and {van der Walt}, St{\'e}fan J. and {Gommers}, Ralf and {Virtanen}, Pauli and {Cournapeau}, David and {Wieser}, Eric and {Taylor}, Julian and {Berg}, Sebastian and {Smith}, Nathaniel J. and {Kern}, Robert and {Picus}, Matti and {Hoyer}, Stephan and {van Kerkwijk}, Marten H. and {Brett}, Matthew and {Haldane}, Allan and {del R{\'\i}o}, Jaime Fern{\'a}ndez and {Wiebe}, Mark and {Peterson}, Pearu and {G{\'e}rard-Marchant}, Pierre and {Sheppard}, Kevin and {Reddy}, Tyler and {Weckesser}, Warren and {Abbasi}, Hameer and {Gohlke}, Christoph and {Oliphant}, Travis E.},
        title = "{Array programming with NumPy}",
      journal = {\nat},
         year = 2020,
        month = sep,
       volume = {585},
       number = {7825},
        pages = {357-362},
          doi = {10.1038/s41586-020-2649-2},
archivePrefix = {arXiv},
       eprint = {2006.10256},
 primaryClass = {cs.MS},
       adsurl = {https://ui.adsabs.harvard.edu/abs/2020Natur.585..357H}
}

@ARTICLE{SciPy2020,
       author = {{Virtanen}, Pauli and {Gommers}, Ralf and {Oliphant}, Travis E. and {Haberland}, Matt and {Reddy}, Tyler and {Cournapeau}, David and {Burovski}, Evgeni and {Peterson}, Pearu and {Weckesser}, Warren and {Bright}, Jonathan and {van der Walt}, St{\'e}fan J. and {Brett}, Matthew and {Wilson}, Joshua and {Millman}, K. Jarrod and {Mayorov}, Nikolay and {Nelson}, Andrew R.~J. and {Jones}, Eric and {Kern}, Robert and {Larson}, Eric and {Carey}, C.~J. and {Polat}, {\.I}lhan and {Feng}, Yu and {Moore}, Eric W. and {VanderPlas}, Jake and {Laxalde}, Denis and {Perktold}, Josef and {Cimrman}, Robert and {Henriksen}, Ian and {Quintero}, E.~A. and {Harris}, Charles R. and {Archibald}, Anne M. and {Ribeiro}, Ant{\^o}nio H. and {Pedregosa}, Fabian and {van Mulbregt}, Paul and {SciPy 1. 0 Contributors}},
        title = "{SciPy 1.0: fundamental algorithms for scientific computing in Python}",
      journal = {Nature Methods},
         year = 2020,
        month = feb,
       volume = {17},
        pages = {261-272},
          doi = {10.1038/s41592-019-0686-2},
archivePrefix = {arXiv},
       eprint = {1907.10121},
 primaryClass = {cs.MS},
       adsurl = {https://ui.adsabs.harvard.edu/abs/2020NatMe..17..261V}
}

@ARTICLE{Hunter2007,
       author = {{Hunter}, John D.},
        title = "{Matplotlib: A 2D Graphics Environment}",
      journal = {Computing in Science and Engineering},
         year = 2007,
        month = may,
       volume = {9},
       number = {3},
        pages = {90-95},
          doi = {10.1109/MCSE.2007.55},
       adsurl = {https://ui.adsabs.harvard.edu/abs/2007CSE.....9...90H}
}

@ARTICLE{Costa2025,
       author = {{Costa}, G. and {Shepherd}, K.~G. and {Bressan}, A. and {Addari}, F. and {Chen}, Y. and {Fu}, X. and {Volpato}, G. and {Nguyen}, C.~T. and {Girardi}, L. and {Marigo}, P. and {Mazzi}, A. and {Pastorelli}, G. and {Trabucchi}, M. and {Bossini}, D. and {Zaggia}, S.},
        title = "{Evolutionary tracks, ejecta, and ionizing photons from intermediate-mass to very massive stars with PARSEC}",
      journal = {\aap},
         year = 2025,
        month = feb,
       volume = {694},
          eid = {A193},
        pages = {A193},
          doi = {10.1051/0004-6361/202452573},
archivePrefix = {arXiv},
       eprint = {2501.12917},
 primaryClass = {astro-ph.SR},
       adsurl = {https://ui.adsabs.harvard.edu/abs/2025A&A...694A.193C}
}

@ARTICLE{Giacobbo_Mapelli2020,
       author = {{Giacobbo}, Nicola and {Mapelli}, Michela},
        title = "{Revising Natal Kick Prescriptions in Population Synthesis Simulations}",
      journal = {\apj},
         year = 2020,
        month = mar,
       volume = {891},
       number = {2},
          eid = {141},
        pages = {141},
          doi = {10.3847/1538-4357/ab7335},
archivePrefix = {arXiv},
       eprint = {1909.06385},
 primaryClass = {astro-ph.HE},
       adsurl = {https://ui.adsabs.harvard.edu/abs/2020ApJ...891..141G}
}

@ARTICLE{Liu2023,
       author = {{Liu}, Boyuan and {Sartorio}, Nina S. and {Izzard}, Robert G. and {Fialkov}, Anastasia},
        title = "{Population synthesis of Be X-ray binaries: metallicity dependence of total X-ray outputs}",
      journal = {\mnras},
         year = 2024,
        month = jan,
       volume = {527},
       number = {3},
        pages = {5023-5048},
          doi = {10.1093/mnras/stad3475},
archivePrefix = {arXiv},
       eprint = {2308.06154},
 primaryClass = {astro-ph.HE},
       adsurl = {https://ui.adsabs.harvard.edu/abs/2024MNRAS.527.5023L}
}

@ARTICLE{Sana2012,
       author = {{Sana}, H. and {de Mink}, S.~E. and {de Koter}, A. and {Langer}, N. and {Evans}, C.~J. and {Gieles}, M. and {Gosset}, E. and {Izzard}, R.~G. and {Le Bouquin}, J. -B. and {Schneider}, F.~R.~N.},
        title = "{Binary Interaction Dominates the Evolution of Massive Stars}",
      journal = {Science},
         year = 2012,
        month = jul,
       volume = {337},
       number = {6093},
        pages = {444},
          doi = {10.1126/science.1223344},
archivePrefix = {arXiv},
       eprint = {1207.6397},
 primaryClass = {astro-ph.SR},
       adsurl = {https://ui.adsabs.harvard.edu/abs/2012Sci...337..444S}
}

@ARTICLE{Moe_DiStefano2017,
       author = {{Moe}, Maxwell and {Di Stefano}, Rosanne},
        title = "{Mind Your Ps and Qs: The Interrelation between Period (P) and Mass-ratio (Q) Distributions of Binary Stars}",
      journal = {\apjs},
         year = 2017,
        month = jun,
       volume = {230},
       number = {2},
          eid = {15},
        pages = {15},
          doi = {10.3847/1538-4365/aa6fb6},
archivePrefix = {arXiv},
       eprint = {1606.05347},
 primaryClass = {astro-ph.SR},
       adsurl = {https://ui.adsabs.harvard.edu/abs/2017ApJS..230...15M}
}

@ARTICLE{Schurmann2025,
       author = {{Sch{\"u}rmann}, C. and {Xu}, X. -T. and {Langer}, N. and {Lennon}, D. and {Kruckow}, M.~U. and {Antoniadis}, J. and {Haberl}, F. and {Herrero}, A. and {Kramer}, M. and {Schootemeijer}, A. and {Shenar}, T. and {Tauris}, T.~M. and {Wang}, C.},
        title = "{Populations of evolved massive binary stars in the Small Magellanic Cloud II: Predictions from rapid binary evolution}",
      journal = {arXiv e-prints},
         year = 2025,
        month = mar,
          eid = {arXiv:2503.23878},
        pages = {arXiv:2503.23878},
          doi = {10.48550/arXiv.2503.23878},
archivePrefix = {arXiv},
       eprint = {2503.23878},
 primaryClass = {astro-ph.SR},
       adsurl = {https://ui.adsabs.harvard.edu/abs/2025arXiv250323878S}
}

@ARTICLE{Rivinius2013,
       author = {{Rivinius}, Thomas and {Carciofi}, Alex C. and {Martayan}, Christophe},
        title = "{Classical Be stars. Rapidly rotating B stars with viscous Keplerian decretion disks}",
      journal = {\aapr},
         year = 2013,
        month = oct,
       volume = {21},
          eid = {69},
        pages = {69},
          doi = {10.1007/s00159-013-0069-0},
archivePrefix = {arXiv},
       eprint = {1310.3962},
 primaryClass = {astro-ph.SR},
       adsurl = {https://ui.adsabs.harvard.edu/abs/2013A&ARv..21...69R}
}

@ARTICLE{Sana13,
       author = {{Sana}, H. and {de Koter}, A. and {de Mink}, S.~E. and {Dunstall}, P.~R. and {Evans}, C.~J. and {H{\'e}nault-Brunet}, V. and {Ma{\'\i}z Apell{\'a}niz}, J. and {Ram{\'\i}rez-Agudelo}, O.~H. and {Taylor}, W.~D. and {Walborn}, N.~R. and {Clark}, J.~S. and {Crowther}, P.~A. and {Herrero}, A. and {Gieles}, M. and {Langer}, N. and {Lennon}, D.~J. and {Vink}, J.~S.},
        title = "{The VLT-FLAMES Tarantula Survey. VIII. Multiplicity properties of the O-type star population}",
      journal = {\aap},
         year = 2013,
        month = feb,
       volume = {550},
          eid = {A107},
        pages = {A107},
          doi = {10.1051/0004-6361/201219621},
archivePrefix = {arXiv},
       eprint = {1209.4638},
 primaryClass = {astro-ph.SR},
       adsurl = {https://ui.adsabs.harvard.edu/abs/2013A&A...550A.107S}
}

@ARTICLE{Gaudin25,
       author = {{Gaudin}, Thomas M. and {Kennea}, Jamie A. and {Coe}, M.~J. and {Evans}, Phil A.},
        title = "{Identification of New Candidate Be/X-Ray Binary Systems in the Small Magellanic Cloud via Analysis of the S-CUBED Source Catalog}",
      journal = {\apj},
         year = 2025,
        month = jul,
       volume = {988},
       number = {1},
          eid = {52},
        pages = {52},
          doi = {10.3847/1538-4357/addfd1},
archivePrefix = {arXiv},
       eprint = {2505.24766},
 primaryClass = {astro-ph.HE},
       adsurl = {https://ui.adsabs.harvard.edu/abs/2025ApJ...988...52G}
}

@ARTICLE{Haberl16,
       author = {{Haberl}, F. and {Sturm}, R.},
        title = "{High-mass X-ray binaries in the Small Magellanic Cloud}",
      journal = {\aap},
         year = 2016,
        month = feb,
       volume = {586},
          eid = {A81},
        pages = {A81},
          doi = {10.1051/0004-6361/201527326},
archivePrefix = {arXiv},
       eprint = {1511.00445},
 primaryClass = {astro-ph.GA},
       adsurl = {https://ui.adsabs.harvard.edu/abs/2016A&A...586A..81H}
}

@ARTICLE{Fortin23,
       author = {{Fortin}, Francis and {Garc{\'\i}a}, Federico and {Simaz Bunzel}, Adolfo and {Chaty}, Sylvain},
        title = "{A catalogue of high-mass X-ray binaries in the Galaxy: from the INTEGRAL to the Gaia era}",
      journal = {\aap},
         year = 2023,
        month = mar,
       volume = {671},
          eid = {A149},
        pages = {A149},
          doi = {10.1051/0004-6361/202245236},
archivePrefix = {arXiv},
       eprint = {2302.02656},
 primaryClass = {astro-ph.HE},
       adsurl = {https://ui.adsabs.harvard.edu/abs/2023A&A...671A.149F}
}

@ARTICLE{Dai2006,
       author = {{Dai}, Hai-Lang and {Liu}, Xi-Wei and {Li}, Xiang-Dong},
        title = "{Exploration of the P$_{s}$-P$_{orb}$ Relation for Wind-fed X-Ray Pulsars}",
      journal = {\apj},
         year = 2006,
        month = dec,
       volume = {653},
       number = {2},
        pages = {1410-1416},
          doi = {10.1086/508735},
archivePrefix = {arXiv},
       eprint = {astro-ph/0608621},
 primaryClass = {astro-ph},
       adsurl = {https://ui.adsabs.harvard.edu/abs/2006ApJ...653.1410D}
}

@ARTICLE{Raguzova2005,
       author = {{Raguzova}, N.~V. and {Popov}, S.~B.},
        title = "{Be X-ray binaries and candidates}",
      journal = {Astronomical and Astrophysical Transactions},
         year = 2005,
        month = jun,
       volume = {24},
       number = {3},
        pages = {151-185},
          doi = {10.1080/10556790500497311},
archivePrefix = {arXiv},
       eprint = {astro-ph/0505275},
 primaryClass = {astro-ph},
       adsurl = {https://ui.adsabs.harvard.edu/abs/2005A&AT...24..151R}
}

@ARTICLE{Disberg2025,
       author = {{Disberg}, Paul and {Mandel}, Ilya},
        title = "{The Kick Velocity Distribution of Isolated Neutron Stars}",
      journal = {\apjl},
         year = 2025,
        month = aug,
       volume = {989},
       number = {1},
          eid = {L8},
        pages = {L8},
          doi = {10.3847/2041-8213/adf286},
archivePrefix = {arXiv},
       eprint = {2505.22102},
 primaryClass = {astro-ph.HE},
       adsurl = {https://ui.adsabs.harvard.edu/abs/2025ApJ...989L...8D}
}

@ARTICLE{Hobbs2005,
       author = {{Hobbs}, G. and {Lorimer}, D.~R. and {Lyne}, A.~G. and {Kramer}, M.},
        title = "{A statistical study of 233 pulsar proper motions}",
      journal = {\mnras},
         year = 2005,
        month = jul,
       volume = {360},
       number = {3},
        pages = {974-992},
          doi = {10.1111/j.1365-2966.2005.09087.x},
archivePrefix = {arXiv},
       eprint = {astro-ph/0504584},
 primaryClass = {astro-ph},
       adsurl = {https://ui.adsabs.harvard.edu/abs/2005MNRAS.360..974H}
}

@article{kiel2008,
    author = {Kiel, Paul D. and Hurley, Jarrod R. and Bailes, Matthew and Murray, James R.},
    title = "{Populating the Galaxy with pulsars – I. Stellar and binary evolution}",
    journal = {Monthly Notices of the Royal Astronomical Society},
    volume = {388},
    number = {1},
    pages = {393-415},
    year = {2008},
    month = {07},
    issn = {0035-8711},
    doi = {10.1111/j.1365-2966.2008.13402.x},
    url = {https://doi.org/10.1111/j.1365-2966.2008.13402.x},
    eprint = {https://academic.oup.com/mnras/article-pdf/388/1/393/18722056/mnras0388-0393.pdf},
}

@ARTICLE{chattopadhyay2020,
       author = {{Chattopadhyay}, Debatri and {Stevenson}, Simon and {Hurley}, Jarrod R. and {Rossi}, Luca J. and {Flynn}, Chris},
        title = "{Modelling double neutron stars: radio and gravitational waves}",
      journal = {\mnras},
         year = 2020,
        month = may,
       volume = {494},
       number = {2},
        pages = {1587-1610},
          doi = {10.1093/mnras/staa756},
archivePrefix = {arXiv},
       eprint = {1912.02415},
 primaryClass = {astro-ph.HE},
       adsurl = {https://ui.adsabs.harvard.edu/abs/2020MNRAS.494.1587C}
}

@ARTICLE{illarionov1975,
       author = {{Illarionov}, A.~F. and {Sunyaev}, R.~A.},
        title = "{Why the Number of Galactic X-ray Stars Is so Small?}",
      journal = {\aap},
         year = 1975,
        month = feb,
       volume = {39},
        pages = {185},
       adsurl = {https://ui.adsabs.harvard.edu/abs/1975A&A....39..185I}
}

@article{konar1997,
    author = {Konar, Sushan and Bhattacharya, Dipankar},
    title = "{Magnetic field evolution of accreting neutron stars}",
    journal = {Monthly Notices of the Royal Astronomical Society},
    volume = {284},
    number = {2},
    pages = {311-317},
    year = {1997},
    month = {01},
    issn = {0035-8711},
    doi = {10.1093/mnras/284.2.311},
    url = {https://doi.org/10.1093/mnras/284.2.311},
    eprint = {https://academic.oup.com/mnras/article-pdf/284/2/311/18199846/284-2-311.pdf},
}

@ARTICLE{Cecilia2023,
       author = {{Sgalletta}, Cecilia and {Iorio}, Giuliano and {Mapelli}, Michela and {Artale}, M. Celeste and {Boco}, Lumen and {Chattopadhyay}, Debatri and {Lapi}, Andrea and {Possenti}, Andrea and {Rinaldi}, Stefano and {Spera}, Mario},
        title = "{Binary neutron star populations in the Milky Way}",
      journal = {\mnras},
         year = 2023,
        month = dec,
       volume = {526},
       number = {2},
        pages = {2210-2229},
          doi = {10.1093/mnras/stad2768},
archivePrefix = {arXiv},
       eprint = {2305.04955},
 primaryClass = {astro-ph.HE},
       adsurl = {https://ui.adsabs.harvard.edu/abs/2023MNRAS.526.2210S}
}

@ARTICLE{Nova2025,
       author = {{Nova}, Sola S. and {Richardson}, Noel D. and {Labadie-Bartz}, Jonathan and {Garcia Flores}, Samantha},
        title = "{The Birth of Be Star Disks II. A High-Resolution Spectroscopic Campaign and TESS Observations of an Outburst of the Classical Be star {\ensuremath{\lambda}} Pavonis}",
      journal = {arXiv e-prints},
         year = 2025,
        month = dec,
          eid = {arXiv:2512.08214},
        pages = {arXiv:2512.08214},
          doi = {10.48550/arXiv.2512.08214},
archivePrefix = {arXiv},
       eprint = {2512.08214},
 primaryClass = {astro-ph.SR},
       adsurl = {https://ui.adsabs.harvard.edu/abs/2025arXiv251208214N}
}

@ARTICLE{Kaaret2017,
       author = {{Kaaret}, Philip and {Feng}, Hua and {Roberts}, Timothy P.},
        title = "{Ultraluminous X-Ray Sources}",
      journal = {\araa},
         year = 2017,
        month = aug,
       volume = {55},
       number = {1},
        pages = {303-341},
          doi = {10.1146/annurev-astro-091916-055259},
archivePrefix = {arXiv},
       eprint = {1703.10728},
 primaryClass = {astro-ph.HE},
       adsurl = {https://ui.adsabs.harvard.edu/abs/2017ARA&A..55..303K}
}

@ARTICLE{Fabrika2021,
       author = {{Fabrika}, S.~N. and {Atapin}, K.~E. and {Vinokurov}, A.~S. and {Sholukhova}, O.~N.},
        title = "{Ultraluminous X-Ray Sources}",
      journal = {Astrophysical Bulletin},
         year = 2021,
        month = jan,
       volume = {76},
       number = {1},
        pages = {6-38},
          doi = {10.1134/S1990341321010077},
archivePrefix = {arXiv},
       eprint = {2105.10537},
 primaryClass = {astro-ph.GA},
       adsurl = {https://ui.adsabs.harvard.edu/abs/2021AstBu..76....6F}
}

@ARTICLE{King2023,
       author = {{King}, Andrew and {Lasota}, Jean-Pierre and {Middleton}, Matthew},
        title = "{Ultraluminous X-ray sources}",
      journal = {\nar},
         year = 2023,
        month = jun,
       volume = {96},
          eid = {101672},
        pages = {101672},
          doi = {10.1016/j.newar.2022.101672},
archivePrefix = {arXiv},
       eprint = {2302.10605},
 primaryClass = {astro-ph.HE},
       adsurl = {https://ui.adsabs.harvard.edu/abs/2023NewAR..9601672K}
}

@ARTICLE{Marchant2017,
       author = {{Marchant}, Pablo and {Langer}, Norbert and {Podsiadlowski}, Philipp and {Tauris}, Thomas M. and {de Mink}, Selma and {Mandel}, Ilya and {Moriya}, Takashi J.},
        title = "{Ultra-luminous X-ray sources and neutron-star-black-hole mergers from very massive close binaries at low metallicity}",
      journal = {\aap},
         year = 2017,
        month = aug,
       volume = {604},
          eid = {A55},
        pages = {A55},
          doi = {10.1051/0004-6361/201630188},
archivePrefix = {arXiv},
       eprint = {1705.04734},
 primaryClass = {astro-ph.HE},
       adsurl = {https://ui.adsabs.harvard.edu/abs/2017A&A...604A..55M}
}

@ARTICLE{Wiktorowicz2017,
       author = {{Wiktorowicz}, Grzegorz and {Sobolewska}, Ma{\l}gorzata and {Lasota}, Jean-Pierre and {Belczynski}, Krzysztof},
        title = "{The Origin of the Ultraluminous X-Ray Sources}",
      journal = {\apj},
         year = 2017,
        month = sep,
       volume = {846},
       number = {1},
          eid = {17},
        pages = {17},
          doi = {10.3847/1538-4357/aa821d},
archivePrefix = {arXiv},
       eprint = {1705.06155},
 primaryClass = {astro-ph.HE},
       adsurl = {https://ui.adsabs.harvard.edu/abs/2017ApJ...846...17W}
}

@ARTICLE{Wiktorowicz2019,
       author = {{Wiktorowicz}, Grzegorz and {Wyrzykowski}, {\L}ukasz and {Chruslinska}, Martyna and {Klencki}, Jakub and {Rybicki}, Krzysztof A. and {Belczynski}, Krzysztof},
        title = "{Populations of Stellar-mass Black Holes from Binary Systems}",
      journal = {\apj},
         year = 2019,
        month = nov,
       volume = {885},
       number = {1},
          eid = {1},
        pages = {1},
          doi = {10.3847/1538-4357/ab45e6},
archivePrefix = {arXiv},
       eprint = {1907.11431},
 primaryClass = {astro-ph.HE},
       adsurl = {https://ui.adsabs.harvard.edu/abs/2019ApJ...885....1W}
}

@ARTICLE{Wiktorowicz2021,
       author = {{Wiktorowicz}, Grzegorz and {Lasota}, Jean-Pierre and {Belczynski}, Krzysztof and {Lu}, Youjun and {Liu}, Jifeng and {I{\l}kiewicz}, Krystian},
        title = "{Wind-powered Ultraluminous X-ray Sources}",
      journal = {\apj},
         year = 2021,
        month = sep,
       volume = {918},
       number = {2},
          eid = {60},
        pages = {60},
          doi = {10.3847/1538-4357/ac0cf7},
archivePrefix = {arXiv},
       eprint = {2103.02026},
 primaryClass = {astro-ph.HE},
       adsurl = {https://ui.adsabs.harvard.edu/abs/2021ApJ...918...60W}
}

@ARTICLE{Shao2019,
       author = {{Shao}, Yong and {Li}, Xiang-Dong and {Dai}, Zi-Gao},
        title = "{A Population of Neutron Star Ultraluminous X-Ray Sources with a Helium Star Companion}",
      journal = {\apj},
         year = 2019,
        month = dec,
       volume = {886},
       number = {2},
          eid = {118},
        pages = {118},
          doi = {10.3847/1538-4357/ab4d50},
archivePrefix = {arXiv},
       eprint = {1910.06590},
 primaryClass = {astro-ph.HE},
       adsurl = {https://ui.adsabs.harvard.edu/abs/2019ApJ...886..118S}
}

@ARTICLE{Shao2020,
       author = {{Shao}, Yong and {Li}, Xiang-Dong},
        title = "{Population Synthesis of Black Hole X-Ray Binaries}",
      journal = {\apj},
         year = 2020,
        month = aug,
       volume = {898},
       number = {2},
          eid = {143},
        pages = {143},
          doi = {10.3847/1538-4357/aba118},
archivePrefix = {arXiv},
       eprint = {2006.15961},
 primaryClass = {astro-ph.HE},
       adsurl = {https://ui.adsabs.harvard.edu/abs/2020ApJ...898..143S}
}

@ARTICLE{Abdusalam2020,
       author = {{Abdusalam}, K. and {Ablimit}, Iminhaji and {Hashim}, P. and {L{\"u}}, G.-L. and {Mardini}, M.~K. and {Wang}, Z.-J.},
        title = "{Formation and Evolution of Ultraluminous X-Ray Pulsar Binaries to Pulsar-Neutron Star and Pulsar-White Dwarf Systems}",
      journal = {\apj},
         year = 2020,
        month = oct,
       volume = {902},
       number = {2},
          eid = {125},
        pages = {125},
          doi = {10.3847/1538-4357/abb5a8},
archivePrefix = {arXiv},
       eprint = {2009.13245},
 primaryClass = {astro-ph.HE},
       adsurl = {https://ui.adsabs.harvard.edu/abs/2020ApJ...902..125A}
}

@ARTICLE{Kuranov2021,
       author = {{Kuranov}, A.~G. and {Postnov}, K.~A. and {Yungelson}, L.~R.},
        title = "{Populations of Ultraluminous X-ray Sources in Galaxies: Origin and Evolution}",
      journal = {Astronomy Letters},
         year = 2021,
        month = dec,
       volume = {47},
       number = {12},
        pages = {831-855},
          doi = {10.1134/S1063773721120021},
archivePrefix = {arXiv},
       eprint = {2112.14833},
 primaryClass = {astro-ph.HE},
       adsurl = {https://ui.adsabs.harvard.edu/abs/2021AstL...47..831K}
}

@ARTICLE{Misra2020,
       author = {{Misra}, D. and {Fragos}, T. and {Tauris}, T.~M. and {Zapartas}, E. and {Aguilera-Dena}, D.~R.},
        title = "{The origin of pulsating ultra-luminous X-ray sources: Low- and intermediate-mass X-ray binaries containing neutron star accretors}",
      journal = {\aap},
         year = 2020,
        month = oct,
       volume = {642},
          eid = {A174},
        pages = {A174},
          doi = {10.1051/0004-6361/202038070},
archivePrefix = {arXiv},
       eprint = {2004.01205},
 primaryClass = {astro-ph.HE},
       adsurl = {https://ui.adsabs.harvard.edu/abs/2020A&A...642A.174M}
}

@ARTICLE{Mao2026,
       author = {{Mao}, Ying-Han and {Li}, Xiang-Dong},
        title = "{How Beaming Shapes the Demographics of Ultraluminous X-ray Sources?}",
      journal = {arXiv e-prints},
         year = 2026,
        month = jan,
          eid = {arXiv:2601.09425},
        pages = {arXiv:2601.09425},
          doi = {10.48550/arXiv.2601.09425},
archivePrefix = {arXiv},
       eprint = {2601.09425},
 primaryClass = {astro-ph.HE},
       adsurl = {https://ui.adsabs.harvard.edu/abs/2026arXiv260109425M}
}

@ARTICLE{King2001,
       author = {{King}, A.~R. and {Davies}, M.~B. and {Ward}, M.~J. and {Fabbiano}, G. and {Elvis}, M.},
        title = "{Ultraluminous X-Ray Sources in External Galaxies}",
      journal = {\apjl},
         year = 2001,
        month = may,
       volume = {552},
       number = {2},
        pages = {L109-L112},
          doi = {10.1086/320343},
archivePrefix = {arXiv},
       eprint = {astro-ph/0104333},
 primaryClass = {astro-ph},
       adsurl = {https://ui.adsabs.harvard.edu/abs/2001ApJ...552L.109K}
}

@ARTICLE{King2009,
       author = {{King}, A.~R.},
        title = "{Masses, beaming and Eddington ratios in ultraluminous X-ray sources}",
      journal = {\mnras},
         year = 2009,
        month = feb,
       volume = {393},
       number = {1},
        pages = {L41-L44},
          doi = {10.1111/j.1745-3933.2008.00594.x},
archivePrefix = {arXiv},
       eprint = {0811.1473},
 primaryClass = {astro-ph},
       adsurl = {https://ui.adsabs.harvard.edu/abs/2009MNRAS.393L..41K}
}

@ARTICLE{King2017,
       author = {{King}, Andrew and {Lasota}, Jean-Pierre and {Klu{\'z}niak}, W{\l}odek},
        title = "{Pulsing ULXs: tip of the iceberg?}",
      journal = {\mnras},
         year = 2017,
        month = jun,
       volume = {468},
       number = {1},
        pages = {L59-L62},
          doi = {10.1093/mnrasl/slx020},
archivePrefix = {arXiv},
       eprint = {1702.00808},
 primaryClass = {astro-ph.HE},
       adsurl = {https://ui.adsabs.harvard.edu/abs/2017MNRAS.468L..59K}
}

@ARTICLE{King2019,
       author = {{King}, Andrew and {Lasota}, Jean-Pierre},
        title = "{No magnetars in ULXs}",
      journal = {\mnras},
         year = 2019,
        month = may,
       volume = {485},
       number = {3},
        pages = {3588-3594},
          doi = {10.1093/mnras/stz720},
archivePrefix = {arXiv},
       eprint = {1903.03624},
 primaryClass = {astro-ph.HE},
       adsurl = {https://ui.adsabs.harvard.edu/abs/2019MNRAS.485.3588K}
}

@ARTICLE{King2024,
       author = {{King}, Andrew and {Lasota}, Jean-Pierre},
        title = "{SN 2022jli: The ultraluminous birth of a low-mass X-ray binary}",
      journal = {\aap},
         year = 2024,
        month = feb,
       volume = {682},
          eid = {L22},
        pages = {L22},
          doi = {10.1051/0004-6361/202349002},
archivePrefix = {arXiv},
       eprint = {2402.09509},
 primaryClass = {astro-ph.HE},
       adsurl = {https://ui.adsabs.harvard.edu/abs/2024A&A...682L..22K}
}

@INCOLLECTION{Gilfanov2022,
       author = {{Gilfanov}, Marat and {Fabbiano}, Giuseppina and {Lehmer}, Bret and {Zezas}, Andreas},
        title = "{X-Ray Binaries in External Galaxies}",
    booktitle = {Handbook of X-ray and Gamma-ray Astrophysics},
         year = 2022,
       editor = {{Bambi}, Cosimo and {Sangangelo}, Andrea},
          eid = {105},
        pages = {105},
          doi = {10.1007/978-981-16-4544-0_108-1},
       adsurl = {https://ui.adsabs.harvard.edu/abs/2022hxga.book..105G}
}

@article{lorimer2008binary,
  title={Binary and millisecond pulsars},
  author={Lorimer, Duncan R},
  journal={Living reviews in relativity},
  volume={11},
  number={1},
  pages={8},
  year={2008},
  publisher={Springer}
}

@ARTICLE{Xu2019,
       author = {{Xu}, Xiao-Tian and {Li}, Xiang-Dong},
        title = "{On the Bimodal Spin-period Distribution of Be/X-Ray Pulsars}",
      journal = {\apj},
         year = 2019,
        month = feb,
       volume = {872},
       number = {1},
          eid = {102},
        pages = {102},
          doi = {10.3847/1538-4357/aafee0},
archivePrefix = {arXiv},
       eprint = {1901.04707},
 primaryClass = {astro-ph.HE},
       adsurl = {https://ui.adsabs.harvard.edu/abs/2019ApJ...872..102X}
}

@ARTICLE{Kennea2021,
       author = {{Kennea}, J.~A. and {Coe}, M.~J. and {Evans}, P.~A. and {Townsend}, L.~J. and {Campbell}, Z.~A. and {Udalski}, A.},
        title = "{Swift J011511.0-725611: discovery of a rare Be star/white dwarf binary system in the SMC}",
      journal = {\mnras},
         year = 2021,
        month = nov,
       volume = {508},
       number = {1},
        pages = {781-788},
          doi = {10.1093/mnras/stab2632},
archivePrefix = {arXiv},
       eprint = {2109.05307},
 primaryClass = {astro-ph.HE},
       adsurl = {https://ui.adsabs.harvard.edu/abs/2021MNRAS.508..781K}
}

@ARTICLE{Hohle2010,
       author = {{Hohle}, M.~M. and {Neuh{\"a}user}, R. and {Schutz}, B.~F.},
        title = "{Masses and luminosities of O- and B-type stars and red supergiants}",
      journal = {Astronomische Nachrichten},
         year = 2010,
        month = apr,
       volume = {331},
       number = {4},
        pages = {349},
          doi = {10.1002/asna.200911355},
archivePrefix = {arXiv},
       eprint = {1003.2335},
 primaryClass = {astro-ph.SR},
       adsurl = {https://ui.adsabs.harvard.edu/abs/2010AN....331..349H}
}

@ARTICLE{Vieira2017,
       author = {{Vieira}, R.~G. and {Carciofi}, A.~C. and {Bjorkman}, J.~E. and {Rivinius}, Th. and {Baade}, D. and {R{\'\i}mulo}, L.~R.},
        title = "{The life cycles of Be viscous decretion discs: time-dependent modelling of infrared continuum observations}",
      journal = {\mnras},
         year = 2017,
        month = jan,
       volume = {464},
       number = {3},
        pages = {3071-3089},
          doi = {10.1093/mnras/stw2542},
archivePrefix = {arXiv},
       eprint = {1707.02861},
 primaryClass = {astro-ph.SR},
       adsurl = {https://ui.adsabs.harvard.edu/abs/2017MNRAS.464.3071V}
}

@ARTICLE{Tetarenko2016,
       author = {{Tetarenko}, B.~E. and {Sivakoff}, G.~R. and {Heinke}, C.~O. and {Gladstone}, J.~C.},
        title = "{WATCHDOG: A Comprehensive All-sky Database of Galactic Black Hole X-ray Binaries}",
      journal = {\apjs},
         year = 2016,
        month = feb,
       volume = {222},
       number = {2},
          eid = {15},
        pages = {15},
          doi = {10.3847/0067-0049/222/2/15},
archivePrefix = {arXiv},
       eprint = {1512.00778},
 primaryClass = {astro-ph.HE},
       adsurl = {https://ui.adsabs.harvard.edu/abs/2016ApJS..222...15T}
}

@ARTICLE{Orosz2007,
       author = {{Orosz}, Jerome A. and {McClintock}, Jeffrey E. and {Narayan}, Ramesh and {Bailyn}, Charles D. and {Hartman}, Joel D. and {Macri}, Lucas and {Liu}, Jiefeng and {Pietsch}, Wolfgang and {Remillard}, Ronald A. and {Shporer}, Avi and {Mazeh}, Tsevi},
        title = "{A 15.65-solar-mass black hole in an eclipsing binary in the nearby spiral galaxy M 33}",
      journal = {\nat},
         year = 2007,
        month = oct,
       volume = {449},
       number = {7164},
        pages = {872-875},
          doi = {10.1038/nature06218},
archivePrefix = {arXiv},
       eprint = {0710.3165},
 primaryClass = {astro-ph},
       adsurl = {https://ui.adsabs.harvard.edu/abs/2007Natur.449..872O}
}

@ARTICLE{Schootemeijer2021,
       author = {{Schootemeijer}, A. and {Langer}, N. and {Lennon}, D. and {Evans}, C.~J. and {Crowther}, P.~A. and {Geen}, S. and {Howarth}, I. and {de Koter}, A. and {Menten}, K.~M. and {Vink}, J.~S.},
        title = "{A dearth of young and bright massive stars in the Small Magellanic Cloud}",
      journal = {\aap},
         year = 2021,
        month = feb,
       volume = {646},
          eid = {A106},
        pages = {A106},
          doi = {10.1051/0004-6361/202038789},
archivePrefix = {arXiv},
       eprint = {2012.05913},
 primaryClass = {astro-ph.GA},
       adsurl = {https://ui.adsabs.harvard.edu/abs/2021A&A...646A.106S}
}

@ARTICLE{Rubele2015,
       author = {{Rubele}, Stefano and {Girardi}, L{\'e}o and {Kerber}, Leandro and {Cioni}, Maria-Rosa L. and {Piatti}, Andr{\'e}s E. and {Zaggia}, Simone and {Bekki}, Kenji and {Bressan}, Alessandro and {Clementini}, Gisella and {de Grijs}, Richard and {Emerson}, Jim P. and {Groenewegen}, Martin A.~T. and {Ivanov}, Valentin D. and {Marconi}, Marcella and {Marigo}, Paola and {Moretti}, Maria-Ida and {Ripepi}, Vincenzo and {Subramanian}, Smitha and {Tatton}, Benjamin L. and {van Loon}, Jacco Th.},
        title = "{The VMC survey - XIV. First results on the look-back time star formation rate tomography of the Small Magellanic Cloud}",
      journal = {\mnras},
         year = 2015,
        month = may,
       volume = {449},
       number = {1},
        pages = {639-661},
          doi = {10.1093/mnras/stv141},
archivePrefix = {arXiv},
       eprint = {1501.05347},
 primaryClass = {astro-ph.SR},
       adsurl = {https://ui.adsabs.harvard.edu/abs/2015MNRAS.449..639R}
}

@ARTICLE{Misra2023b,
       author = {{Misra}, Devina and {Kovlakas}, Konstantinos and {Fragos}, Tassos and {Lazzarini}, Margaret and {Bavera}, Simone S. and {Lehmer}, Bret D. and {Zezas}, Andreas and {Zapartas}, Emmanouil and {Xing}, Zepei and {Andrews}, Jeff J. and {Dotter}, Aaron and {Rocha}, Kyle Akira and {Srivastava}, Philipp M. and {Sun}, Meng},
        title = "{X-ray luminosity function of high-mass X-ray binaries: Studying the signatures of different physical processes using detailed binary evolution calculations}",
      journal = {\aap},
         year = 2023,
        month = apr,
       volume = {672},
          eid = {A99},
        pages = {A99},
          doi = {10.1051/0004-6361/202244929},
archivePrefix = {arXiv},
       eprint = {2209.05505},
 primaryClass = {astro-ph.HE},
       adsurl = {https://ui.adsabs.harvard.edu/abs/2023A&A...672A..99M}
}

@ARTICLE{Bavera2020,
       author = {{Bavera}, Simone S. and {Fragos}, Tassos and {Qin}, Ying and {Zapartas}, Emmanouil and {Neijssel}, Coenraad J. and {Mandel}, Ilya and {Batta}, Aldo and {Gaebel}, Sebastian M. and {Kimball}, Chase and {Stevenson}, Simon},
        title = "{The origin of spin in binary black holes. Predicting the distributions of the main observables of Advanced LIGO}",
      journal = {\aap},
         year = 2020,
        month = mar,
       volume = {635},
          eid = {A97},
        pages = {A97},
          doi = {10.1051/0004-6361/201936204},
archivePrefix = {arXiv},
       eprint = {1906.12257},
 primaryClass = {astro-ph.HE},
       adsurl = {https://ui.adsabs.harvard.edu/abs/2020A&A...635A..97B}
}

@ARTICLE{Coe2020,
       author = {{Coe}, M.~J. and {Kennea}, J.~A. and {Evans}, P.~A. and {Udalski}, A.},
        title = "{Swift J004427.3-734801 - a probable Be/white dwarf system in the Small Magellanic Cloud}",
      journal = {\mnras},
         year = 2020,
        month = sep,
       volume = {497},
       number = {1},
        pages = {L50-L55},
          doi = {10.1093/mnrasl/slaa112},
archivePrefix = {arXiv},
       eprint = {2005.02891},
 primaryClass = {astro-ph.HE},
       adsurl = {https://ui.adsabs.harvard.edu/abs/2020MNRAS.497L..50C}
}

@ARTICLE{Gaudin2024,
       author = {{Gaudin}, T.~M. and {Coe}, M.~J. and {Kennea}, J.~A. and {Monageng}, I.~M. and {Buckley}, D.~A.~H. and {Udalski}, A. and {Evans}, P.~A.},
        title = "{CXOU J005245.0-722844: discovery of a Be star/white dwarf binary system in the SMC via a very fast, super-Eddington X-ray outburst event}",
      journal = {\mnras},
         year = 2024,
        month = nov,
       volume = {534},
       number = {3},
        pages = {1937-1948},
          doi = {10.1093/mnras/stae2176},
archivePrefix = {arXiv},
       eprint = {2408.01388},
 primaryClass = {astro-ph.HE},
       adsurl = {https://ui.adsabs.harvard.edu/abs/2024MNRAS.534.1937G}
}

@BOOK{Lang1992,
       author = {{Lang}, Kenneth R.},
        title = "{Astrophysical Data I. Planets and Stars.}",
         year = 1992,
       adsurl = {https://ui.adsabs.harvard.edu/abs/1992adps.book.....L}
}

@ARTICLE{Hurley2000,
       author = {{Hurley}, Jarrod R. and {Pols}, Onno R. and {Tout}, Christopher A.},
        title = "{Comprehensive analytic formulae for stellar evolution as a function of mass and metallicity}",
      journal = {\mnras},
         year = 2000,
        month = jul,
       volume = {315},
       number = {3},
        pages = {543-569},
          doi = {10.1046/j.1365-8711.2000.03426.x},
archivePrefix = {arXiv},
       eprint = {astro-ph/0001295},
 primaryClass = {astro-ph},
       adsurl = {https://ui.adsabs.harvard.edu/abs/2000MNRAS.315..543H}
}

@ARTICLE{Langer2020,
       author = {{Langer}, N. and {Sch{\"u}rmann}, C. and {Stoll}, K. and {Marchant}, P. and {Lennon}, D.~J. and {Mahy}, L. and {de Mink}, S.~E. and {Quast}, M. and {Riedel}, W. and {Sana}, H. and {Schneider}, P. and {Schootemeijer}, A. and {Wang}, C. and {Almeida}, L.~A. and {Bestenlehner}, J.~M. and {Bodensteiner}, J. and {Castro}, N. and {Clark}, S. and {Crowther}, P.~A. and {Dufton}, P. and {Evans}, C.~J. and {Fossati}, L. and {Gr{\"a}fener}, G. and {Grassitelli}, L. and {Grin}, N. and {Hastings}, B. and {Herrero}, A. and {de Koter}, A. and {Menon}, A. and {Patrick}, L. and {Puls}, J. and {Renzo}, M. and {Sander}, A.~A.~C. and {Schneider}, F.~R.~N. and {Sen}, K. and {Shenar}, T. and {Sim{\'o}n-D{\'\i}as}, S. and {Tauris}, T.~M. and {Tramper}, F. and {Vink}, J.~S. and {Xu}, X.-T.},
        title = "{Properties of OB star-black hole systems derived from detailed binary evolution models}",
      journal = {\aap},
         year = 2020,
        month = jun,
       volume = {638},
          eid = {A39},
        pages = {A39},
          doi = {10.1051/0004-6361/201937375},
archivePrefix = {arXiv},
       eprint = {1912.09826},
 primaryClass = {astro-ph.SR},
       adsurl = {https://ui.adsabs.harvard.edu/abs/2020A&A...638A..39L}
}

@ARTICLE{Zahn1977,
       author = {{Zahn}, J.-P.},
        title = "{Tidal friction in close binary systems.}",
      journal = {\aap},
         year = 1977,
        month = may,
       volume = {57},
        pages = {383-394},
       adsurl = {https://ui.adsabs.harvard.edu/abs/1977A&A....57..383Z}
}

@ARTICLE{Lechien2025,
       author = {{Lechien}, Thibault and {de Mink}, Selma E. and {Valli}, Ruggero and {Rubio}, Amanda C. and {van Son}, Lieke A.~C. and {Klement}, Robert and {Jin}, Harim and {Pols}, Onno},
        title = "{Binary Stars Take What They Get: Evidence for Efficient Mass Transfer from Stripped Stars with Rapidly Rotating Companions}",
      journal = {\apjl},
         year = 2025,
        month = sep,
       volume = {990},
       number = {2},
          eid = {L51},
        pages = {L51},
          doi = {10.3847/2041-8213/adfdd4},
archivePrefix = {arXiv},
       eprint = {2505.14780},
 primaryClass = {astro-ph.SR},
       adsurl = {https://ui.adsabs.harvard.edu/abs/2025ApJ...990L..51L}
}

@ARTICLE{Tout1997,
       author = {{Tout}, Christopher A. and {Aarseth}, Sverre J. and {Pols}, Onno R. and {Eggleton}, Peter P.},
        title = "{Rapid binary star evolution for N-body simulations and population synthesis}",
      journal = {\mnras},
         year = 1997,
        month = nov,
       volume = {291},
       number = {4},
        pages = {732-748},
          doi = {10.1093/mnras/291.4.732},
       adsurl = {https://ui.adsabs.harvard.edu/abs/1997MNRAS.291..732T}
}

@ARTICLE{Hurley2002,
       author = {{Hurley}, Jarrod R. and {Tout}, Christopher A. and {Pols}, Onno R.},
        title = "{Evolution of binary stars and the effect of tides on binary populations}",
      journal = {\mnras},
         year = 2002,
        month = feb,
       volume = {329},
       number = {4},
        pages = {897-928},
          doi = {10.1046/j.1365-8711.2002.05038.x},
archivePrefix = {arXiv},
       eprint = {astro-ph/0201220},
 primaryClass = {astro-ph},
       adsurl = {https://ui.adsabs.harvard.edu/abs/2002MNRAS.329..897H}
}

@ARTICLE{Packet1981,
       author = {{Packet}, W.},
        title = "{On the spin-up of the mass accreting component in a close binary system}",
      journal = {\aap},
         year = 1981,
        month = sep,
       volume = {102},
       number = {1},
        pages = {17-19},
       adsurl = {https://ui.adsabs.harvard.edu/abs/1981A&A...102...17P}
}

@INPROCEEDINGS{Langer2003,
       author = {{Langer}, N. and {Wellstein}, S. and {Petrovic}, J.},
        title = "{On the evolution of massive close binaries}",
    booktitle = {A Massive Star Odyssey: From Main Sequence to Supernova},
         year = 2003,
       editor = {{van der Hucht}, Karel and {Herrero}, Artemio and {Esteban}, C{\'e}sar},
       series = {IAU Symposium},
       volume = {212},
        month = jan,
        pages = {275},
       adsurl = {https://ui.adsabs.harvard.edu/abs/2003IAUS..212..275L}
}

@ARTICLE{Petrovic2005a,
       author = {{Petrovic}, J. and {Langer}, N. and {van der Hucht}, K.~A.},
        title = "{Constraining the mass transfer in massive binaries through progenitor evolution models of Wolf-Rayet+O binaries}",
      journal = {\aap},
         year = 2005,
        month = jun,
       volume = {435},
       number = {3},
        pages = {1013-1030},
          doi = {10.1051/0004-6361:20042368},
archivePrefix = {arXiv},
       eprint = {astro-ph/0504242},
 primaryClass = {astro-ph},
       adsurl = {https://ui.adsabs.harvard.edu/abs/2005A&A...435.1013P}
}

@ARTICLE{Ghodla2023,
       author = {{Ghodla}, Sohan and {Eldridge}, J.~J. and {Stanway}, Elizabeth R. and {Stevance}, H{\'e}lo{\"\i}se F.},
        title = "{Evaluating chemically homogeneous evolution in stellar binaries: electromagnetic implications - ionizing photons, SLSN-I, GRB, Ic-BL}",
      journal = {\mnras},
         year = 2023,
        month = jan,
       volume = {518},
       number = {1},
        pages = {860-877},
          doi = {10.1093/mnras/stac3177},
archivePrefix = {arXiv},
       eprint = {2208.03999},
 primaryClass = {astro-ph.SR},
       adsurl = {https://ui.adsabs.harvard.edu/abs/2023MNRAS.518..860G}
}

@ARTICLE{Paczynki1972,
       author = {{Paczy{\'n}ski}, B. and {Sienkiewicz}, R.},
        title = "{Evolution of Close Binaries VIII. Mass Exchange on the Dynamical Time Scale}",
      journal = {\actaa},
         year = 1972,
        month = jan,
       volume = {22},
        pages = {73-91},
       adsurl = {https://ui.adsabs.harvard.edu/abs/1972AcA....22...73P}
}

@ARTICLE{Artymowicz1994,
       author = {{Artymowicz}, Pawel and {Lubow}, Stephen H.},
        title = "{Dynamics of Binary-Disk Interaction. I. Resonances and Disk Gap Sizes}",
      journal = {\apj},
         year = 1994,
        month = feb,
       volume = {421},
        pages = {651},
          doi = {10.1086/173679},
       adsurl = {https://ui.adsabs.harvard.edu/abs/1994ApJ...421..651A}
}

@ARTICLE{Vinciguerra2020,
       author = {{Vinciguerra}, Serena and {Neijssel}, Coenraad J. and {Vigna-G{\'o}mez}, Alejandro and {Mandel}, Ilya and {Podsiadlowski}, Philipp and {Maccarone}, Thomas J. and {Nicholl}, Matt and {Kingdon}, Samuel and {Perry}, Alice and {Salemi}, Francesco},
        title = "{Be X-ray binaries in the SMC as indicators of mass-transfer efficiency}",
      journal = {\mnras},
         year = 2020,
        month = nov,
       volume = {498},
       number = {4},
        pages = {4705-4720},
          doi = {10.1093/mnras/staa2177},
archivePrefix = {arXiv},
       eprint = {2003.00195},
 primaryClass = {astro-ph.HE},
       adsurl = {https://ui.adsabs.harvard.edu/abs/2020MNRAS.498.4705V}
}

@ARTICLE{Popham1991,
       author = {{Popham}, Robert and {Narayan}, Ramesh},
        title = "{Does Accretion Cease When a Star Approaches Breakup?}",
      journal = {\apj},
         year = 1991,
        month = apr,
       volume = {370},
        pages = {604},
          doi = {10.1086/169847},
       adsurl = {https://ui.adsabs.harvard.edu/abs/1991ApJ...370..604P}
}

@ARTICLE{Paczynski1991,
       author = {{Paczynski}, Bohdan},
        title = "{A Polytropic Model of an Accretion Disk, a Boundary Layer, and a Star}",
      journal = {\apj},
         year = 1991,
        month = apr,
       volume = {370},
        pages = {597},
          doi = {10.1086/169846},
       adsurl = {https://ui.adsabs.harvard.edu/abs/1991ApJ...370..597P}
}

@ARTICLE{Langer1998,
       author = {{Langer}, N.},
        title = "{Coupled mass and angular momentum loss of massive main sequence stars}",
      journal = {\aap},
         year = 1998,
        month = jan,
       volume = {329},
        pages = {551-558},
       adsurl = {https://ui.adsabs.harvard.edu/abs/1998A&A...329..551L}
}

@ARTICLE{Rocha2024,
       author = {{Rocha}, Kyle Akira and {Kalogera}, Vicky and {Doctor}, Zoheyr and {Andrews}, Jeff J. and {Sun}, Meng and {Gossage}, Seth and {Bavera}, Simone S. and {Fragos}, Tassos and {Kovlakas}, Konstantinos and {Kruckow}, Matthias U. and {Misra}, Devina and {Srivastava}, Philipp M. and {Xing}, Zepei and {Zapartas}, Emmanouil},
        title = "{To Be or Not To Be: The Role of Rotation in Modeling Galactic Be X-Ray Binaries}",
      journal = {\apj},
         year = 2024,
        month = aug,
       volume = {971},
       number = {2},
          eid = {133},
        pages = {133},
          doi = {10.3847/1538-4357/ad5955},
archivePrefix = {arXiv},
       eprint = {2403.07172},
 primaryClass = {astro-ph.HE},
       adsurl = {https://ui.adsabs.harvard.edu/abs/2024ApJ...971..133R}
}

@MISC{Reynolds2021,
       author = {{Reynolds}, Mark and {Degenaar}, Nathalie and {Miller}, Jon Matthew and {Walton}, Dominic and {van den Eijnden}, Jakob},
        title = "{Probing Super-Eddington Outflows via Accreting Galactic BeXRBs}",
 howpublished = {HST Proposal. Cycle 29, ID. \#16844},
         year = 2021,
        month = sep,
        pages = {16844},
       adsurl = {https://ui.adsabs.harvard.edu/abs/2021hst..prop16844R}
}

@ARTICLE{Huang2010,
       author = {{Huang}, Wenjin and {Gies}, D.~R. and {McSwain}, M.~V.},
        title = "{A Stellar Rotation Census of B Stars: From ZAMS to TAMS}",
      journal = {\apj},
         year = 2010,
        month = oct,
       volume = {722},
       number = {1},
        pages = {605-619},
          doi = {10.1088/0004-637X/722/1/605},
archivePrefix = {arXiv},
       eprint = {1008.1761},
 primaryClass = {astro-ph.SR},
       adsurl = {https://ui.adsabs.harvard.edu/abs/2010ApJ...722..605H}
}

@article{Abac2025,
    author = "Abac, A. G. and others",
    collaboration = "LIGO Scientific, VIRGO, KAGRA",
    title = "{GWTC-4.0: Updating the Gravitational-Wave Transient Catalog with Observations from the First Part of the Fourth LIGO-Virgo-KAGRA Observing Run}",
    eprint = "2508.18082",
    archivePrefix = "arXiv",
    primaryClass = "gr-qc",
    reportNumber = "LIGO-P2400386",
    doi = "10.3847/2041-8213/ae2c74",
    journal = "Astrophys. J. Lett.",
    volume = "1004",
    number = "2",
    pages = "L22",
    year = "2026"
}

@INCOLLECTION{Mapelli2021,
       author = {{Mapelli}, Michela},
        title = "{Formation Channels of Single and Binary Stellar-Mass Black Holes}",
    booktitle = {Handbook of Gravitational Wave Astronomy},
         year = 2021,
       editor = {{Bambi}, Cosimo and {Katsanevas}, Stavros and {Kokkotas}, Konstantinos D.},
          eid = {16},
        pages = {16},
          doi = {10.1007/978-981-15-4702-7_16-1},
       adsurl = {https://ui.adsabs.harvard.edu/abs/2021hgwa.bookE..16M}
}

@ARTICLE{Postnov2014,
       author = {{Postnov}, Konstantin A. and {Yungelson}, Lev R.},
        title = "{The Evolution of Compact Binary Star Systems}",
      journal = {Living Reviews in Relativity},
         year = 2014,
        month = dec,
       volume = {17},
       number = {1},
          eid = {3},
        pages = {3},
          doi = {10.12942/lrr-2014-3},
archivePrefix = {arXiv},
       eprint = {1403.4754},
 primaryClass = {astro-ph.HE},
       adsurl = {https://ui.adsabs.harvard.edu/abs/2014LRR....17....3P}
}

@ARTICLE{Schneider2021,
       author = {{Schneider}, F.~R.~N. and {Podsiadlowski}, Ph. and {M{\"u}ller}, B.},
        title = "{Pre-supernova evolution, compact-object masses, and explosion properties of stripped binary stars}",
      journal = {\aap},
         year = 2021,
        month = jan,
       volume = {645},
          eid = {A5},
        pages = {A5},
          doi = {10.1051/0004-6361/202039219},
archivePrefix = {arXiv},
       eprint = {2008.08599},
 primaryClass = {astro-ph.SR},
       adsurl = {https://ui.adsabs.harvard.edu/abs/2021A&A...645A...5S}
}

@ARTICLE{Bavera2021,
       author = {{Bavera}, Simone S. and {Fragos}, Tassos and {Zevin}, Michael and {Berry}, Christopher P.~L. and {Marchant}, Pablo and {Andrews}, Jeff J. and {Coughlin}, Scott and {Dotter}, Aaron and {Kovlakas}, Konstantinos and {Misra}, Devina and {Serra-Perez}, Juan G. and {Qin}, Ying and {Rocha}, Kyle A. and {Rom{\'a}n-Garza}, Jaime and {Tran}, Nam H. and {Zapartas}, Emmanouil},
        title = "{The impact of mass-transfer physics on the observable properties of field binary black hole populations}",
      journal = {\aap},
         year = 2021,
        month = mar,
       volume = {647},
          eid = {A153},
        pages = {A153},
          doi = {10.1051/0004-6361/202039804},
archivePrefix = {arXiv},
       eprint = {2010.16333},
 primaryClass = {astro-ph.HE},
       adsurl = {https://ui.adsabs.harvard.edu/abs/2021A&A...647A.153B}
}

@ARTICLE{Kruckow2016,
       author = {{Kruckow}, M.~U. and {Tauris}, T.~M. and {Langer}, N. and {Sz{\'e}csi}, D. and {Marchant}, P. and {Podsiadlowski}, Ph.},
        title = "{Common-envelope ejection in massive binary stars. Implications for the progenitors of GW150914 and GW151226}",
      journal = {\aap},
         year = 2016,
        month = nov,
       volume = {596},
          eid = {A58},
        pages = {A58},
          doi = {10.1051/0004-6361/201629420},
archivePrefix = {arXiv},
       eprint = {1610.04417},
 primaryClass = {astro-ph.SR},
       adsurl = {https://ui.adsabs.harvard.edu/abs/2016A&A...596A..58K}
}

@ARTICLE{Boesky2024,
       author = {{Boesky}, Adam P. and {Broekgaarden}, Floor S. and {Berger}, Edo},
        title = "{Investigating the Cosmological Rate of Compact Object Mergers from Isolated Massive Binary Stars}",
      journal = {\apj},
         year = 2024,
        month = nov,
       volume = {976},
       number = {1},
          eid = {24},
        pages = {24},
          doi = {10.3847/1538-4357/ad7fe3},
archivePrefix = {arXiv},
       eprint = {2405.01630},
 primaryClass = {astro-ph.HE},
       adsurl = {https://ui.adsabs.harvard.edu/abs/2024ApJ...976...24B}
}

@ARTICLE{Mandel2022,
       author = {{Mandel}, Ilya and {Broekgaarden}, Floor S.},
        title = "{Rates of compact object coalescences}",
      journal = {Living Reviews in Relativity},
         year = 2022,
        month = dec,
       volume = {25},
       number = {1},
          eid = {1},
        pages = {1},
          doi = {10.1007/s41114-021-00034-3},
archivePrefix = {arXiv},
       eprint = {2107.14239},
 primaryClass = {astro-ph.HE},
       adsurl = {https://ui.adsabs.harvard.edu/abs/2022LRR....25....1M}
}

@ARTICLE{Reig2011,
       author = {{Reig}, Pablo},
        title = "{Be/X-ray binaries}",
      journal = {\apss},
         year = 2011,
        month = mar,
       volume = {332},
       number = {1},
        pages = {1-29},
          doi = {10.1007/s10509-010-0575-8},
archivePrefix = {arXiv},
       eprint = {1101.5036},
 primaryClass = {astro-ph.HE},
       adsurl = {https://ui.adsabs.harvard.edu/abs/2011Ap&SS.332....1R}
}

@ARTICLE{Negueruela1998,
       author = {{Negueruela}, Ignacio},
        title = "{On the nature of Be/X-ray binaries}",
      journal = {\aap},
         year = 1998,
        month = oct,
       volume = {338},
        pages = {505-510},
          doi = {10.48550/arXiv.astro-ph/9807158},
archivePrefix = {arXiv},
       eprint = {astro-ph/9807158},
 primaryClass = {astro-ph},
       adsurl = {https://ui.adsabs.harvard.edu/abs/1998A&A...338..505N}
}

@ARTICLE{Okazaki2001,
       author = {{Okazaki}, A.~T. and {Negueruela}, I.},
        title = "{A natural explanation for periodic X-ray outbursts in Be/X-ray binaries}",
      journal = {\aap},
         year = 2001,
        month = oct,
       volume = {377},
        pages = {161-174},
          doi = {10.1051/0004-6361:20011083},
archivePrefix = {arXiv},
       eprint = {astro-ph/0108037},
 primaryClass = {astro-ph},
       adsurl = {https://ui.adsabs.harvard.edu/abs/2001A&A...377..161O}
}

@ARTICLE{Shao2014,
       author = {{Shao}, Yong and {Li}, Xiang-Dong},
        title = "{On the Formation of Be Stars through Binary Interaction}",
      journal = {\apj},
         year = 2014,
        month = nov,
       volume = {796},
       number = {1},
          eid = {37},
        pages = {37},
          doi = {10.1088/0004-637X/796/1/37},
archivePrefix = {arXiv},
       eprint = {1410.0100},
 primaryClass = {astro-ph.HE},
       adsurl = {https://ui.adsabs.harvard.edu/abs/2014ApJ...796...37S}
}

@ARTICLE{Pols1991,
       author = {{Pols}, O.~R. and {Cote}, J. and {Waters}, L.~B.~F.~M. and {Heise}, J.},
        title = "{The formation of Be stars through close binary evolution.}",
      journal = {\aap},
         year = 1991,
        month = jan,
       volume = {241},
        pages = {419},
       adsurl = {https://ui.adsabs.harvard.edu/abs/1991A&A...241..419P}
}

@ARTICLE{Hastings2021,
       author = {{Hastings}, B. and {Langer}, N. and {Wang}, C. and {Schootemeijer}, A. and {Milone}, A.~P.},
        title = "{Stringent upper limit on Be star fractions produced by binary interaction}",
      journal = {\aap},
         year = 2021,
        month = sep,
       volume = {653},
          eid = {A144},
        pages = {A144},
          doi = {10.1051/0004-6361/202141269},
archivePrefix = {arXiv},
       eprint = {2106.12263},
 primaryClass = {astro-ph.SR},
       adsurl = {https://ui.adsabs.harvard.edu/abs/2021A&A...653A.144H}
}

@ARTICLE{Muller2026,
       author = {{M{\"u}ller-Horn}, Johanna and {Ramachandran}, Varsha and {El-Badry}, Kareem and {Sander}, Andreas A.~C. and {Bodensteiner}, Julia and {Gies}, Douglas R. and {G{\"o}tberg}, Ylva and {Rivinius}, Thomas and {Shenar}, Tomer and {Sch{\"o}sser}, Elisa C. and {Wang}, Luqian and {Bieryla}, Allyson and {Buchhave}, Lars A. and {Latham}, David W.},
        title = "{Ultraviolet spectroscopy reveals a hot and luminous companion to the Be star + black hole candidate MWC 656}",
      journal = {arXiv e-prints},
         year = 2026,
        month = jan,
          eid = {arXiv:2601.14403},
        pages = {arXiv:2601.14403},
          doi = {10.48550/arXiv.2601.14403},
archivePrefix = {arXiv},
       eprint = {2601.14403},
 primaryClass = {astro-ph.SR},
       adsurl = {https://ui.adsabs.harvard.edu/abs/2026arXiv260114403M}
}

@ARTICLE{Coe2015,
       author = {{Coe}, M.~J. and {Kirk}, J.},
        title = "{Catalogue of Be/X-ray binary systems in the Small Magellanic Cloud: X-ray, optical and IR properties}",
      journal = {\mnras},
         year = 2015,
        month = sep,
       volume = {452},
       number = {1},
        pages = {969-977},
          doi = {10.1093/mnras/stv1283},
archivePrefix = {arXiv},
       eprint = {1506.01920},
 primaryClass = {astro-ph.HE},
       adsurl = {https://ui.adsabs.harvard.edu/abs/2015MNRAS.452..969C}
}

@ARTICLE{Choudhury2018,
       author = {{Choudhury}, S. and {Subramaniam}, A. and {Cole}, A.~A. and {Sohn}, Y.-J.},
        title = "{Photometric metallicity map of the Small Magellanic Cloud}",
      journal = {\mnras},
         year = 2018,
        month = apr,
       volume = {475},
       number = {4},
        pages = {4279-4297},
          doi = {10.1093/mnras/sty087},
archivePrefix = {arXiv},
       eprint = {1801.03403},
 primaryClass = {astro-ph.GA},
       adsurl = {https://ui.adsabs.harvard.edu/abs/2018MNRAS.475.4279C}
}

@ARTICLE{Douna2015,
       author = {{Douna}, V.~M. and {Pellizza}, L.~J. and {Mirabel}, I.~F. and {Pedrosa}, S.~E.},
        title = "{Metallicity dependence of high-mass X-ray binary populations}",
      journal = {\aap},
         year = 2015,
        month = jul,
       volume = {579},
          eid = {A44},
        pages = {A44},
          doi = {10.1051/0004-6361/201525617},
archivePrefix = {arXiv},
       eprint = {1505.05483},
 primaryClass = {astro-ph.GA},
       adsurl = {https://ui.adsabs.harvard.edu/abs/2015A&A...579A..44D}
}

@ARTICLE{Lian2023,
       author = {{Lian}, Jianhui and {Bergemann}, Maria and {Pillepich}, Annalisa and {Zasowski}, Gail and {Lane}, Richard R.},
        title = "{The integrated metallicity profile of the Milky Way}",
      journal = {Nature Astronomy},
         year = 2023,
        month = aug,
       volume = {7},
        pages = {951-958},
          doi = {10.1038/s41550-023-01977-z},
archivePrefix = {arXiv},
       eprint = {2306.14100},
 primaryClass = {astro-ph.GA},
       adsurl = {https://ui.adsabs.harvard.edu/abs/2023NatAs...7..951L}
}

@ARTICLE{Zhang2004,
       author = {{Zhang}, Fan and {Li}, X.-D. and {Wang}, Z.-R.},
        title = "{Where Are the Be/Black Hole Binaries?}",
      journal = {\apj},
         year = 2004,
        month = mar,
       volume = {603},
       number = {2},
        pages = {663-668},
          doi = {10.1086/381540},
archivePrefix = {arXiv},
       eprint = {astro-ph/0311523},
 primaryClass = {astro-ph},
       adsurl = {https://ui.adsabs.harvard.edu/abs/2004ApJ...603..663Z}
}

@ARTICLE{Linden2009,
       author = {{Linden}, T. and {Sepinsky}, J.~F. and {Kalogera}, V. and {Belczynski}, K.},
        title = "{Probing Electron-Capture Supernovae: X-Ray Binaries in Starbursts}",
      journal = {\apj},
         year = 2009,
        month = jul,
       volume = {699},
       number = {2},
        pages = {1573-1577},
          doi = {10.1088/0004-637X/699/2/1573},
archivePrefix = {arXiv},
       eprint = {0807.1097},
 primaryClass = {astro-ph},
       adsurl = {https://ui.adsabs.harvard.edu/abs/2009ApJ...699.1573L}
}

@ARTICLE{Xing2021,
       author = {{Xing}, Ze-Pei and {Li}, Xiang-Dong},
        title = "{Population Synthesis of Neutron Star X-Ray Binaries Associated with Supernova Remnants}",
      journal = {\apj},
         year = 2021,
        month = oct,
       volume = {920},
       number = {2},
          eid = {67},
        pages = {67},
          doi = {10.3847/1538-4357/ac16e1},
archivePrefix = {arXiv},
       eprint = {2107.09325},
 primaryClass = {astro-ph.HE},
       adsurl = {https://ui.adsabs.harvard.edu/abs/2021ApJ...920...67X}
}

@ARTICLE{Zuo2014,
       author = {{Zuo}, Zhao-Yu and {Li}, Xiang-Dong and {Gu}, Qiu-Sheng},
        title = "{Population synthesis on high-mass X-ray binaries: prospects and constraints from the universal X-ray luminosity function}",
      journal = {\mnras},
         year = 2014,
        month = jan,
       volume = {437},
       number = {2},
        pages = {1187-1198},
          doi = {10.1093/mnras/stt1918},
archivePrefix = {arXiv},
       eprint = {1310.5424},
 primaryClass = {astro-ph.HE},
       adsurl = {https://ui.adsabs.harvard.edu/abs/2014MNRAS.437.1187Z}
}

@ARTICLE{Mondal2020,
       author = {{Mondal}, Samaresh and {Belczy{\'n}ski}, Krzysztof and {Wiktorowicz}, Grzegorz and {Lasota}, Jean-Pierre and {King}, Andrew R.},
        title = "{The connection between merging double compact objects and the ultraluminous X-ray sources}",
      journal = {\mnras},
         year = 2020,
        month = jan,
       volume = {491},
       number = {2},
        pages = {2747-2759},
          doi = {10.1093/mnras/stz3227},
archivePrefix = {arXiv},
       eprint = {1909.04435},
 primaryClass = {astro-ph.HE},
       adsurl = {https://ui.adsabs.harvard.edu/abs/2020MNRAS.491.2747M}
}

@ARTICLE{Romero2023,
       author = {{Romero-Shaw}, Isobel and {Hirai}, Ryosuke and {Bahramian}, Arash and {Willcox}, Reinhold and {Mandel}, Ilya},
        title = "{Rapid population synthesis of black hole high-mass X-ray binaries: implications for binary stellar evolution}",
      journal = {\mnras},
         year = 2023,
        month = sep,
       volume = {524},
       number = {1},
        pages = {245-259},
          doi = {10.1093/mnras/stad1732},
archivePrefix = {arXiv},
       eprint = {2303.05375},
 primaryClass = {astro-ph.HE},
       adsurl = {https://ui.adsabs.harvard.edu/abs/2023MNRAS.524..245R}
}

@ARTICLE{Davies2015,
       author = {{Davies}, Ben and {Kudritzki}, Rolf-Peter and {Gazak}, Zach and {Plez}, Bertrand and {Bergemann}, Maria and {Evans}, Chris and {Patrick}, Lee},
        title = "{Red Supergiants as Cosmic Abundance Probes: The Magellanic Clouds}",
      journal = {\apj},
         year = 2015,
        month = jun,
       volume = {806},
       number = {1},
          eid = {21},
        pages = {21},
          doi = {10.1088/0004-637X/806/1/21},
archivePrefix = {arXiv},
       eprint = {1504.03694},
 primaryClass = {astro-ph.GA},
       adsurl = {https://ui.adsabs.harvard.edu/abs/2015ApJ...806...21D}
}

@ARTICLE{Asplund2009,
       author = {{Asplund}, Martin and {Grevesse}, Nicolas and {Sauval}, A. Jacques and {Scott}, Pat},
        title = "{The Chemical Composition of the Sun}",
      journal = {\araa},
         year = 2009,
        month = sep,
       volume = {47},
       number = {1},
        pages = {481-522},
          doi = {10.1146/annurev.astro.46.060407.145222},
archivePrefix = {arXiv},
       eprint = {0909.0948},
 primaryClass = {astro-ph.SR},
       adsurl = {https://ui.adsabs.harvard.edu/abs/2009ARA&A..47..481A}
}

@BOOK{Chaty2022,
       author = {{Chaty}, Sylvain},
        title = "{Accreting Binaries; Nature, formation, and evolution}",
         year = 2022,
          doi = {10.1088/2514-3433/ac595f},
       adsurl = {https://ui.adsabs.harvard.edu/abs/2022abn..book.....C}
}

@ARTICLE{Marchant2024,
       author = {{Marchant}, Pablo and {Bodensteiner}, Julia},
        title = "{The Evolution of Massive Binary Stars}",
      journal = {\araa},
         year = 2024,
        month = sep,
       volume = {62},
       number = {1},
        pages = {21-61},
          doi = {10.1146/annurev-astro-052722-105936},
archivePrefix = {arXiv},
       eprint = {2311.01865},
 primaryClass = {astro-ph.SR},
       adsurl = {https://ui.adsabs.harvard.edu/abs/2024ARA&A..62...21M}
}

@ARTICLE{Igoshev2021,
       author = {{Igoshev}, Andrei P. and {Chruslinska}, Martyna and {Dorozsmai}, Andris and {Toonen}, Silvia},
        title = "{Combined analysis of neutron star natal kicks using proper motions and parallax measurements for radio pulsars and Be X-ray binaries}",
      journal = {\mnras},
         year = 2021,
        month = dec,
       volume = {508},
       number = {3},
        pages = {3345-3364},
          doi = {10.1093/mnras/stab2734},
archivePrefix = {arXiv},
       eprint = {2109.10362},
 primaryClass = {astro-ph.HE},
       adsurl = {https://ui.adsabs.harvard.edu/abs/2021MNRAS.508.3345I}
}

@ARTICLE{Vigna2025,
       author = {{Vigna-G{\'o}mez}, Alejandro},
        title = "{The impact of natal kicks on black hole binaries}",
      journal = {\aap},
         year = 2025,
        month = sep,
       volume = {701},
          eid = {L3},
        pages = {L3},
          doi = {10.1051/0004-6361/202556051},
archivePrefix = {arXiv},
       eprint = {2507.07573},
 primaryClass = {astro-ph.HE},
       adsurl = {https://ui.adsabs.harvard.edu/abs/2025A&A...701L...3V}
}

@ARTICLE{Nagarajan2025,
       author = {{Nagarajan}, Pranav and {El-Badry}, Kareem},
        title = "{Mixed Origins: Strong Natal Kicks for Some Black Holes and None for Others}",
      journal = {\pasp},
         year = 2025,
        month = mar,
       volume = {137},
       number = {3},
          eid = {034203},
        pages = {034203},
          doi = {10.1088/1538-3873/adb6d6},
archivePrefix = {arXiv},
       eprint = {2411.16847},
 primaryClass = {astro-ph.GA},
       adsurl = {https://ui.adsabs.harvard.edu/abs/2025PASP..137c4203N}
}

@ARTICLE{Panoglou2016,
       author = {{Panoglou}, Despina and {Carciofi}, Alex C. and {Vieira}, Rodrigo G. and {Cyr}, Isabelle H. and {Jones}, Carol E. and {Okazaki}, Atsuo T. and {Rivinius}, Thomas},
        title = "{Be discs in binary systems - I. Coplanar orbits}",
      journal = {\mnras},
         year = 2016,
        month = sep,
       volume = {461},
       number = {3},
        pages = {2616-2629},
          doi = {10.1093/mnras/stw1508},
archivePrefix = {arXiv},
       eprint = {1605.06674},
 primaryClass = {astro-ph.SR},
       adsurl = {https://ui.adsabs.harvard.edu/abs/2016MNRAS.461.2616P}
}

@ARTICLE{Panoglou2018,
       author = {{Panoglou}, Despina and {Faes}, Daniel M. and {Carciofi}, Alex C. and {Okazaki}, Atsuo T. and {Baade}, Dietrich and {Rivinius}, Thomas and {Borges Fernandes}, Marcelo},
        title = "{Be discs in coplanar circular binaries: Phase-locked variations of emission lines}",
      journal = {\mnras},
         year = 2018,
        month = jan,
       volume = {473},
       number = {3},
        pages = {3039-3050},
          doi = {10.1093/mnras/stx2497},
archivePrefix = {arXiv},
       eprint = {1704.06751},
 primaryClass = {astro-ph.SR},
       adsurl = {https://ui.adsabs.harvard.edu/abs/2018MNRAS.473.3039P}
}

@INPROCEEDINGS{Rivinius2019,
       author = {{Rivinius}, Thomas},
        title = "{Be stars in the X-ray binary context}",
    booktitle = {High-mass X-ray Binaries: Illuminating the Passage from Massive Binaries to Merging Compact Objects},
         year = 2019,
       editor = {{Oskinova}, Lidia M. and {Bozzo}, Enrico and {Bulik}, Tomasz and {Gies}, Douglas R.},
       series = {IAU Symposium},
       volume = {346},
        month = dec,
        pages = {105-113},
          doi = {10.1017/S1743921318008207},
       adsurl = {https://ui.adsabs.harvard.edu/abs/2019IAUS..346..105R}
}

@ARTICLE{Antoniou2016,
       author = {{Antoniou}, V. and {Zezas}, A.},
        title = "{Star formation history and X-ray binary populations: the case of the Large Magellanic Cloud}",
      journal = {\mnras},
         year = 2016,
        month = jun,
       volume = {459},
       number = {1},
        pages = {528-553},
          doi = {10.1093/mnras/stw167},
archivePrefix = {arXiv},
       eprint = {1603.08011},
 primaryClass = {astro-ph.HE},
       adsurl = {https://ui.adsabs.harvard.edu/abs/2016MNRAS.459..528A}
}

@ARTICLE{Kaltenbrunner2026,
       author = {{Kaltenbrunner}, D. and {Maitra}, C. and {Haberl}, F. and {Bodensteiner}, J. and {Bogensberger}, D. and {Buckley}, D.~A.~H. and {Cioni}, M.~R.~L. and {Greiner}, J. and {Monageng}, I. and {Udalski}, A. and {Vasilopoulos}, G. and {Willer}, R.},
        title = "{A comprehensive catalogue of high-mass X-ray binaries in the Large Magellanic Cloud detected during the first eROSITA all-sky survey}",
      journal = {arXiv e-prints},
         year = 2026,
        month = feb,
          eid = {arXiv:2602.08152},
        pages = {arXiv:2602.08152},
          doi = {10.48550/arXiv.2602.08152},
archivePrefix = {arXiv},
       eprint = {2602.08152},
 primaryClass = {astro-ph.HE},
       adsurl = {https://ui.adsabs.harvard.edu/abs/2026arXiv260208152K}
}

@ARTICLE{Deschamps2013,
       author = {{Deschamps}, R. and {Siess}, L. and {Davis}, P.~J. and {Jorissen}, A.},
        title = "{Critically-rotating accretors and non-conservative evolution in Algols}",
      journal = {\aap},
         year = 2013,
        month = sep,
       volume = {557},
          eid = {A40},
        pages = {A40},
          doi = {10.1051/0004-6361/201321509},
archivePrefix = {arXiv},
       eprint = {1306.1348},
 primaryClass = {astro-ph.SR},
       adsurl = {https://ui.adsabs.harvard.edu/abs/2013A&A...557A..40D}
}

@ARTICLE{Wang2026,
       author = {{Wang}, Chen and {Lau}, Mike Y.~M. and {Li}, Xiang-Dong and {Langer}, Norbert and {de Mink}, Selma E. and {Valli}, Ruggero and {Justham}, Stephen and {Xu}, Xiao-Tian and {Klencki}, Jakub and {Ryu}, Taeho},
        title = "{Thermal-timescale accretion does not always yield critical rotation in mass gainers}",
      journal = {arXiv e-prints},
         year = 2026,
        month = jan,
          eid = {arXiv:2601.08508},
        pages = {arXiv:2601.08508},
          doi = {10.48550/arXiv.2601.08508},
archivePrefix = {arXiv},
       eprint = {2601.08508},
 primaryClass = {astro-ph.SR},
       adsurl = {https://ui.adsabs.harvard.edu/abs/2026arXiv260108508W}
}

@ARTICLE{Brown2018,
       author = {{Brown}, R.~O. and {Ho}, W.~C.~G. and {Coe}, M.~J. and {Okazaki}, A.~T.},
        title = "{Simulating the X-ray luminosity of Be X-ray binaries: the case for black holes versus neutron stars}",
      journal = {\mnras},
         year = 2018,
        month = jul,
       volume = {477},
       number = {4},
        pages = {4810-4816},
          doi = {10.1093/mnras/sty973},
archivePrefix = {arXiv},
       eprint = {1804.05749},
 primaryClass = {astro-ph.HE},
       adsurl = {https://ui.adsabs.harvard.edu/abs/2018MNRAS.477.4810B}
}

@inproceedings{Berger2014,
  title={Kolmogorov–Smirnov Test: Overview},
  author={Vance W. Berger and YanYan Zhou},
  year={2014},
  url={https://api.semanticscholar.org/CorpusID:118053132},
  booktitle={Wiley statsref: Statistics reference online}
}

@ARTICLE{Valli2025,
       author = {{Valli}, Ruggero and {de Mink}, Selma E. and {Justham}, Stephen and {Callister}, Thomas and {Johnston}, Cole and {Kresse}, Daniel and {Langer}, Norbert and {Rubio}, Amanda C. and {Vigna-G{\'o}mez}, Alejandro and {Wang}, Chen},
        title = "{Evidence of polar and ultralow supernova kicks from the orbits of Be X-ray binaries}",
      journal = {arXiv e-prints},
         year = 2025,
        month = may,
          eid = {arXiv:2505.08857},
        pages = {arXiv:2505.08857},
          doi = {10.48550/arXiv.2505.08857},
archivePrefix = {arXiv},
       eprint = {2505.08857},
 primaryClass = {astro-ph.HE},
       adsurl = {https://ui.adsabs.harvard.edu/abs/2025arXiv250508857V}
}

@ARTICLE{Bao2025,
       author = {{Bao}, Yuchen and {Li}, Zhenwei and {Ge}, Hongwei and {Chen}, Xuefei and {Han}, Zhanwen},
        title = "{A Be Star + He Star Binary as an Indicator of a Binary Mass Transfer Phase}",
      journal = {\apj},
         year = 2025,
        month = jul,
       volume = {987},
       number = {2},
          eid = {210},
        pages = {210},
          doi = {10.3847/1538-4357/addfd8},
archivePrefix = {arXiv},
       eprint = {2506.02662},
 primaryClass = {astro-ph.SR},
       adsurl = {https://ui.adsabs.harvard.edu/abs/2025ApJ...987..210B}
}

@ARTICLE{Rast2025,
       author = {{Rast}, Rina G. and {Jones}, Carol E. and {Suffak}, Mark and {Labadie-Bartz}, Jonathan and {ud-Doula}, Asif and {Carciofi}, Alex C. and {Quigley}, Peter and {Neiner}, Coralie and {Drake}, Jeremy J.},
        title = "{Predicted observational effects of rapid rotation for Be stars}",
      journal = {\apss},
         year = 2025,
        month = oct,
       volume = {370},
       number = {10},
          eid = {120},
        pages = {120},
          doi = {10.1007/s10509-025-04512-w},
archivePrefix = {arXiv},
       eprint = {2510.21640},
 primaryClass = {astro-ph.SR},
       adsurl = {https://ui.adsabs.harvard.edu/abs/2025Ap&SS.370..120R}
}

@ARTICLE{Martinez2025,
       author = {{Martinez}, Mark and {O'Grady}, Anna and {Breivik}, Katelyn and {Chen}, Gina},
        title = "{Properties of Core Collapse Supernovae from Binary Population Synthesis}",
      journal = {arXiv e-prints},
         year = 2025,
        month = nov,
          eid = {arXiv:2511.23285},
        pages = {arXiv:2511.23285},
          doi = {10.48550/arXiv.2511.23285},
archivePrefix = {arXiv},
       eprint = {2511.23285},
 primaryClass = {astro-ph.SR},
       adsurl = {https://ui.adsabs.harvard.edu/abs/2025arXiv251123285M}
}

@ARTICLE{McNeill2025,
       author = {{McNeill}, Lucy O. and {Hirai}, Ryosuke},
        title = "{Mass transfer stability for AM CVn binaries with white dwarf donors}",
      journal = {arXiv e-prints},
         year = 2025,
        month = dec,
          eid = {arXiv:2512.01377},
        pages = {arXiv:2512.01377},
          doi = {10.48550/arXiv.2512.01377},
archivePrefix = {arXiv},
       eprint = {2512.01377},
 primaryClass = {astro-ph.SR},
       adsurl = {https://ui.adsabs.harvard.edu/abs/2025arXiv251201377M}
}

@ARTICLE{Inayoshi2017,
       author = {{Inayoshi}, Kohei and {Hirai}, Ryosuke and {Kinugawa}, Tomoya and {Hotokezaka}, Kenta},
        title = "{Formation pathway of Population III coalescing binary black holes through stable mass transfer}",
      journal = {\mnras},
         year = 2017,
        month = jul,
       volume = {468},
       number = {4},
        pages = {5020-5032},
          doi = {10.1093/mnras/stx757},
archivePrefix = {arXiv},
       eprint = {1701.04823},
 primaryClass = {astro-ph.HE},
       adsurl = {https://ui.adsabs.harvard.edu/abs/2017MNRAS.468.5020I}
}

@ARTICLE{Xu2025bbh,
       author = {{Xu}, Xiao-Tian and {Langer}, Norbert and {Klencki}, Jakub and {Wang}, Chen and {Li}, Xiang-Dong},
        title = "{Stable mass transfer in massive binaries leading to merging black holes}",
      journal = {arXiv e-prints},
         year = 2025,
        month = dec,
          eid = {arXiv:2512.20054},
        pages = {arXiv:2512.20054},
          doi = {10.48550/arXiv.2512.20054},
archivePrefix = {arXiv},
       eprint = {2512.20054},
 primaryClass = {astro-ph.SR},
       adsurl = {https://ui.adsabs.harvard.edu/abs/2025arXiv251220054X}
}

@ARTICLE{Xu2025,
       author = {{Xu}, X.-T. and {Sch{\"u}rmann}, C. and {Langer}, N. and {Wang}, C. and {Schootemeijer}, A. and {Shenar}, T. and {Ercolino}, A. and {Haberl}, F. and {Hastings}, B. and {Jin}, H. and {Kramer}, M. and {Lennon}, D. and {Marchant}, P. and {Sen}, K. and {Tauris}, T.~M. and {de Mink}, S.~E.},
        title = "{Populations of evolved massive binary stars in the Small Magellanic Cloud: I. Predictions from detailed evolution models}",
      journal = {\aap},
         year = 2025,
        month = dec,
       volume = {704},
          eid = {A218},
        pages = {A218},
          doi = {10.1051/0004-6361/202554786},
archivePrefix = {arXiv},
       eprint = {2503.23876},
 primaryClass = {astro-ph.SR},
       adsurl = {https://ui.adsabs.harvard.edu/abs/2025A&A...704A.218X}
}

@ARTICLE{Lailey2026,
       author = {{Lailey}, B.~D. and {Sigut}, T.~A.~A.},
        title = "{Inclination Bias in Techniques Used to Identify Be Star Candidates}",
      journal = {arXiv e-prints},
         year = 2026,
        month = jan,
          eid = {arXiv:2601.22014},
        pages = {arXiv:2601.22014},
          doi = {10.48550/arXiv.2601.22014},
archivePrefix = {arXiv},
       eprint = {2601.22014},
 primaryClass = {astro-ph.SR},
       adsurl = {https://ui.adsabs.harvard.edu/abs/2026arXiv260122014L}
}

@ARTICLE{Picco2024,
       author = {{Picco}, A. and {Marchant}, P. and {Sana}, H. and {Nelemans}, G.},
        title = "{Forming merging double compact objects with stable mass transfer}",
      journal = {\aap},
         year = 2024,
        month = jan,
       volume = {681},
          eid = {A31},
        pages = {A31},
          doi = {10.1051/0004-6361/202347090},
archivePrefix = {arXiv},
       eprint = {2309.05736},
 primaryClass = {astro-ph.SR},
       adsurl = {https://ui.adsabs.harvard.edu/abs/2024A&A...681A..31P}
}

@ARTICLE{Li2026,
       author = {{Li}, Zhenwei and {Jia}, Shi and {Wei}, Dandan and {Ge}, Hongwei and {Chen}, Hailiang and {Zhang}, Yangyang and {Chen}, Xuefei and {Han}, Zhanwen},
        title = "{Formation of Be Stars via Wind Accretion: Case Study on Black Hole + Be Star Binaries}",
      journal = {\apjl},
         year = 2026,
        month = jan,
       volume = {996},
       number = {2},
          eid = {L42},
        pages = {L42},
          doi = {10.3847/2041-8213/ae3008},
archivePrefix = {arXiv},
       eprint = {2512.18565},
 primaryClass = {astro-ph.SR},
       adsurl = {https://ui.adsabs.harvard.edu/abs/2026ApJ...996L..42L}
}

@ARTICLE{Belczynski2009,
       author = {{Belczynski}, Krzysztof and {Ziolkowski}, Janusz},
        title = "{On the Apparent Lack of Be X-Ray Binaries with Black Holes}",
      journal = {\apj},
         year = 2009,
        month = dec,
       volume = {707},
       number = {2},
        pages = {870-877},
          doi = {10.1088/0004-637X/707/2/870},
archivePrefix = {arXiv},
       eprint = {0907.4990},
 primaryClass = {astro-ph.GA},
       adsurl = {https://ui.adsabs.harvard.edu/abs/2009ApJ...707..870B}
}

@BOOK{Scott1992,
       author = {{Scott}, D.~W.},
        title = "{Multivariate Density Estimation}",
         year = 1992,
         publisher={Wiley Online Library},
       adsurl = {https://ui.adsabs.harvard.edu/abs/1992mde..book.....S}
}

@ARTICLE{Xing2026,
       author = {{Xing}, Zepei and {Fragos}, Tassos and {Kalogera}, Vicky and {Gossage}, Seth and {Akira Rocha}, Kyle and {Zapartas}, Emmanouil},
        title = "{Disk-Regulated Mass Transfer Between Rotating Non-Degenerate Stars: Insights from Be and sdOB Binaries}",
      journal = {arXiv e-prints},
         year = 2026,
        month = feb,
          eid = {arXiv:2602.06259},
        pages = {arXiv:2602.06259},
          doi = {10.48550/arXiv.2602.06259},
archivePrefix = {arXiv},
       eprint = {2602.06259},
 primaryClass = {astro-ph.SR},
       adsurl = {https://ui.adsabs.harvard.edu/abs/2026arXiv260206259X}
}

@ARTICLE{Cranmer2005,
       author = {{Cranmer}, Steven R.},
        title = "{A Statistical Study of Threshold Rotation Rates for the Formation of Disks around Be Stars}",
      journal = {\apj},
         year = 2005,
        month = nov,
       volume = {634},
       number = {1},
        pages = {585-601},
          doi = {10.1086/491696},
archivePrefix = {arXiv},
       eprint = {astro-ph/0507718},
 primaryClass = {astro-ph},
       adsurl = {https://ui.adsabs.harvard.edu/abs/2005ApJ...634..585C}
}

@ARTICLE{Graczyk2020,
       author = {{Graczyk}, Dariusz and {Pietrzy{\'n}ski}, Grzegorz and {Thompson}, Ian B. and {Gieren}, Wolfgang and {Zgirski}, Bart{\l}omiej and {Villanova}, Sandro and {G{\'o}rski}, Marek and {Wielg{\'o}rski}, Piotr and {Karczmarek}, Paulina and {Narloch}, Weronika and {Pilecki}, Bogumi{\l} and {Taormina}, Monica and {Smolec}, Rados{\l}aw and {Suchomska}, Ksenia and {Gallenne}, Alexandre and {Nardetto}, Nicolas and {Storm}, Jesper and {Kudritzki}, Rolf-Peter and {Ka{\l}uszy{\'n}ski}, Miko{\l}aj and {Pych}, Wojciech},
        title = "{A Distance Determination to the Small Magellanic Cloud with an Accuracy of Better than Two Percent Based on Late-type Eclipsing Binary Stars}",
      journal = {\apj},
         year = 2020,
        month = nov,
       volume = {904},
       number = {1},
          eid = {13},
        pages = {13},
          doi = {10.3847/1538-4357/abbb2b},
archivePrefix = {arXiv},
       eprint = {2010.08754},
 primaryClass = {astro-ph.GA},
       adsurl = {https://ui.adsabs.harvard.edu/abs/2020ApJ...904...13G}
}

@ARTICLE{Osaki1986,
       author = {{Osaki}, Y.},
        title = "{Connection between nonradial pulsations and stellar winds in massive stars. I - Nonradial pulsation theory of massive stars}",
      journal = {\pasp},
         year = 1986,
        month = jan,
       volume = {98},
        pages = {30-32},
          doi = {10.1086/131713},
       adsurl = {https://ui.adsabs.harvard.edu/abs/1986PASP...98...30O}
}

@INPROCEEDINGS{Baade1988,
       author = {{Baade}, D.},
        title = "{Nonradial Pulsations and the be Phenomenon}",
    booktitle = {The Impact of Very High S/N Spectroscopy on Stellar Physics},
         year = 1988,
       editor = {{Cayrel de Strobel}, G. and {Spite}, Monique},
       series = {IAU Symposium},
       volume = {132},
        month = jan,
        pages = {217},
       adsurl = {https://ui.adsabs.harvard.edu/abs/1988IAUS..132..217B}
}

@ARTICLE{Semaan2018,
       author = {{Semaan}, T. and {Hubert}, A.~M. and {Zorec}, J. and {Guti{\'e}rrez-Soto}, J. and {Fr{\'e}mat}, Y. and {Martayan}, C. and {Fabregat}, J. and {Eggenberger}, P.},
        title = "{Study of a sample of faint Be stars in the exofield of CoRoT. II. Pulsation and outburst events: Time series analysis of photometric variations}",
      journal = {\aap},
         year = 2018,
        month = jun,
       volume = {613},
          eid = {A70},
        pages = {A70},
          doi = {10.1051/0004-6361/201629243},
       adsurl = {https://ui.adsabs.harvard.edu/abs/2018A&A...613A..70S}
}

@ARTICLE{Staritsin2026,
       author = {{Staritsin}, Evgeny},
        title = "{Formation of classical Be-stars of the early spectral subclass in the case of nonconservative mass transfer in close binary systems}",
      journal = {arXiv e-prints},
         year = 2026,
        month = apr,
          eid = {arXiv:2604.20352},
        pages = {arXiv:2604.20352},
archivePrefix = {arXiv},
       eprint = {2604.20352},
 primaryClass = {astro-ph.SR},
       adsurl = {https://ui.adsabs.harvard.edu/abs/2026arXiv260420352S}
}

@ARTICLE{Xu2026,
       author = {{Xu}, Xiao-Tian and {Podsiadlowski}, Philipp and {Langer}, Norbert and {Wang}, Xue-Feng and {Li}, Xiang-Dong and {Heger}, Alexander and {Mackey}, Jonathan and {Gr{\"a}fener}, G{\"o}tz and {Jin}, Harim},
        title = "{Evolution of wide O star binaries through their LBV stage. Population synthesis with mass-ejection-driven orbital evolution}",
      journal = {arXiv e-prints},
         year = 2026,
        month = mar,
          eid = {arXiv:2603.14840},
        pages = {arXiv:2603.14840},
          doi = {10.48550/arXiv.2603.14840},
archivePrefix = {arXiv},
       eprint = {2603.14840},
 primaryClass = {astro-ph.SR},
       adsurl = {https://ui.adsabs.harvard.edu/abs/2026arXiv260314840X}
}

@ARTICLE{deWit2006,
       author = {{de Wit}, W.~J. and {Lamers}, H.~J.~G.~L.~M. and {Marquette}, J.~B. and {Beaulieu}, J.~P.},
        title = "{The remarkable light and colour variability of Small Magellanic Cloud Be stars}",
      journal = {\aap},
         year = 2006,
        month = sep,
       volume = {456},
       number = {3},
        pages = {1027-1035},
          doi = {10.1051/0004-6361:20065137},
archivePrefix = {arXiv},
       eprint = {astro-ph/0606270},
 primaryClass = {astro-ph},
       adsurl = {https://ui.adsabs.harvard.edu/abs/2006A&A...456.1027D}
}

@BOOK{Tauris2023,
       author = {{Tauris}, Thomas M. and {van den Heuvel}, Edward P.~J.},
        title = "{Physics of Binary Star Evolution. From Stars to X-ray Binaries and Gravitational Wave Sources}",
         year = 2023,
          doi = {10.48550/arXiv.2305.09388},
       adsurl = {https://ui.adsabs.harvard.edu/abs/2023pbse.book.....T}
}

@ARTICLE{deMink2007,
       author = {{de Mink}, S.~E. and {Pols}, O.~R. and {Hilditch}, R.~W.},
        title = "{Efficiency of mass transfer in massive close binaries. Tests from double-lined eclipsing binaries in the SMC}",
      journal = {\aap},
         year = 2007,
        month = jun,
       volume = {467},
       number = {3},
        pages = {1181-1196},
          doi = {10.1051/0004-6361:20067007},
archivePrefix = {arXiv},
       eprint = {astro-ph/0703480},
 primaryClass = {astro-ph},
       adsurl = {https://ui.adsabs.harvard.edu/abs/2007A&A...467.1181D}
}

@ARTICLE{Hut1981,
       author = {{Hut}, P.},
        title = "{Tidal evolution in close binary systems.}",
      journal = {\aap},
         year = 1981,
        month = jun,
       volume = {99},
        pages = {126-140},
       adsurl = {https://ui.adsabs.harvard.edu/abs/1981A&A....99..126H}
}

@ARTICLE{Ge2023,
       author = {{Ge}, Hongwei and {Tout}, Christopher A. and {Chen}, Xuefei and {Sarkar}, Arnab and {Walton}, Dominic J. and {Han}, Zhanwen},
        title = "{Criteria for Dynamical Timescale Mass Transfer of Metal-poor Intermediate-mass Stars}",
      journal = {\apj},
         year = 2023,
        month = mar,
       volume = {945},
       number = {1},
          eid = {7},
        pages = {7},
          doi = {10.3847/1538-4357/acb7e9},
archivePrefix = {arXiv},
       eprint = {2302.00183},
 primaryClass = {astro-ph.SR},
       adsurl = {https://ui.adsabs.harvard.edu/abs/2023ApJ...945....7G}
}

@ARTICLE{Ge2015,
       author = {{Ge}, Hongwei and {Webbink}, Ronald F. and {Chen}, Xuefei and {Han}, Zhanwen},
        title = "{Adiabatic Mass Loss in Binary Stars. II. From Zero-age Main Sequence to the Base of the Giant Branch}",
      journal = {\apj},
         year = 2015,
        month = oct,
       volume = {812},
       number = {1},
          eid = {40},
        pages = {40},
          doi = {10.1088/0004-637X/812/1/40},
archivePrefix = {arXiv},
       eprint = {1507.04843},
 primaryClass = {astro-ph.SR},
       adsurl = {https://ui.adsabs.harvard.edu/abs/2015ApJ...812...40G}
}

@ARTICLE{Ge2020,
       author = {{Ge}, Hongwei and {Webbink}, Ronald F. and {Chen}, Xuefei and {Han}, Zhanwen},
        title = "{Adiabatic Mass Loss in Binary Stars. III. From the Base of the Red Giant Branch to the Tip of the Asymptotic Giant Branch}",
      journal = {\apj},
         year = 2020,
        month = aug,
       volume = {899},
       number = {2},
          eid = {132},
        pages = {132},
          doi = {10.3847/1538-4357/aba7b7},
archivePrefix = {arXiv},
       eprint = {2007.09848},
 primaryClass = {astro-ph.SR},
       adsurl = {https://ui.adsabs.harvard.edu/abs/2020ApJ...899..132G}
}

@ARTICLE{Echeveste2026,
       author = {{Echeveste}, M. and {Escobar}, G.~J. and {Iorio}, G. and {Pancino}, E. and {Mapelli}, M. and {Alvarez Garay}, D. and {Avdeeva}, A. and {Leitinger}, E. and {Nedhath}, S. and {Rani}, S. and {Reggiani}, E. and {Sanna}, N. and {Saracino}, S. and {Steinbauer}, L. and {Turchi}, A.},
        title = "{The role of mass transfer efficiency in stability criteria: Implementation in SEVN and a test on blue stragglers and binary compact objects}",
      journal = {arXiv e-prints},
         year = 2026,
        month = jun,
          eid = {arXiv:2606.30303},
        pages = {arXiv:2606.30303},
archivePrefix = {arXiv},
       eprint = {2606.30303},
 primaryClass = {astro-ph.SR},
       adsurl = {https://ui.adsabs.harvard.edu/abs/2026arXiv260630303E}
}

@ARTICLE{Munoz2019,
       author = {{Mu{\~n}oz}, Diego J. and {Miranda}, Ryan and {Lai}, Dong},
        title = "{Hydrodynamics of Circumbinary Accretion: Angular Momentum Transfer and Binary Orbital Evolution}",
      journal = {\apj},
         year = 2019,
        month = jan,
       volume = {871},
       number = {1},
          eid = {84},
        pages = {84},
          doi = {10.3847/1538-4357/aaf867},
archivePrefix = {arXiv},
       eprint = {1810.04676},
 primaryClass = {astro-ph.HE},
       adsurl = {https://ui.adsabs.harvard.edu/abs/2019ApJ...871...84M}
}

@ARTICLE{Munoz2020,
       author = {{Mu{\~n}oz}, Diego J. and {Lai}, Dong and {Kratter}, Kaitlin and {Miranda}, Ryan},
        title = "{Circumbinary Accretion from Finite and Infinite Disks}",
      journal = {\apj},
         year = 2020,
        month = feb,
       volume = {889},
       number = {2},
          eid = {114},
        pages = {114},
          doi = {10.3847/1538-4357/ab5d33},
archivePrefix = {arXiv},
       eprint = {1910.04763},
 primaryClass = {astro-ph.HE},
       adsurl = {https://ui.adsabs.harvard.edu/abs/2020ApJ...889..114M}
}

@ARTICLE{Dittmann2022,
       author = {{Dittmann}, Alexander J. and {Ryan}, Geoffrey},
        title = "{A survey of disc thickness and viscosity in circumbinary accretion: Binary evolution, variability, and disc morphology}",
      journal = {\mnras},
         year = 2022,
        month = jul,
       volume = {513},
       number = {4},
        pages = {6158-6176},
          doi = {10.1093/mnras/stac935},
archivePrefix = {arXiv},
       eprint = {2201.07816},
 primaryClass = {astro-ph.HE},
       adsurl = {https://ui.adsabs.harvard.edu/abs/2022MNRAS.513.6158D}
}

@ARTICLE{Wang2022,
       author = {{Wang}, Lile and {Li}, Xinyu},
        title = "{Dynamical Effects of Colliding Outflows in Binary Systems}",
      journal = {\apj},
         year = 2022,
        month = jun,
       volume = {932},
       number = {2},
          eid = {108},
        pages = {108},
          doi = {10.3847/1538-4357/ac6ce6},
archivePrefix = {arXiv},
       eprint = {2203.02615},
 primaryClass = {astro-ph.SR},
       adsurl = {https://ui.adsabs.harvard.edu/abs/2022ApJ...932..108W}
}

@ARTICLE{Lai2023,
       author = {{Lai}, Dong and {Mu{\~n}oz}, Diego J.},
        title = "{Circumbinary Accretion: From Binary Stars to Massive Binary Black Holes}",
      journal = {\araa},
         year = 2023,
        month = aug,
       volume = {61},
        pages = {517-560},
          doi = {10.1146/annurev-astro-052622-022933},
archivePrefix = {arXiv},
       eprint = {2211.00028},
 primaryClass = {astro-ph.HE},
       adsurl = {https://ui.adsabs.harvard.edu/abs/2023ARA&A..61..517L}
}

@ARTICLE{Zainab2026,
       author = {{Zainab}, Aafia and {Thalhammer}, Philipp and {Zalot}, Nico and {Avakyan}, Artur and {Stierhof}, Jakob and {Sokolova-Lapa}, Ekaterina and {Doroshenko}, Victor and {Grinberg}, Victoria and {Kretschmar}, Peter and {Lipunova}, Galina and {Islam}, Nazma and {Schreiber}, Matthias R. and {Kirsch}, Christian and {H\"ammerich}, Steven and {Weber}, Philipp and {Ballhausen}, Ralf and {Escorial}, Alicia Rouc and {Coley}, Joel and {Rothschild}, Richard and {Pottschmidt}, Katja and {Wilms}, Joern.},
        title = "{The low luminosity end of Galactic HMXBs with eROSITA: Establishing a luminosity floor for accreting BeXRBs}",
      journal = {arXiv e-prints},
         year = 2026,
        month = aug,
          eid = {arXiv:2608.13259},
        pages = {arXiv:2608.13259},
          doi = {10.48550//arXiv.2608.13259},
archivePrefix = {arXiv},
       eprint = {2608.13259},
 primaryClass = {astro-ph.HE},
       adsurl = {https://ui.adsabs.harvard.edu/abs/arXiv:2608.13259}
}

\appendix

\section{Likelihood calculation}\label{app:likelihood}

Here, we describe the procedure used to compute the likelihood of each synthetic model with respect to the observed BeXRB population. The comparison is performed in two steps: first, we evaluate the likelihood associated with the observed distributions of the relevant observables, and second we combine this with the likelihood associated with the total number of systems.

\subsection{Reconstruction of the model distributions}

For each model, we first reconstruct the PDF of the observables using a Gaussian KDE. Each synthetic system is weighted by $\Delta t_i \times {\rm SFR}(t_i)/M_{\rm tot}$, where $\Delta t_i$ is the duration of the phase, $t_i$ is the corresponding stellar age, ${\rm SFR}(t)$ is the star formation rate of the SMC as a function of look-back time $t$, and $M_{\rm tot}=M_{\rm BPS}/f_{\rm imf}$ is the total stellar mass underlying the simulated binaries that have total mass of $M_{\rm BPS}$. Using these weights we construct KDE estimates of both the joint distribution of orbital period and optical magnitude, $f_{2D}(\pi, m_V)$, and the orbital period distribution $f_{\pi}(\pi)$, with $\pi \equiv \log(P_{\rm orb}\ \rm [day])$. These distributions represent the probability densities predicted by each synthetic population. The bandwidth factor of the Gaussian KDE is determined following the Scott's rule \citep{Scott1992}. 

\subsection{Synthetic likelihood distribution}

To evaluate the statistical significance of the observed sample, we generate a synthetic likelihood distribution by drawing random realizations from the model population. The random sampling is performed with probabilities proportional to the associated weights.

For the one–dimensional $P_{\rm orb}$ comparison, this procedure is as follows. For each trial we draw $N$ systems from the synthetic catalog, where $N=49$ corresponds to the total number of observed orbital periods in the SMC, and compute the likelihood
\begin{align}
\log \mathcal{L} = \frac{1}{N} \sum_{i=1}^{N} \log \frac{f_{\pi}(\pi_i)}{\Delta m_V}\ ,
\end{align}
where $f_{\pi}$ is multiplied with a factor of $1/\Delta m_V=1/6$ to capture the uniform prior of $m_{V}\in[12,18]$, ensuring that the one-dimensional likelihood can be directly compared with the two-dimensional likelihood (see below). 

The two–dimensional case requires a slightly more elaborate treatment because the observational dataset contains three types of measurements: systems with both $P_{\rm orb}$ and $m_V$, systems with only $m_V$, and systems with only $P_{\rm orb}$. We therefore construct synthetic realizations that reproduce the same observational structure of the SMC BeXRB population, drawing $N_{2D}=47$ joint values, $N_{m_V}=53$ values with only $m_V$, and $N_{P_{\rm orb}}=2$ values with only $P_{\rm orb}$. The corresponding likelihood is then written as
\begin{equation}
\begin{aligned}
\log \mathcal{L} = \frac{1}{N_{\rm tot}} \Bigg[
&\sum_{i=1}^{N_{2D}}
\log f_{2D}(\pi_i, m_{V,i}) 
+
\sum_{j=1}^{N_{m_V}}
\log \frac{f_{m_V}(m_{V,j})}{\Delta \pi} \\
&+
\sum_{k=1}^{N_{P_{\rm orb}}}
\log \frac{f_{\pi}(\pi_k)}{\Delta m_V}
\Bigg]\ .
\end{aligned}
\end{equation}
where $N_{\rm tot}=N_{2D}+N_{m_V}+N_{P_{\rm orb}}=102$. The normalization factors $\Delta \pi=3.2$ and $\Delta m_V=6$ ensure a fair comparison between the one– and two–dimensional contributions, corresponding to uniform priors $\pi\in[0,3.2]$ and $m_{V}\in[12,18]$ for unmeasured quantities.

Repeating this procedure for a large number of trials produces the synthetic likelihood distribution expected from the model.

\subsection{Observed likelihood and statistical comparison}
We then compute the likelihood of the observed dataset $\mathcal{L}_{\rm obs}$ using the same expressions as described in the previous subsection. The position of this value within the synthetic likelihood distribution determines the statistical regime of the model. In practice, we calculate the fraction of synthetic realizations that is less probable than the observed sample as
\begin{align}
    p_{\rm dist}=\begin{cases}
        2F(\mathcal{L}_{\rm obs})\ ,\quad &F(\mathcal{L}_{\rm obs})<0.5
         \ ,\\
        2[1-F(\mathcal{L}_{\rm obs})]\ ,\quad &F(\mathcal{L}_{\rm obs})\ge0.5\ ,
    \end{cases} 
\end{align}
where $F(\mathcal{L})$ is the CDF of $\mathcal{L}$ constructed from 2000 random realizations.
This quantity can be interpreted as a p-value describing the compatibility between the model and the observations.  Here, we approximate the peak of the likelihood distribution as the median for simplicity, as we are only concerned with the tails. 

\subsection{Likelihood of the total number of systems}

In addition to the PDF comparison, we also evaluate the likelihood associated with the total number of BeXRBs predicted by each model. If a model predicts $N_{\rm theo}$ systems and the observed catalog contains $N_{\rm obs}$ systems, we approximate the uncertainty in the theoretical prediction due to the uncertainty on the SFR of the SMC based on \citet[][see their appendix A]{Liu2023} with a normal distribution $\mathcal{N}$ centered on $\mu=N_{\rm theo}$ of width $\sigma = 0.2\times N_{\rm theo}$. 
The likelihood associated with the observed number is then
\begin{align}
\log \mathcal{L}_N = \log \left[ \mathcal{N}(N_{\rm obs}\,|\,\mu=N_{\rm theo},\sigma=0.2\times N_{\rm theo})
\right]\ .
\end{align}

\subsection{Combined likelihood}

Finally, the likelihood obtained from the PDF comparison is combined with the likelihood associated with the total number of systems. The combined likelihood is written as
\begin{align}
\log \mathcal{L}_{\rm tot} = \log \mathcal{L}_{\rm dist} + \log \mathcal{L}_N\ .
\end{align}

For completeness we also combine the associated p-values using Fisher’s method, providing a single statistic that quantifies the global agreement between each synthetic model and the observed BeXRB population. If $p_{\rm dist}$ and $p_{N}$ denote the p-values obtained from the distribution comparison and from the number-count comparison, respectively, the Fisher statistic is
\begin{align}
    \chi^2=-2\left[ \ln(p_{\rm dist}) + \ln(p_{N}) \right]\ ,
\end{align}
which follows a $\chi^2$ distribution with four degrees of freedom. The combined p-value is therefore
\begin{align}
    p_{\rm comb}=1-F_{\chi^2}(\chi^2; 4)\ ,
\end{align}
where $F_{\chi^2}$ denotes the CDF of the $\chi^2$ distribution. The models with $p_{\rm comb}\geq0.01$ are regarded as good models that have high enough chances to reproduce observations.

\section{Additional results for the SMC}\label{app:SMC}

We present here additional results for the SMC. We first show some additional distribution maps predicted by the overall best-fitting model (considering the joint $P_{\rm orb}-m_V$ distribution). We then examine models that are incompatible with the observations, highlighting the physical assumptions that lead to discrepancies. Next, we explore the impact of the adopted NS accretion prescription on the observable population. Finally, we assess the effect of alternative initial conditions on our results.

\subsection{Additional distribution maps}\label{app:SMC_extra}

Similar to Fig.~\ref{fig:P_m_V_smc}, we present additional distributions predicted by the overall best-fitting SMC model, based on the initial conditions from \citetalias{Sana2012}, a MT efficiency of $f_{\rm MT}=0.6$, isotropic re-emission mass loss, always-stable RLOF and recycling of AM excess.

%%%%%%%%%%%%FIGURE%%%%%%%%%%%%%%
\begin{figure}
    \centering
    \includegraphics[width=0.5\textwidth]{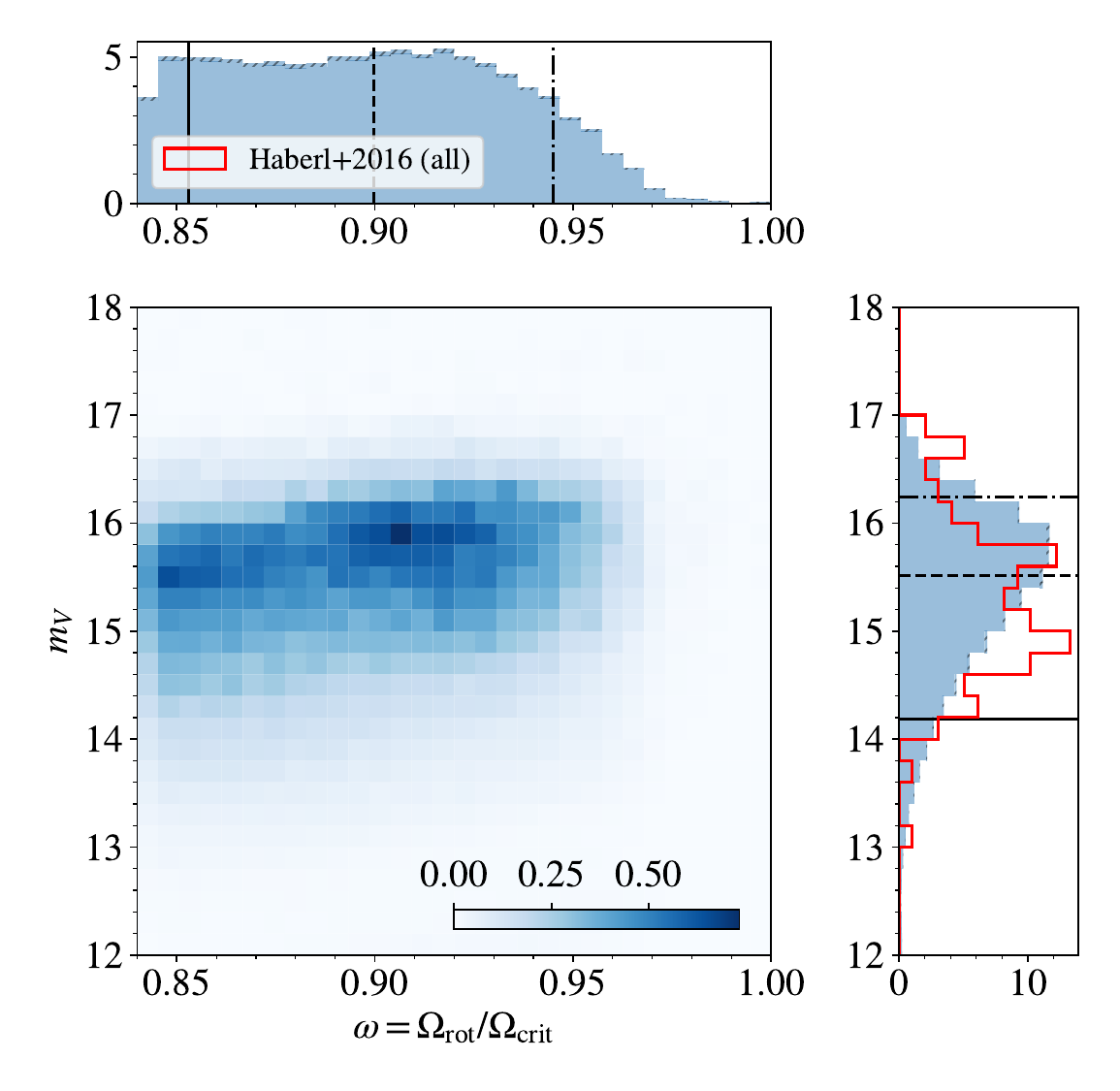}
    \caption{Same as Fig.~\ref{fig:P_m_V_smc} but for the stellar spin $\omega = \Omega_{\rm rot}/\Omega_{\rm crit}$ and $V$-band magnitude $m_V$ distributions, where $\Omega_{\rm rot}$ is the angular rotation rate and $\Omega_{\rm crit}$ is the critical angular rotation rate.}
    \label{fig:w_m_V_smc}
\end{figure}
%%%%%%%%%%%%%%%%%%%%%%%%%%%%%%%%%

Fig.~\ref{fig:w_m_V_smc} shows the distribution of stellar spin $\omega=\Omega_{\rm rot}/\Omega_{\rm crit}$ and $V$-band magnitude. The predicted stellar spin distribution of Be stars is intrinsically broad and non-trivial, rather than sharply peaked at a single value. This behavior implies a relatively smooth transition between normal B-type stars and Be stars as a function of rotation, mass, and evolutionary stage, considering that the Be phase threshold spin is also not a sharp discontinuous boundary but depends on temperature and stellar structure according to observations of classical Be stars \citet{Rivinius2013}. 
The combination of MT-induced spin-up, tidal interactions, and subsequent stellar evolution naturally generates a distribution spanning $\omega \sim 0.8-1$.

%%%%%%%%%%%%FIGURE%%%%%%%%%%%%%%
\begin{figure}
    \centering
    \includegraphics[width=0.5\textwidth]{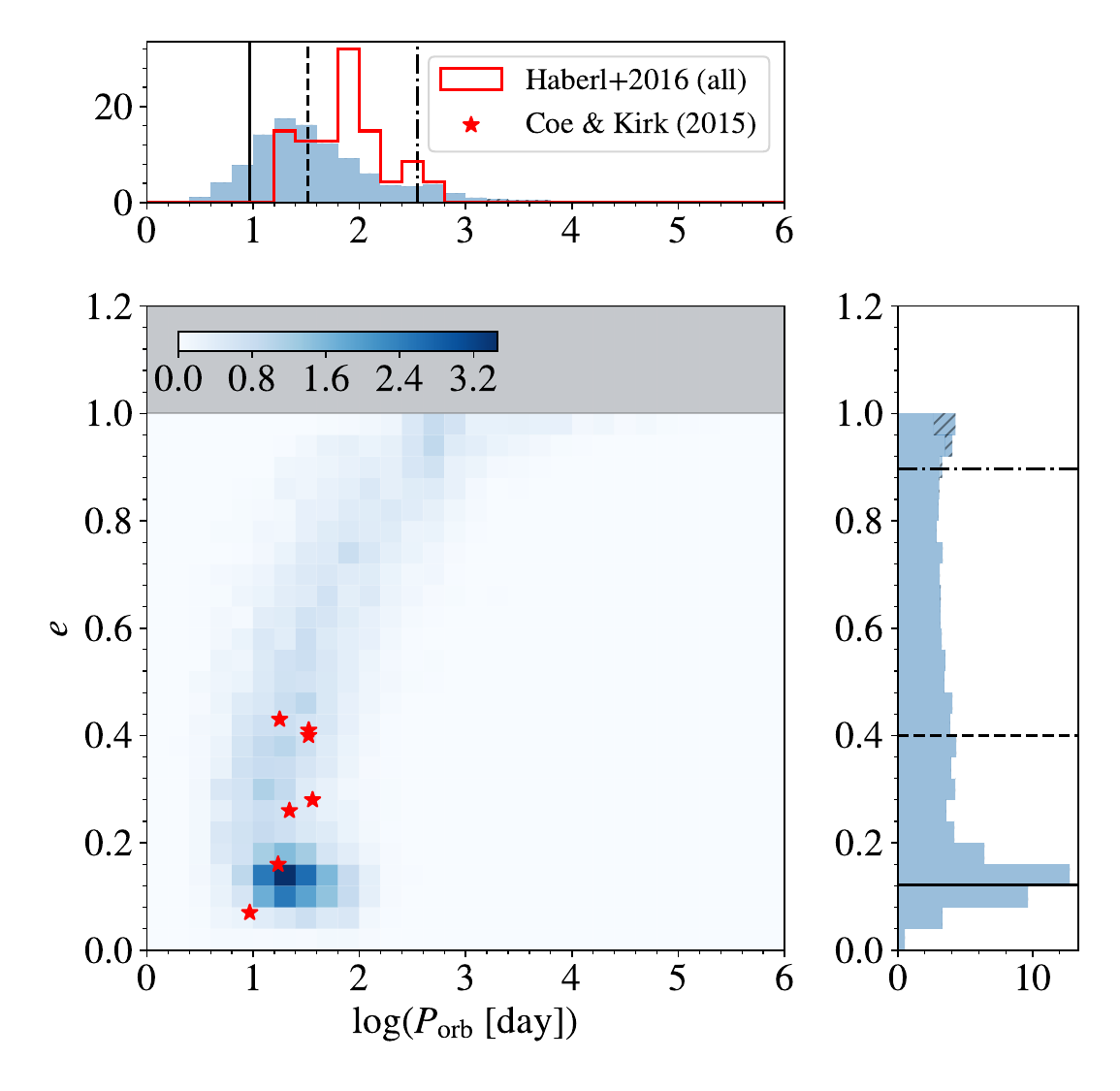}
    \caption{Same as Fig.~\ref{fig:P_m_V_smc} but for $P_{\rm orb}$ and eccentricity $e$ distributions. Observational points with both $P_{\rm orb}$ and $e$ measurements are taken from \citet{Coe2015}.}
    \label{fig:P_e_smc}
\end{figure}
%%%%%%%%%%%%%%%%%%%%%%%%%%%%%%%%%

Fig.~\ref{fig:P_e_smc} presents the predicted orbital period–eccentricity ($P_{\rm orb} - e$) distribution. The synthetic population exhibits the characteristic structure of observed BeXRBs: a concentration of systems at low eccentricities and short-to-intermediate orbital periods, together with a branch extending toward longer orbital periods at increasing eccentricity. This 
trend naturally arises from the interplay between pre-SN orbital configuration and natal kicks. Systems receiving small kicks remain nearly circular and populate the short-period region, whereas larger kicks both widen and give the surviving binaries a non-vanishing eccentricity. This behavior is consistent with theoretical expectations and has been predicted in previous binary-evolution studies \citep[e.g.,][]{Liu2023,Cecilia2023,Rocha2024,Schurmann2025,Li2026}.  

%%%%%%%%%%%%FIGURE%%%%%%%%%%%%%%
\begin{figure}
    \centering
    \includegraphics[width=0.5\textwidth]{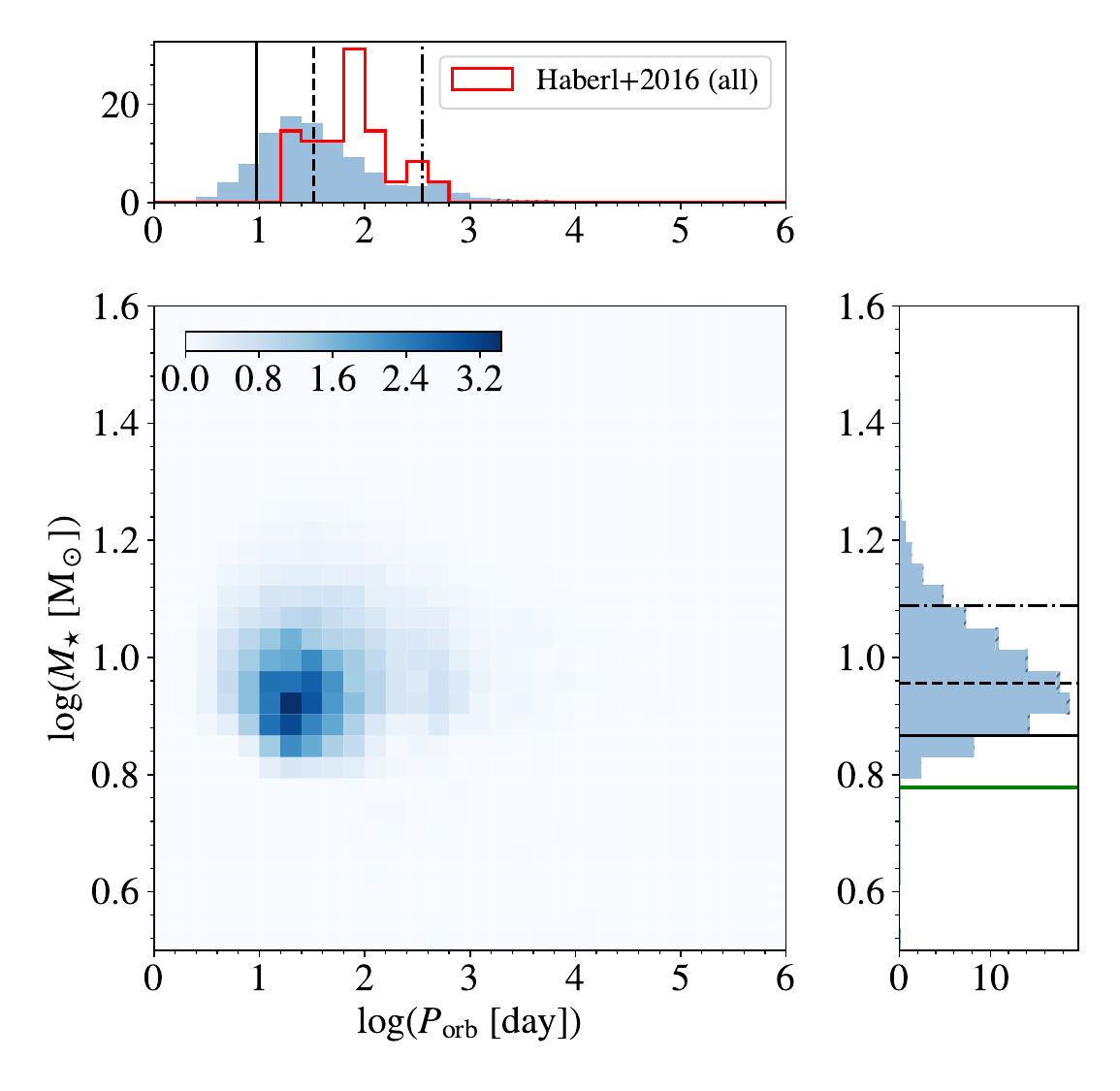}
    \caption{Same as Fig.~\ref{fig:P_m_V_smc} but for $P_{\rm orb}$ and mass of the Be star $M_\star$. The green line at 6 M$_\odot$ in the marginalized mass distribution shows the minimum Be star mass in BeXRBs from observations \citep{Hohle2010,Coe2015,Haberl16,Fortin23}.}
    \label{fig:P_M_smc}
\end{figure}
%%%%%%%%%%%%%%%%%%%%%%%%%%%%%%%%%

Fig.~\ref{fig:P_M_smc} shows the predicted orbital period--Be star mass ($P_{\rm orb}-M_\star$) distribution. We can see that all the Be stars forming BeXRBs have masses above the minimum 6 M$_\odot$ from observations \citep{Hohle2010,Coe2015,Haberl16,Fortin23}, further confirming that our overall best-fitting model can reproduce the Be star properties in BeXRBs. 

%%%%%%%%%%%%FIGURE%%%%%%%%%%%%%%
\begin{figure}
    \centering
    \includegraphics[width=0.5\textwidth]{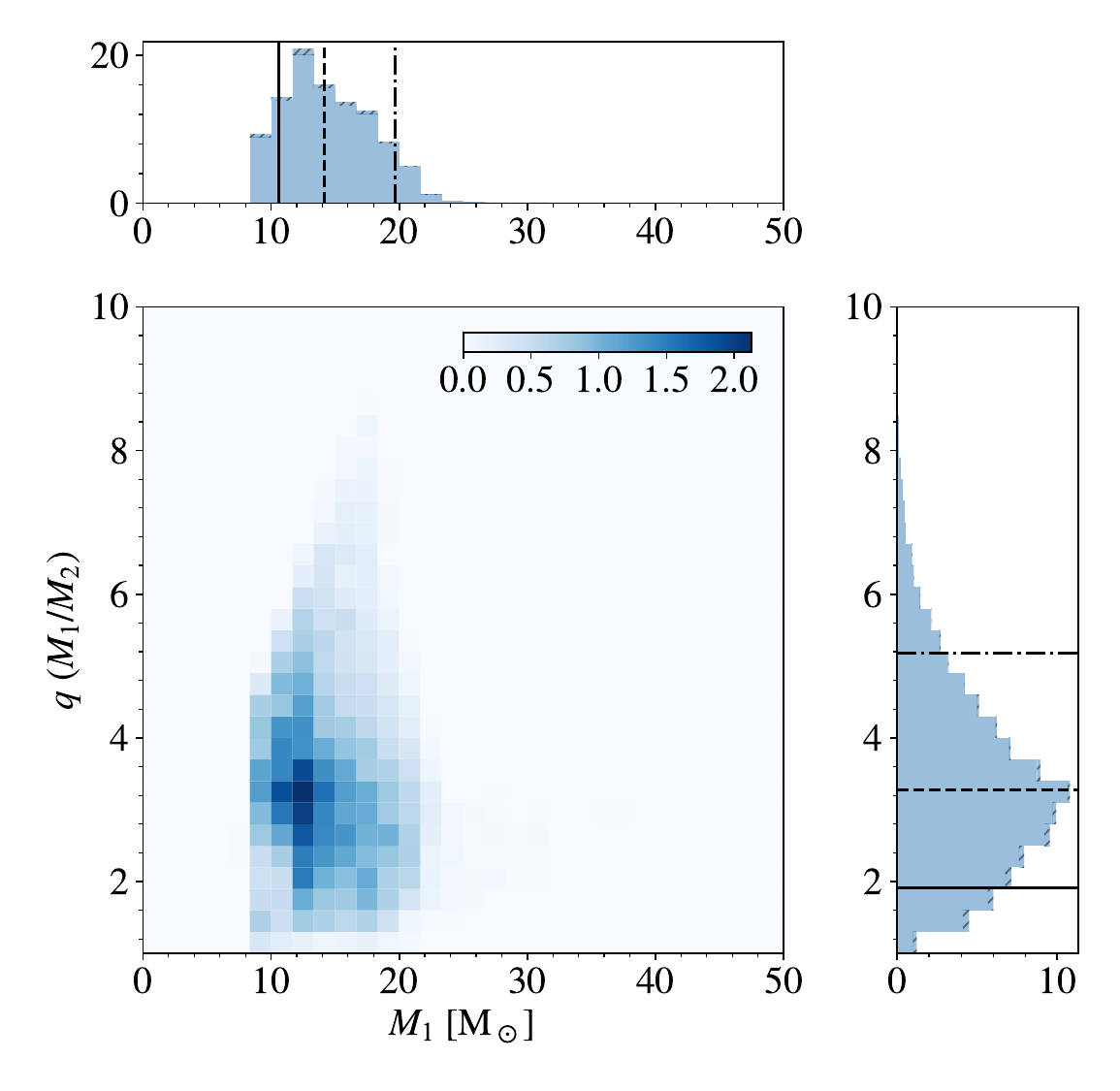}
    \caption{Same as Fig.~\ref{fig:P_m_V_smc} but for initial primary mass $M_1$, i.e. the NS progenitor, and initial mass ratio $q=M_1/M_2$ for progenitors of observable BeXRBs.}
    \label{fig:q_M1_smc}
\end{figure}
%%%%%%%%%%%%%%%%%%%%%%%%%%%%%%%%%

Fig.~\ref{fig:q_M1_smc} shows the initial primary mass $M_1$, i.e. the progenitor of the NS, and the initial mass ratio $q$ for systems that evolve into observable BeXRBs. The distribution indicates that BeXRBs predominantly originate from binaries with primary masses in the range $\sim 10-20\,{\rm M_\odot}$. Hence, the SMC BeXRB population probes the physics of massive binaries in a mass regime complementary to that explored by \citet{Lechien2025}, who focused on systems with initial masses $\sim 6-10\,{\rm M_\odot}$ as progenitors of Be+sdOB binaries. The present work therefore extends previous constraints on MT efficiency and AM evolution to higher-mass binaries that are more directly connected to compact-object formation.

Improved measurements of eccentricities, Be star masses, and rotational velocities will enable a more stringent test of the physical mechanisms governing massive binaries evolution.

\subsection{Models incompatible with observations}\label{app:SMC_bad_models}

%%%%%%%%%%%FIGURE%%%%%%%%%%%%%%%%
\begin{figure}[h!]
    \centering
    \includegraphics[width=0.5\textwidth]{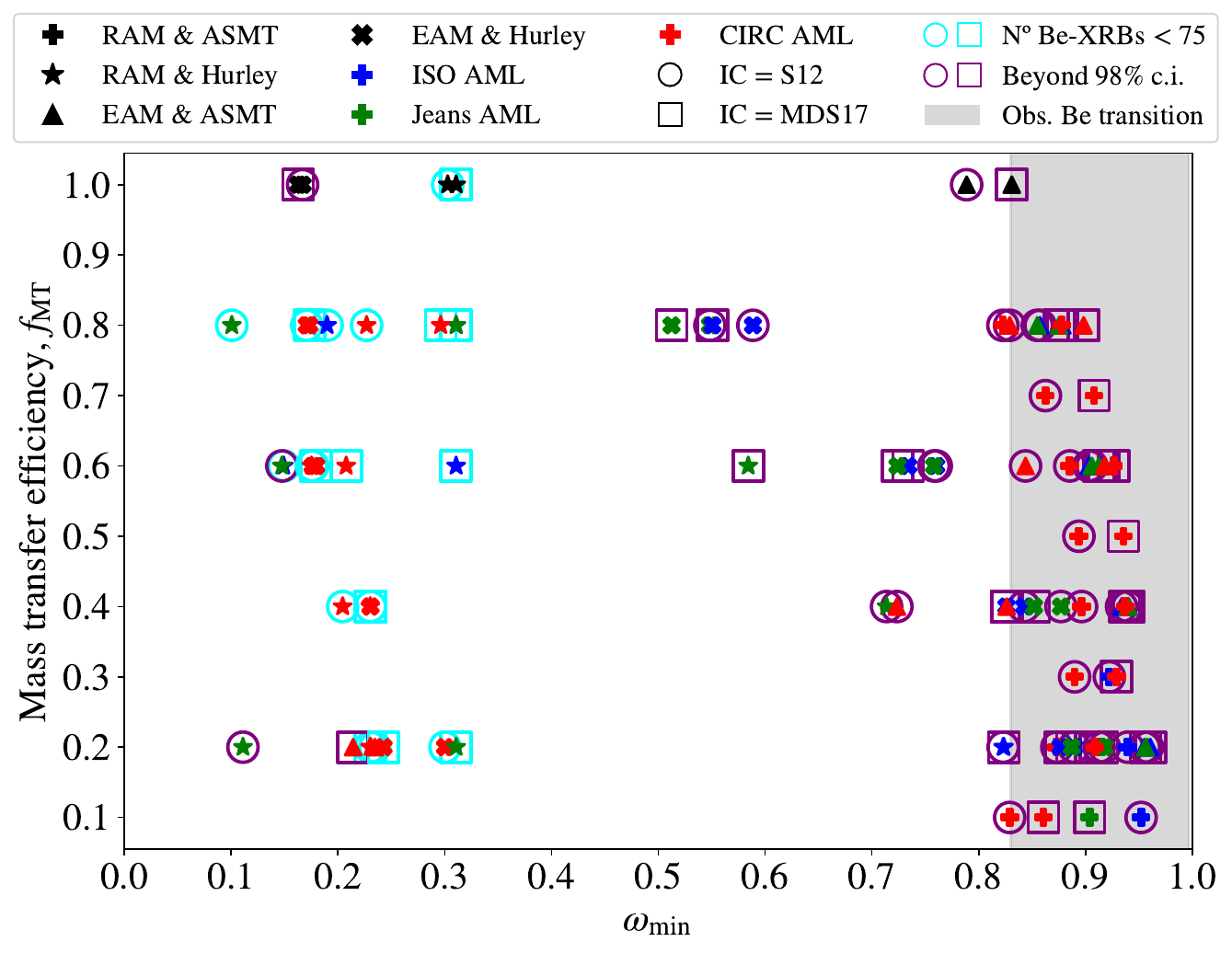}
    \caption{Spin threshold $\omega_{\rm min}$ and $f_{\rm MT}$ of models unable to reproduce the SMC BeXRB population considering the $P_{\rm orb}$ distribution. As in Fig.~\ref{fig:good_smc}, the marker shape encodes the combination of MT stability and AM feedback onto the orbit and the marker color identifies the specific AM loss channel associated with non-accreted mass. For fully conservative MT ($f_{\rm MT}=1$), the AM loss prescription is irrelevant; these models are therefore shown with black symbols. The outer symbol shape identifies the initial conditions, with circular outlines for \citetalias{Sana2012} and squared  outlines for \citetalias{Moe_DiStefano2017}. Points with cyan outer contour  are models that form less than 75 BeXRBs. Points with purple outer contour are outside  the 98\% confidence interval for the $P_{\rm orb}$ distribution, combined with the total number of systems. The grey shaded regions correspond to the minimum spin required to become Be stars from observations (see sec. 3.1 of \citealp{Rivinius2013} and \citealp{Huang2010}).}
    \label{fig:bad_porb_smc} 
\end{figure}
%%%%%%%%%%%%%%%%%%%%%%%%%%%%%%%%%

Fig.~\ref{fig:bad_porb_smc} displays the spin threshold $\omega_{\rm min}$ and MT efficiency $f_{\rm MT}$ for models that are incompatible with the observed SMC BeXRB $P_{\rm orb}$ distribution. 
Two different classes of incompatible models emerge. First, models with $\omega_{\rm min} < 0.5$ systematically underproduce BeXRBs, even accounting for uncertainties in the SMC SFR. These configurations are typically associated with the modified Hurley RLOF stability criterion and, in most cases, with a circumbinary ring AM loss prescription. The low threshold spin allows a large fraction of post-MT systems to enter the Be phase; however, the accompanying orbital evolution and AM loss treatment prevent the formation of a sufficient number of systems within the observable parameter space. 

Second, models with $\omega_{\rm min} > 0.8$ can reproduce the total number of BeXRBs and remain consistent with observational constraints on near-critical rotation, yet fail to match the observed orbital period distribution. In these cases, the predicted $P_{\rm orb}$ distribution is statistically inconsistent with the data despite acceptable number counts. As discussed in Section~\ref{sec:discussion}, this tension arises from the interplay between MT efficiency, AM feedback, and post-SN orbital widening, which together shape the period distribution.

\subsection{Impact of the propeller model}\label{app:SMC_prop}

%%%%%%%%%%%%FIGURE%%%%%%%%%%%%%%
\begin{figure}
    \centering
    \includegraphics[width=0.5\textwidth]{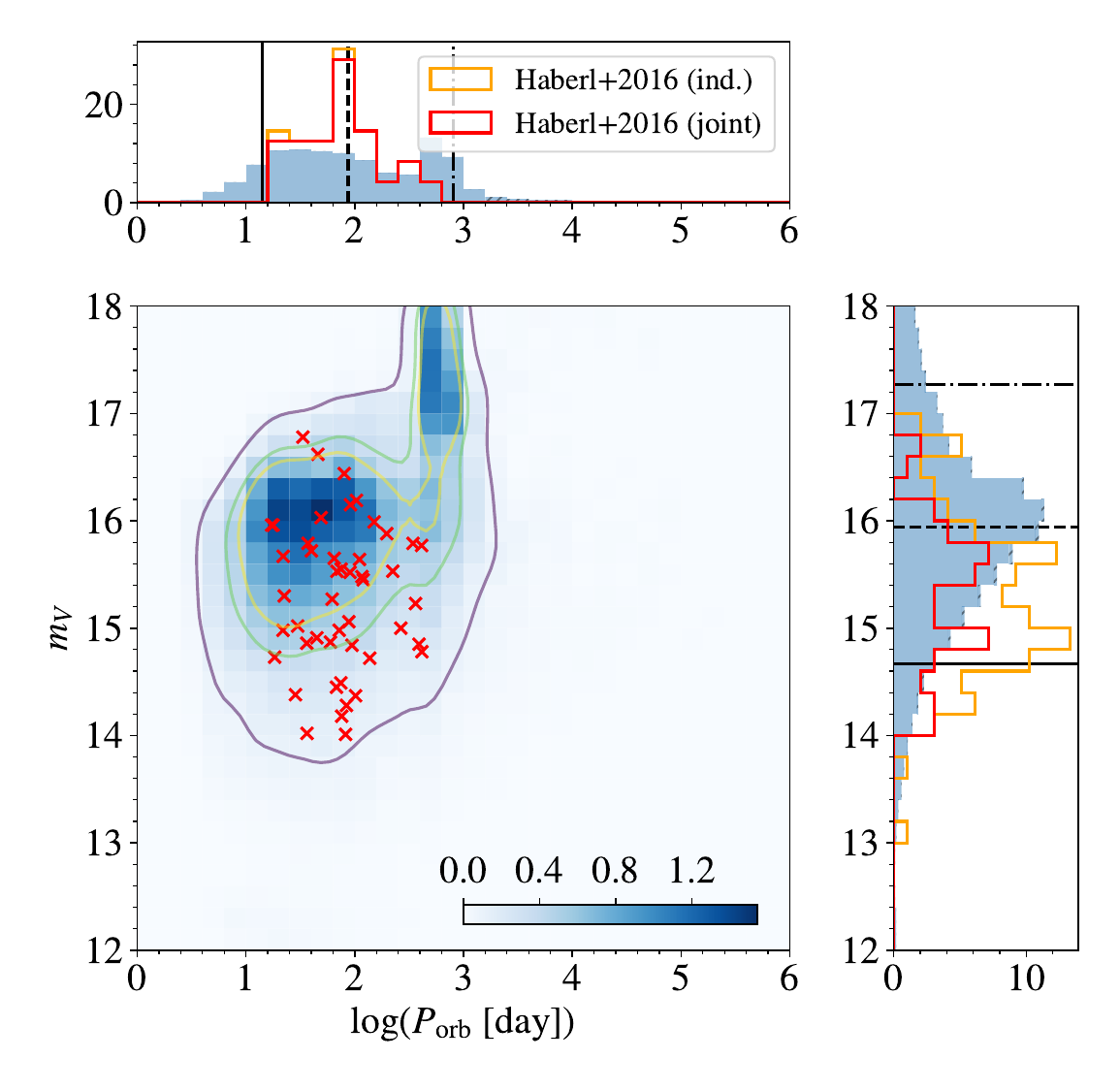}
    \caption{Same as Fig.~\ref{fig:P_m_V_smc} but considering $f_{\rm prop}=10^{-6}$.}
    \label{fig:m_V_P_fprop}
\end{figure}
%%%%%%%%%%%%%%%%%%%%%%%%%%%%%%%%%

%%%%%%%%%%%%FIGURE%%%%%%%%%%%%%%
\begin{figure}
    \centering
    \includegraphics[width=0.5\textwidth]{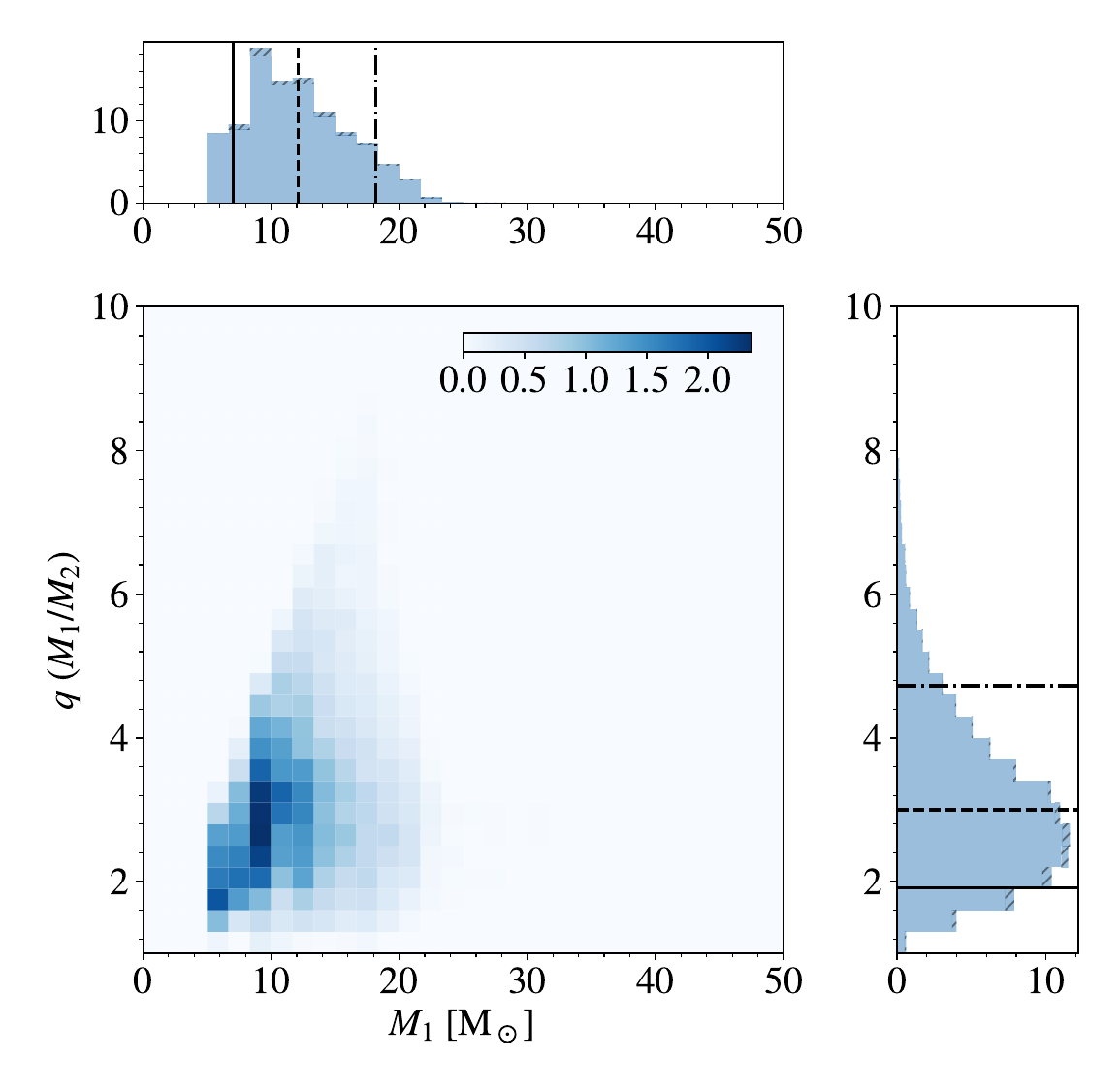}
    \caption{Same as Fig.~\ref{fig:q_M1_smc} but considering $f_{\rm prop}=10^{-6}$.}
    \label{fig:q_M1_fprop}
\end{figure}
%%%%%%%%%%%%%%%%%%%%%%%%%%%%%%%%%

Fig.~\ref{fig:m_V_P_fprop} illustrates the effect of relaxing the propeller prescription. It is analogous to Fig.~\ref{fig:P_m_V_smc} but adopts a less strict treatment with $f_{\rm prop} = 10^{-6}$. A subpopulation emerges at long orbital periods and faint $m_V$. The latter corresponds to low-mass Be stars, what we find is associated with lower initial primary masses as can be seen in Fig.~\ref{fig:q_M1_fprop}. These systems are characterized by low MT rates and would lie in the propeller regime under a fully suppressive prescription. Allowing a small residual accretion fraction enables them to remain above the adopted X-ray luminosity threshold, thereby entering the observable sample.

As $f_{\rm prop}$ increases further, this long-period, low-luminosity subpopulation progressively dominates the synthetic catalog. To recover the 102 observed SMC systems, we must then raise $\omega_{\rm min}$ accordingly; however, doing so shifts the main population away from the observed joint $P_{\rm orb}-m_V$ distribution. Consequently, the models that were previously consistent with the observations become statistically disfavored. This behavior underscores that the inferred constraints on MT efficiency and threshold spin are not independent of the adopted accretion physics: the propeller effect directly regulates the visibility of wide, marginally accreting systems. 

\subsection{Impact of RLOF stability criteria}\label{app:RLOF_stability}

As discussed in the main text, our grid explores two prescriptions for the stability of RLOF: (i) an always-stable model, in which all MT episodes are assumed to remain stable, and (ii) the modified Hurley prescription adopted by \citet{Iorio2023}, in which MT from main-sequence and Hertzsprung-gap donors is always stable, whereas more evolved donors follow the phase-dependent critical mass-ratio criteria from \citet{Hurley2002}. We found that the latter systematically underpredicts the observed number of BeXRBs in the SMC, even when accounting for the uncertainties in the SFR.

To further investigate this result, we have considered two additional stability models. In the first one, referred as \texttt{qc\_fixed}, a constant critical mass ratio $q_{\rm c}$ is adopted independently of the evolutionary stage of the donor. In the second, referred as \texttt{Hurley\_qc\_fixed}, MT from main-sequence and Hertzsprung-gap donors is always stable, while more evolved donors follow the same fixed $q_{\rm c}$. Motivated by previous detailed and rapid binary-evolution studies, we considered $q_{\rm c}=2$ \citep{deMink2007,Shao2014}, $q_{\rm c}=3.3$ \citep{Xu2025}, and $q_{\rm c}=5$ as an optimistic value \citep[][]{Ge2015,Ge2020,Langer2020,Echeveste2026}\footnote{The mass ratio ($M_{\rm BH}/M_{\rm OB}$) distribution of BH-OB star binaries obtained by \citet[see their fig.~4]{Langer2020} with highly non-conservative MT ($f_{\rm MT}\rightarrow 0$) implies a maximum $q_{\rm c}\sim 3\times M_{\rm BH}/M_{\rm OB}\sim 5$. On the other hand, \citet[see their figs.~8 and 9]{Ge2015} found that fully conservative MT ($f_{\rm MT}=1$) has $q_{\rm c}\gtrsim 5$ for $M_{\rm donor}\gtrsim 10\ \rm M_\odot$ and $R_{\rm donor}\gtrsim 300\ \rm R_\odot$ before the giant phase.  
MT is generally less stable during the giant phase, but $q_{\rm c}\gtrsim 5$ still holds for $M_{\rm donor}\gtrsim 20\ \rm M_\odot$ \citep[see their figs.~6 and 7]{Ge2020}. These studies indicate that highly stable MT with $q_{\rm c}\sim 5$ is plausible in the evolution of BeXRB progenitors, although they consider more metal-rich stars in the Large Magellanic Cloud and the MW \citep[which are more/less likely to undergo stable MT before/during the giant phase,][]{Ge2023} compared with our SMC stars and do not cover intermediate MT efficiencies $f_{\rm MT}\sim 0.2-0.7$ favored by our analysis. Recently, \citet{Echeveste2026} studied the dependence of $q_{\rm c}$ on $f_{\rm MT}$ and AM loss prescription, finding $q_{\rm c}>5$ for Case A/B MT with $f_{\rm MT}\lesssim 0.5$ and Jeans AM loss.}.

\begin{figure}
    \centering
    \includegraphics[width=0.5\textwidth]{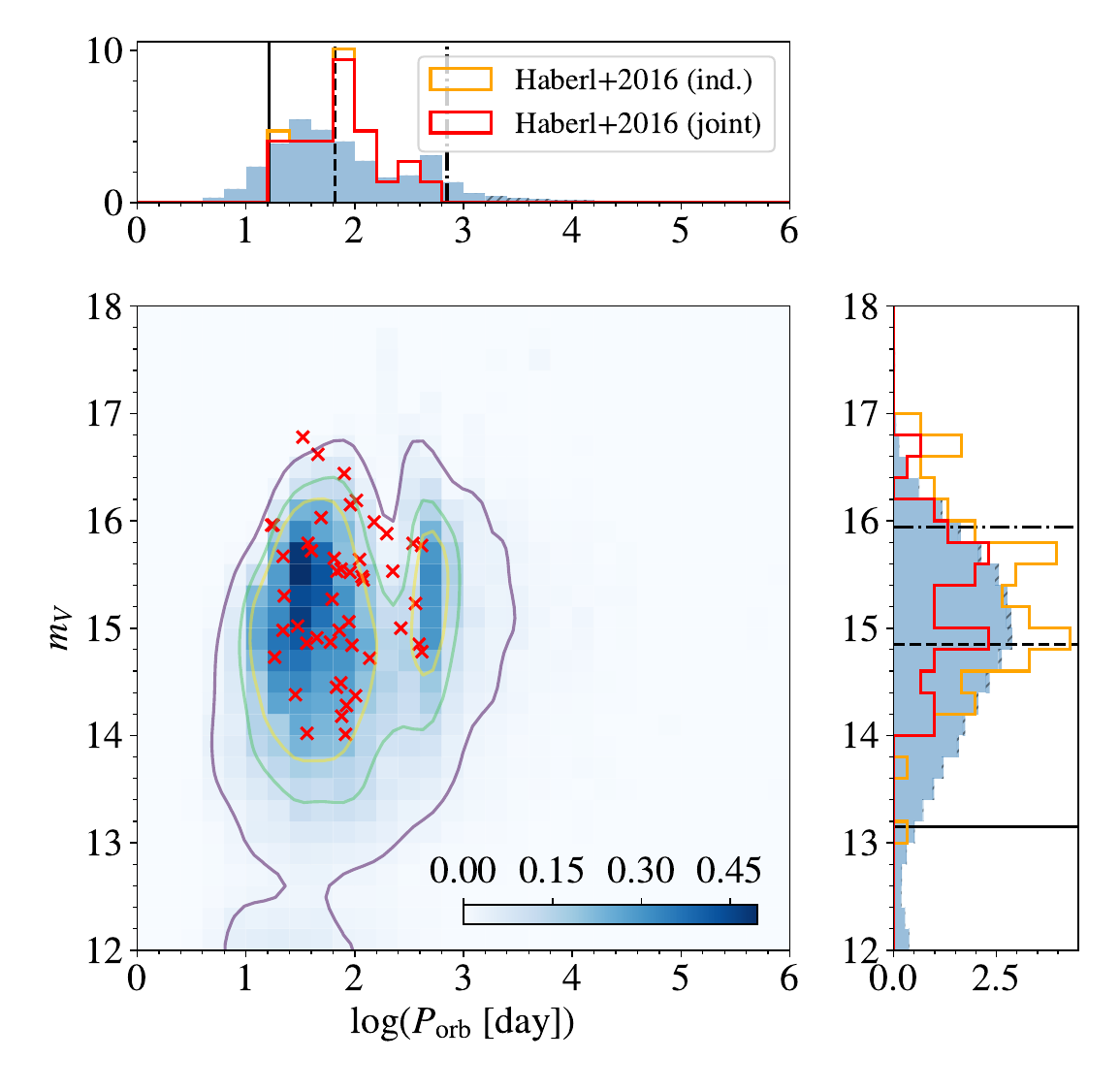}
    \caption{Same as Fig.~\ref{fig:P_m_V_smc} but considering the \texttt{qc\_fixed} RLOF stability model with $q_{\rm c}=3.3$. In this case we are considering the $\omega_{\rm crit}$ from the overall best-fitting model for the SMC.}
    \label{fig:m_V_P_qc}
\end{figure}

For the overall best-fitting SMC model, the predicted number of BeXRBs decreases significantly when more restrictive stability criteria are adopted. Relative to the always-stable model, the total number of systems is reduced by approximately $\sim$96\%, $\sim$68\%, and $\sim$19\% for the \texttt{qc\_fixed} prescription with $q_{\rm c}=2$, 3.3, and 5, respectively. For the \texttt{Hurley\_qc\_fixed} prescription, the corresponding reductions are $\sim$58\%, $\sim$38\%, and $\sim$14\%. In addition, lower values of $q_{\rm c}$ systematically shift the predicted BeXRB population towards longer orbital periods and higher donor masses for the \texttt{qc\_fixed} model, as can be seen in Fig.~\ref{fig:m_V_P_qc}, while the \texttt{Hurley\_qc\_fixed} model mostly modifies the total number of systems.

%%%%%%%%%%%%FIGURE%%%%%%%%%%%%%%
%%%%%%%%%%%%%%%%%%%%%%%%%%%%%%%%%

The origin of this strong sensitivity can be understood from the progenitor properties of BeXRBs. Fig.~\ref{fig:q_M1_smc} shows that most systems originate from binaries with initial primary masses of $\sim10$--$20\,\rm M_\odot$ and initial mass ratios around $q=M_1/M_2\simeq2-5$ in our best-fitting model with always stable MT. The preferred mass ratio range lies precisely in the regime where the outcome is most sensitive to the adopted stability criterion. 
The preference for unequal-mass progenitors with $q>2$ is physically motivated by the subsequent evolution of the system. 
Binaries with small $q=M_1/M_2$, and thus, massive initial secondaries (which later become comparatively massive Be stars) experience strong orbital widening 
due to mass-ratio reversal during RLOF. 
Furthermore, the propeller effect suppresses accretion immediately after the NS formation. Since more massive stars evolve more rapidly, many systems with massive Be stars die before efficient accretion onto their NS companions can occur. The combination of 
orbital widening by mass-ratio reversal, the propeller effect, and Be star lifetime therefore suppresses the contribution from binaries with initially small mass ratios to the BeXRB population. 

The shift toward longer orbital periods observed for decreasing $q_{\rm c}$ in the \texttt{qc\_fixed} model can be understood from the different MT channels leading to BeXRB formation. Close binaries typically initiate RLOF while the donor is still on the main sequence (Case A) or the Hertzsprung gap (Case B), 
whereas wider systems more commonly experience MT in later stages. 
When a fixed $q_{\rm c}$ is applied to all donor evolutionary stages, a significant fraction of close binaries undergo unstable Case A/B MT and merge, particularly for low values of $q_{\rm c}$, while the surviving binaries with $q<q_{\rm c}$ are widened by mass-ratio reversal. This survivor bias shifts the the $P_{\rm orb}$ distribution of BeXRBs towards longer periods. The surviving binaries also contain more massive Be stars from their more massive initial secondaries, as can be seen from their $m_V$ distribution. 
On the other hand, the widening by mass-ratio reversal is much weaker for binaries surviving from post-Hertzsprung-gap MT, in which the fraction of donor mass transferred is smaller due to its limited envelope mass. 
Therefore, when MT from main-sequence and Hertzsprung-gap donors is assumed to remain stable, most Case A/B progenitors survive regardless of the adopted $q_{\rm c}$, and the orbital-period distribution is only weakly affected, with the main impact being a reduction in the total number of systems through unstable post-Hertzsprung-gap MT.

%%%%%%%%%%%%FIGURE%%%%%%%%%%%%%%
\begin{figure}
    \centering
    \includegraphics[width=0.5\textwidth]{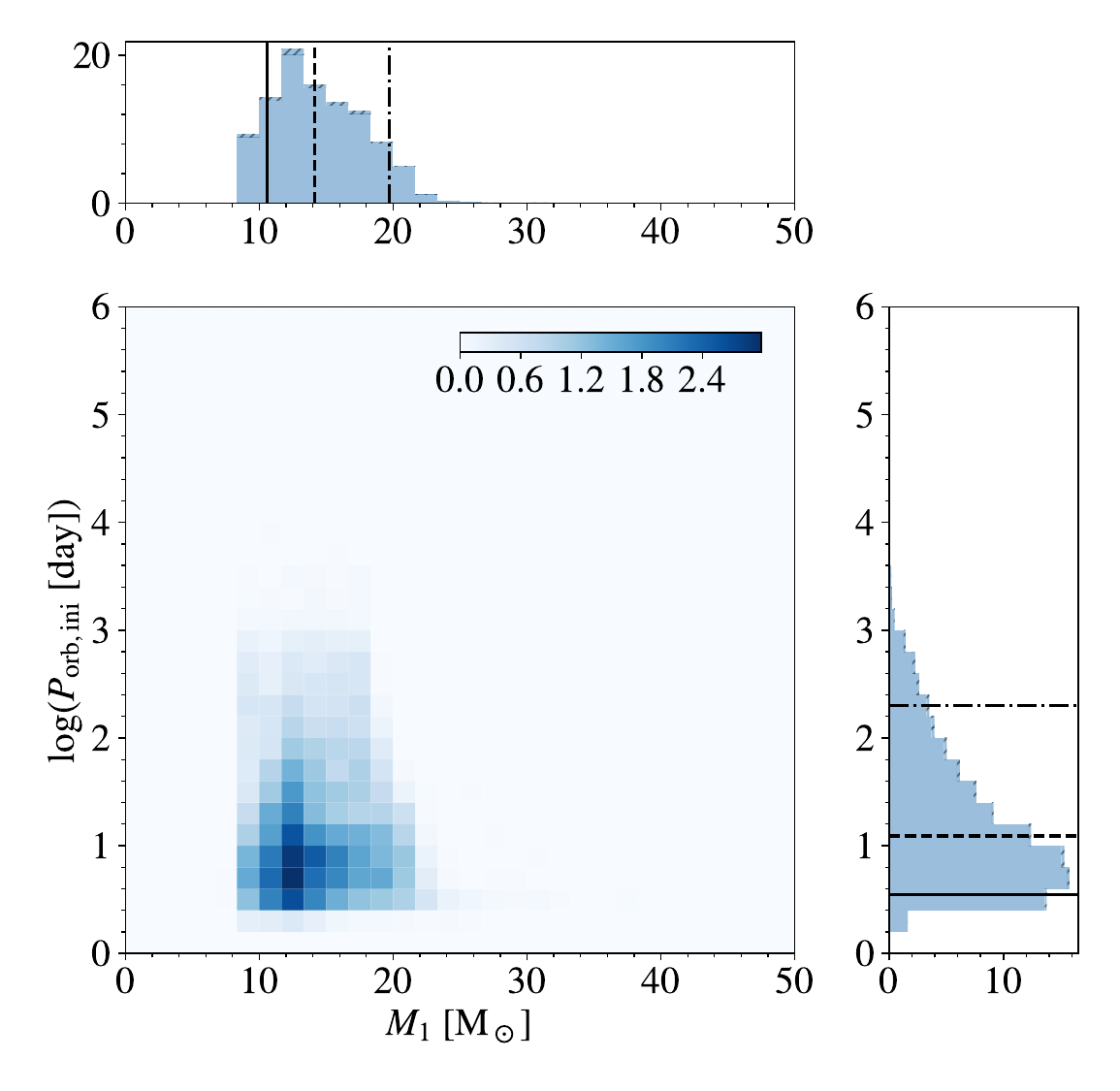}
    \caption{Same as Fig.~\ref{fig:P_m_V_smc} but for initial primary mass $M_1$, i.e. the NS progenitor, and initial orbital period $P_{\rm orb, ini}$ for progenitors of observable BeXRBs.}
    \label{fig:Porb_M1_smc}
\end{figure}
%%%%%%%%%%%%%%%%%%%%%%%%%%%%%%%%%

Finally, Fig.~\ref{fig:Porb_M1_smc} presents the distribution of initial primary mass and initial orbital period for the progenitors of BeXRBs. Compared with the orbital period distribution of BeXRBs that peaks at $P_{\rm orb}\sim 20$~days (see Fig.~\ref{fig:P_m_V_smc}), the distribution of their progenitors is clearly biased by close binaries with a peak at $P_{\rm orb}\sim 5$~days. This indicates that most BeXRBs originate from relatively close binaries that subsequently widen during RLOF by AM recycle and mass ratio reversal.

\subsection{Impact of alternative initial conditions}\label{app:SMC_IC}

In addition to the initial conditions adopted in the main analysis, we explored an alternative prescription to assess the robustness of our conclusions against uncertainties in the primordial binary population. We considered 
the initial conditions of \citet{Sana13} based on observations of the Large Magellanic Cloud. Similar to the case of \citetalias{Sana2012}, we extend the initial orbital period range from $P_{\rm orb}\in [10^{0.15},10^{3.5}]$~days to $P_{\rm orb}\in[10^{0.15}, 10^{5.5}]$~days,  
thus enhancing the binary fraction from 51\% to $69\%$. 

\begin{figure}
    \centering
    \includegraphics[width=0.5\textwidth]{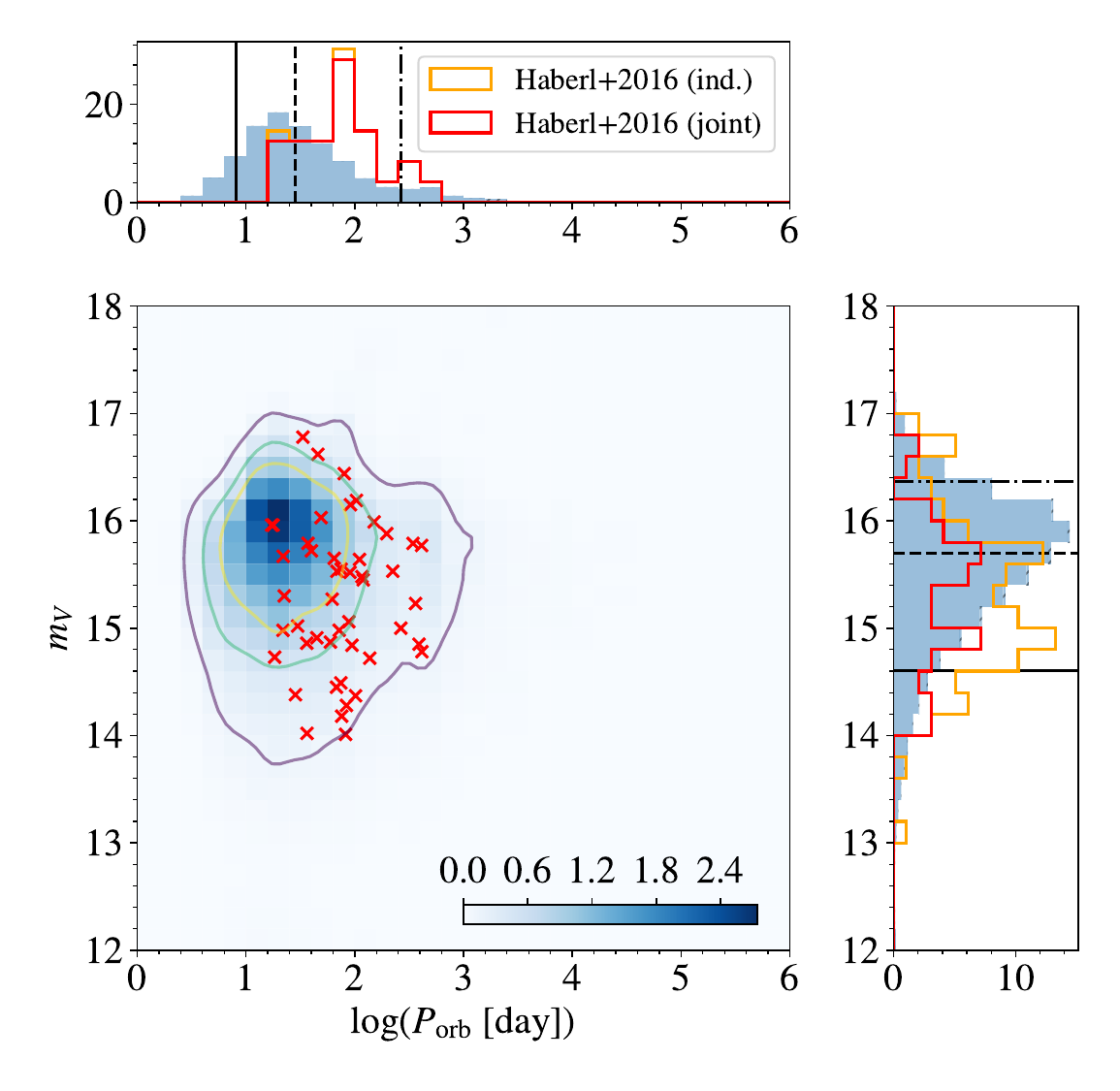}
    \caption{Same as Fig.~\ref{fig:P_m_V_smc} but considering the initial conditions from \citet{Sana13}.}
    \label{fig:m_V_P_sana13}
\end{figure}

For the overall best-fitting model, we find that the initial conditions of \citetalias{Sana2012} provide a better overall agreement with observations than those of \citet{Sana13}. 
Moreover, adopting the initial conditions of \citet{Sana13} leads to a larger population of BeXRBs, increasing the number of systems by $\sim50\%$ relative to \citetalias{Sana2012}. 
The enhancement originates from a greater fraction of systems that experience efficient MT and spin-up. As a result, with the initial conditions of \citet{Sana13}, a higher threshold spin, $\omega_{\rm min}=0.878$ instead of $\omega_{\rm min}=0.842$ for the overall best-fitting model, is required to match the observed sample size. Fig.~\ref{fig:m_V_P_sana13} shows the distribution of BeXRBs in the $P_{\rm orb}-m_V$ space for the initial conditions from \citet{Sana13}, analogous to Fig.~\ref{fig:P_m_V_smc} for \citetalias{Sana2012}. The distribution is very similar to that for \citetalias{Sana2012}, except for a slightly lower fraction of very luminous ($m_V\lesssim 14$) systems.

%%%%%%%%%%%%FIGURE%%%%%%%%%%%%%%
%%%%%%%%%%%%%%%%%%%%%%%%%%%%%%%%%

The primary difference between the two initial conditions sets lies in the mass ratio distribution, parameterized as $f_q \propto q^{\kappa}$. In \citetalias{Sana2012}, $\kappa=-0.1$, corresponding to an almost flat distribution in $q$, whereas \citet{Sana13} adopt $\kappa=-1.0$, favoring low mass ratio systems. In our best-fitting SMC models, the initial mass ratio distribution of progenitors that evolve into BeXRBs has a minimum $q \sim 1.5$, peaks around $q \sim 3.5$, and exhibits a long tail extending to $q \sim 8$ (see Fig.~\ref{fig:q_M1_smc}). Because \citet{Sana13} intrinsically produce more low-$q$ systems, they naturally yield a higher number of binaries entering the BeXRB formation channel. The less massive secondary stars in the \citet{Sana13} initial conditions also reduces the fraction of massive/luminous Be stars in BeXRBs. 

A more exhaustive exploration of the initial conditions from \citet{Sana13} and other initial conditions 
would require a dedicated study and is beyond the scope of the present work. Here, we conclude that 
the main physical constraints derived in this paper 
remain robust against reasonable variations in the initial binary population.

\section{Results for the MW}\label{app:MW}

The observational characterization of BeXRBs in the MW is  less complete than in the SMC. Selection effects associated with distance uncertainties, non-uniform sky coverage, and spatially variable interstellar absorption can introduce strong biases in the observed sample. In addition, donor-star masses are typically inferred from spectral types rather than obtained through direct dynamical measurements, which adds systematic uncertainties in the mass distribution. On the other hand, the more complex nature of the MW precludes a reliable computation of $V$-band magnitudes and existing observational catalogs of Galactic BeXRBs \citep[e.g.,][]{Fortin23} do not provide systematically tabulated $V$-band magnitudes. These limitations hinder a fully consistent  comparison equivalent to that performed for the SMC. Hence, we restrict the comparison to the orbital period distribution, which is the most robustly measured observable. 

We examine whether the models that successfully reproduce the SMC population are also able to describe the MW BeXRB sample. For each binary-evolution model, we adopt the value of $\omega_{\rm min}$ calibrated from the SMC analysis and follow the same statistical procedure described in Section~\ref{sec:obs_constrains} and Appendix~\ref{app:likelihood}. The observational comparison is performed using the catalog compiled by \citet{Fortin23}\footnote{An online version of the catalog is publicly available at \url{https://binary-revolution.github.io/HMXBwebcat/downloads.html}. There are 74 BeXRB candidates in the catalog with class labels of `Be', `$\gamma$~Be' and 'Be/sg'. Among them, 55 binaries with measured orbital periods $P_{\rm orb}<t_{\rm obs}=1.6\times10^3\ \rm days$ are considered in the PDF comparison.}. Although the SFH of the MW is complex and spatially structured, we assume for simplicity a constant SFR of $2\,{\rm M_\odot\,yr^{-1}}$. This choice primarily affects the absolute number of predicted systems, which plays a secondary role in the present orbital period comparison. For the present study we use stellar tracks representative of MW metallicity with $Z=Z_\odot=0.0142$ \citep{Asplund2009}.

%%%%%%%%%%%%FIGURE%%%%%%%%%%%%%%
\begin{figure*}
    \centering
    \includegraphics[width=1\textwidth]{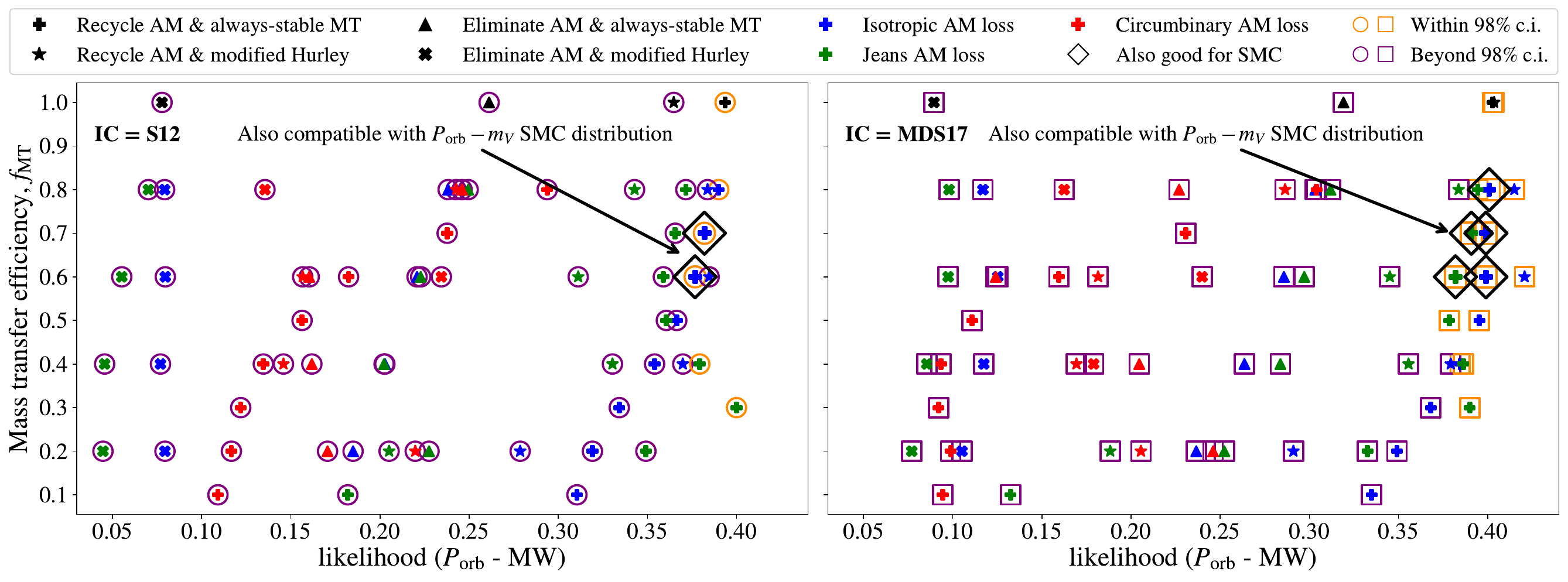}
    \caption{Same as Fig.~\ref{fig:good_smc} but for the likelihood of the $P_{\rm orb}$ distribution of all models producing more than 74 BeXRBs \citep{Fortin23} for the MW. Models with a black diamond contour denote models that are in statistical agreement with both the $P_{\rm orb}$ distribution of the MW and the joint $P_{\rm orb}-m_V$ distribution further combined with the total number of the SMC (see Fig.~\ref{fig:good_smc}).}
    \label{fig:good_mw_joint_smc}
\end{figure*}
%%%%%%%%%%%%%%%%%%%%%%%%%%%%%%%%%

Fig.~\ref{fig:good_mw_joint_smc} is the same as Fig.~\ref{fig:good_smc}, but for the MW case. It shows the likelihood distribution for models that produce more than 74 BeXRBs \citep{Fortin23}. The threshold is chosen to ensure consistency with the observed sample size, acknowledging the incompleteness of the MW catalog. In the same figure, we highlight with a black diamond contour those models that simultaneously reproduce the MW orbital period distribution and, simultaneously, the joint $P_{\rm orb}-m_V$ distribution and total number of the SMC population.

Despite the intrinsic differences between the two galaxies (metallicity, SFH, and observational completeness), we find a non-zero overlap between the best-fitting MW models and the best-fitting SMC models. Remarkably, the overlapping region corresponds to models that were already identified as successful in reproducing the joint SMC constraints. 
This result indicates that the physical prescriptions required to describe the SMC BeXRB population are also compatible with the orbital period distribution of MW BeXRBs.

We therefore conclude that our SMC best-fitting models provide a consistent description of both galaxies. In particular, the combination of moderately non-conservative MT, stable RLOF, AM recycling through tides, and low SN natal kicks is not only sufficient to explain the SMC population but also reproduces the orbital properties of MW BeXRBs

%%%%%%%%%%%%FIGURE%%%%%%%%%%%%%%
\begin{figure}
    \centering
    \includegraphics[width=0.5\textwidth]{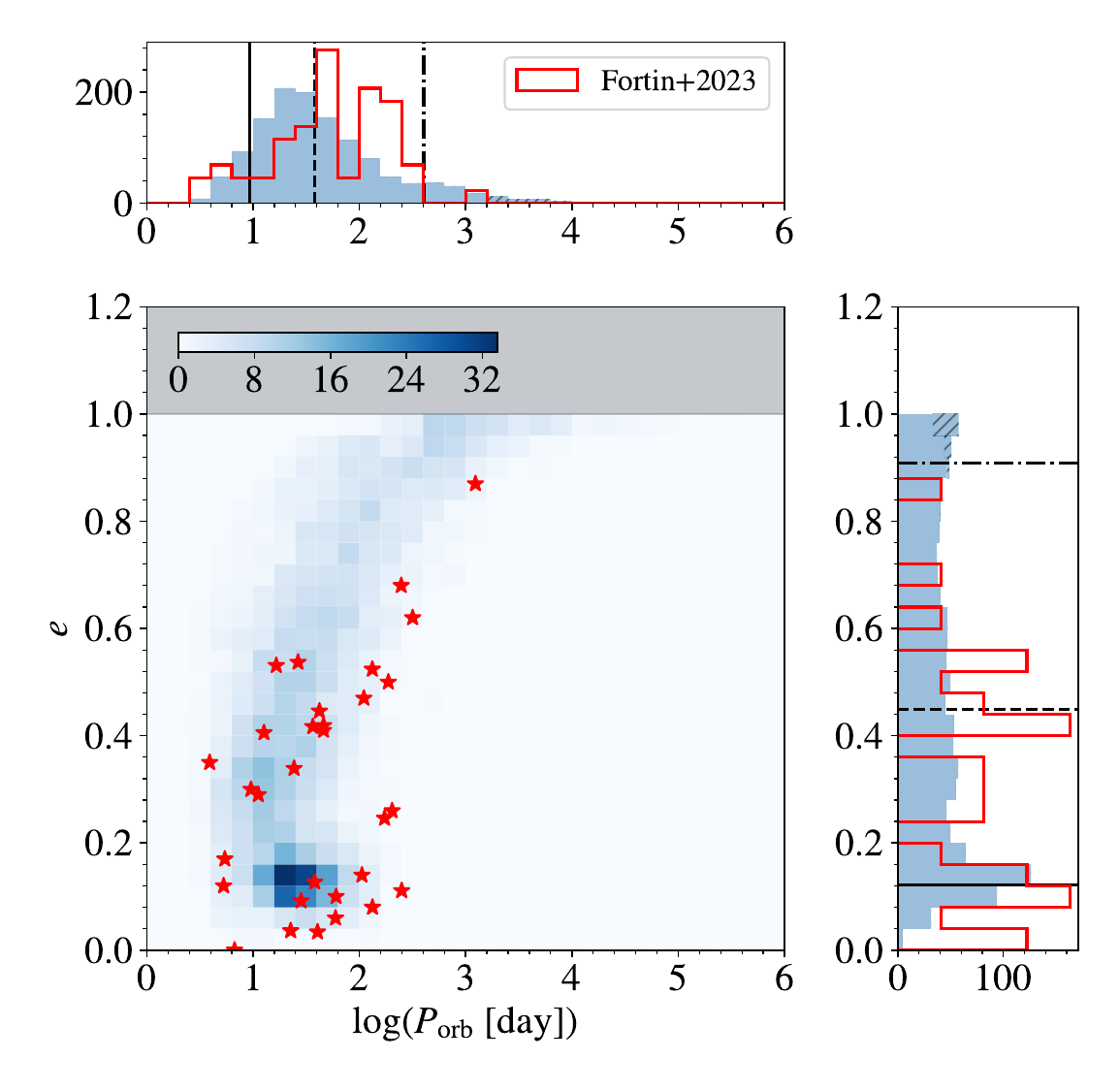}
    \caption{Same as Fig.~\ref{fig:P_m_V_smc} but for $P_{\rm orb}$ and eccentricity $e$  of BeXRBs
    in the MW. The  data are  from \citet{Fortin23}.}
    \label{fig:P_e_MW}
\end{figure}
%%%%%%%%%%%%%%%%%%%%%%%%%%%%%%%%%

Fig.~\ref{fig:P_e_MW}  shows the predicted orbital period--eccentricity ($P_{\rm orb}-e$) distribution for the MW under the SMC overall best-fitting model. The synthetic population exhibits a morphology similar to that obtained for the SMC (Fig.~\ref{fig:P_e_smc}): a concentration of low-eccentricity systems at short-to-intermediate orbital periods and a branch extending toward higher eccentricities at longer periods. This structure arises from the distribution of SN natal kicks and pre-explosion orbital separations. Systems experiencing small kicks remain nearly circular, whereas larger kicks induce both eccentricity growth and orbital expansion. 

%%%%%%%%%%%%FIGURE%%%%%%%%%%%%%%
\begin{figure}
    \centering
    \includegraphics[width=0.5\textwidth]{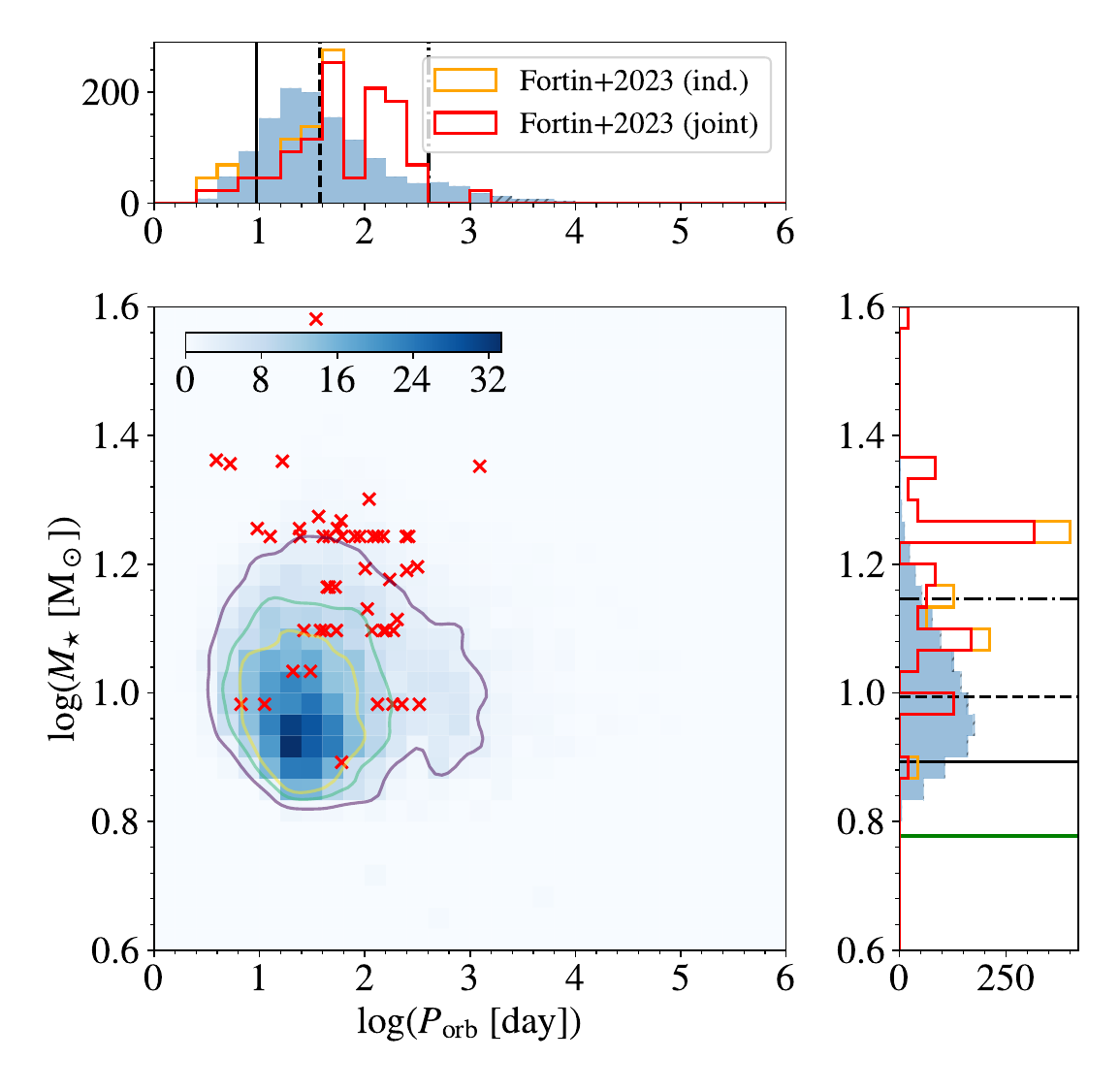}
    \caption{Same as Fig.~\ref{fig:P_m_V_smc} but for $P_{\rm orb}$ and mass of the Be star $M_\star$. 
    The data are from \citet{Fortin23}. The green line at 6 M$_\odot$ in the marginalized distribution shows the minimum Be star mass in BeXRBs from observations \citep{Hohle2010,Coe2015,Haberl16,Fortin23}.}
    \label{fig:P_M_MW}
\end{figure}
%%%%%%%%%%%%%%%%%%%%%%%%%%%%%%%%%

Fig.~\ref{fig:P_M_MW} displays the orbital period–Be star mass ($P_{\rm orb}-M_{\star}$) distribution predicted for the MW. Most observational mass estimates are derived from spectral classifications, which explains the clustering of data points at discrete mass intervals. The model reproduces the overall mass range and the absence of a strong correlation between orbital period and Be star mass within the observed parameter space. In particular, as in Fig.~\ref{fig:P_M_smc}, all the Be stars in BeXRBs have masses above the minimum observed mass of 6 M$_\odot$ \citep{Hohle2010,Coe2015,Haberl16,Fortin23}. 

Under the assumption of a constant SFR of $2\,{\rm M_\odot\,yr^{-1}}$, our best-fitting models predict a total of $\sim1200$ BeXRBs currently present in the MW. This value should be interpreted as an order-of-magnitude estimate. A more realistic treatment including spatially resolved SFH and metallicity evolution would likely modify the exact number (and properties) of BeXRBs (Vivanco Cádiz et al in prep.). However, because the formation timescale of observable BeXRBs is short compared to Galactic chemical-evolution timescales, we do not expect drastic variations in the predicted present-day population.

Overall, the ability to reproduce both the SMC and the MW BeXRB populations with the same combination of binary-evolution parameters significantly strengthens our conclusions. The consistency across two galaxies with different metallicities and observational biases suggests that the inferred constraints on MT efficiency, AM evolution, propeller physics, and SN natal kicks reflect underlying physical processes rather than metallicity-specific systems.

\end{document}